%% file: main.tex
\input{preamble.tex}

\begin{document}
\pagenumbering{roman}
\begin{titlepage}
\singlespacing
\thispagestyle{empty}
\renewcommand{\thefootnote}{\fnsymbol{footnote}}
\begin{center}
{\LARGE The Financing of Debt Relief:\\[4pt] Intermediary Liquidity and\\[2pt] the Value of Borrower Information}\\[22pt]
\ifdefined\ANON\else
{\large Arka P. Bandyopadhyay\footnote{Tandon School of Engineering and Schack Institute of Real Estate, New York University; and Sy Syms School of Business, Yeshiva University. Email: \href{mailto:apb321@nyu.edu}{apb321@nyu.edu}. Phone: (347) 260-0869. I am grateful to Patrick Smith and Dongshin Kim for sharing the Ginnie Mae loan-level data they assembled and for their guidance and mentorship, from which I have benefited significantly. I thank Taha Ahsin, Linda Allen, Xudong An, Archishman Chakraborty, Thomas Chemmanur, Sean Flynn, Andreas Fuster, Erica Jiang, Samuel Kruger, Brittany Lewis, Lilia Maliar, Gonzalo Maturana, Amine Ouazad, Manisha Padi, Uday Rajan, Philipp Schnabl, James Vickery, Liuren Wu, and Yildiray Yildirim for helpful comments and discussions. The paper builds on companion work on servicers' relief decisions \citep{BandyopadhyayDelegated2026,BandyopadhyayInside2026,BandyopadhyayRelief2026} and on \citet{BKS2026}. The accompanying package includes model code, empirical analysis scripts, and aggregate outputs; loan-level inputs are not bundled (Internet Appendix~\ref{ia:verify}). Declarations of interest and of the use of AI-assisted technologies appear before the references. All errors are my own.}}\\[10pt]
{\normalsize October 2026}\\[2pt]
{\small Latest version: \href{https://arkabandyopadhyay.com}{arkabandyopadhyay.com}}
\fi
\end{center}
\vspace{18pt}
\begin{abstract}
\noindent I study how intermediary liquidity shapes cash-intensive mortgage resolution using 1.9 million Ginnie Mae buyout options. During the 2020 disruption, nonbank repurchase rates fell relative to depositories, and selection temporarily shifted against large balances. Repurchases fell approximately \$19 billion below a pre-pandemic-rate benchmark within three months; much of the gap subsequently closed. These patterns are consistent with liquidity rationing, although funding is not directly observed. A model shows that borrower information is most valuable at intermediate funding capacity and that predictive models should be evaluated at the intermediary's own capacity. Repurchase is a step toward resolution, not completed borrower relief.
\end{abstract}
\vspace{10pt}
\end{titlepage}
\doublespacing
\renewcommand{\thefootnote}{\arabic{footnote}}\setcounter{footnote}{0}
\clearpage
\pagenumbering{arabic}
\setcounter{page}{1}

\section{Introduction}
Debt relief can require an intermediary to supply cash before a borrower recovers. A mortgage servicer may need to advance missed payments, repurchase a delinquent loan, or fund a modification before receiving the proceeds of resolution. The intermediary can therefore know which actions are attractive and still lack the resources to take them. This paper studies how that financing constraint changes the allocation and timing of mortgage resolution, and what it implies for the value of borrower information.

The setting is the Ginnie Mae buyout market. After a borrower misses three payments, the issuer can repurchase the loan from its securitization pool at par. Repurchase ends the issuer's obligation to advance payments to security holders but requires the unpaid balance in cash. It can facilitate subsequent modification and redelivery. It does not itself change the borrower's obligations, and the data do not follow borrower outcomes after repurchase. The empirical object is therefore a cash-intensive step in resolution, rather than completed debt relief or its effect on borrower welfare.

I combine a model of portfolio allocation under a funding constraint with \eEmpN{} buyout options at \eEmpIssuers{} issuers from January 2019 to September 2022. The model ranks actions by private value per dollar of cash. When funding binds, a higher shadow value of cash excludes otherwise profitable actions and increases the penalty on cash-intensive opportunities in the exercise decision. When funding is slack, the intermediary instead declines actions with negative private value. These two regimes provide a framework for interpreting the different responses of nonbank and depository issuers to the pandemic and to subsequent changes in mortgage rates.

\begin{figure}[H]
\centering\includegraphics[width=\textwidth]{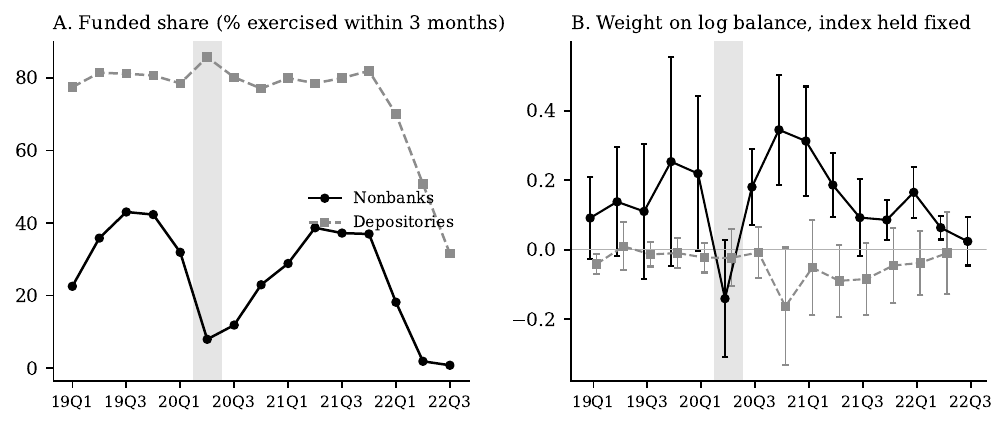}
\caption{Nonbank exercise temporarily shifted away from large balances.\label{fig:tilt}}
\begin{minipage}{0.97\textwidth}\footnotesize\setstretch{1.0}
Panel A plots, by quarter of vesting, the share of Ginnie Mae buyout options exercised within three months at nonbank and depository issuers. Panel B plots the weight on standardized log unpaid balance in a logit of exercise with decision-cell fixed effects (issuer $\times$ vesting month $\times$ forbearance status) that also includes the standardized ranking index; both weights are interacted with quarter indicators, and bars are 95\% confidence intervals clustered by issuer. Holding the ranking index fixed, a negative weight means that larger loans, which need more cash to repurchase, were less likely to be repurchased. The shaded quarter is 2020Q2. The depository estimate for 2022Q3, based on a handful of interior cells (s.e.\ 0.71), is omitted. Source: \texttt{code/empirics/12\_tilt\_quarterly.py} and \texttt{12b\_tilt\_figure.py}.
\end{minipage}
\end{figure}

The central empirical result concerns selection by loan balance. In the second quarter of 2020, nonbanks temporarily shifted exercise away from large-balance loans, conditional on an index of their earlier exercise decisions and on issuer, month, and observed forbearance status. Their coefficient on standardized log balance fell by \eBalNbDFrzAbs{} log-odds in March--June 2020 (s.e.\ \eBalNbDFrzSE), became negative, and recovered in the following quarter. Depositories did not exhibit the same change. Figure~\ref{fig:tilt} displays the quarterly pattern. Because a buyout requires the unpaid balance, this is consistent with an increase in the opportunity cost of cash. The coefficient is not a dollar-valued estimate of the shadow price: the index is a proxy for private value, and unobserved changes in value or collateral eligibility could also affect the balance gradient.

Three further findings help interpret that pattern. First, nonbanks' three-month exercise rate fell from \eNbPreM\% before the pandemic to \eNbFrzM\% in March--June 2020, while depositories' rate rose from \eDepPreM\% to \eDepFrzM\%. Within pool and month, the relative decline was \fPool{} percentage points (s.e.\ \fPoolSE; wild cluster bootstrap $p=\fPoolP$). This is a descriptive contrast: the issuer types did not follow parallel paths throughout the pre-period, and sharing a pool does not equalize loan risk or issuer-specific funding and continuation values.

Second, nonbanks continued to select among options. I estimate an exercise index using only their 2016--2018 decisions and compare subsequent selection with fixed-ranking and random-rationing benchmarks at each decision cell's observed exercise rate. The freeze alone does not distinguish a fixed ranking from a mixture of ranked and random rationing. The second half of 2020 is more informative: the relative response to the index was \eNbRelRel, close to the fixed-ranking prediction of \eNbRelRelPred{} and above the random-rationing benchmark of \eNbRelRelRand. Later exercise of the backlog also remained related to the index. These results concern selection on observed characteristics; they do not establish that the index measures borrower recovery probabilities or private information.

Third, much of the initial repurchase gap narrowed over time. Applying the pre-shock balance-weighted exercise rate to the freeze cohort implies \$\eDolNbShortThree{} billion more repurchases within three months than occurred. At a common horizon of 24 months, the analogous gap was \$\eDolNbShortTwentyfour{} billion. These are benchmark shortfalls, not estimates of lost borrower relief, social losses, or the causal effect of a funding intervention. The later repurchases nevertheless show that a low initial exercise rate need not represent permanent abandonment of the options.

Additional evidence is consistent with the funding interpretation but does not independently identify it. Among \fOrActN{} active nonbanks, greater 2019 origination volume relative to the servicing book predicts a smaller freeze decline in exercise. A log point of origination intensity is associated with a \fOrAFrz{} percentage-point smaller decline (s.e.\ \fOrAFrzSE; wild cluster bootstrap $p=\fOrAP$). The proxy is predetermined but not exogenous; the specification was selected after examining the data, and the association weakens when issuer size is controlled for ($p=\fOrSizeP$). In 2022, as mortgage rates rose, depositories reduced exercise and selected more strongly on the mortgage spread. That pattern is consistent with screening when repurchase values decline, although the data do not directly identify which options have negative value.

The model's main result explains why the value of information depends on funding capacity. With a common cash requirement and calibrated borrower-success probabilities, information helps allocate a scarce budget to the highest-value actions. Its incremental value is concave in the fundable share and peaks where that share equals the probability of receiving good news relative to the prior. Below the peak, information raises the marginal value of liquidity; above it, information lowers that marginal value. At full funding, only screening value remains: information allows the intermediary to decline actions that are unprofitable conditional on the signal. The result concerns a fixed opportunity set and funding technology, not a universal ordering of information across balance sheets.

The same framework gives a criterion for comparing predictive models. At a specified funding capacity, ranking value is proportional to the cumulative accuracy profile above its diagonal. When all actions remain profitable under the signal, AUC measures this value averaged uniformly across funding capacities, rather than value at the capacity of a particular intermediary. AUC can therefore select the wrong model for that intermediary. The empirical exercise does not test this proposition: its index predicts exercise, whereas the theoretical scores predict success under intervention. Extensions show how verifiable information can support lending against the selected portfolio and why private selection need not maximize borrower or total surplus.

The paper contributes to research on mortgage servicing frictions. Securitization, incentives, and organizational capacity affect loss mitigation \citep{PiskorskiSeruVig2010,AdelinoGerardiWillen2013,Agarwal2017,Maturana2017,Kruger2018}. Advancing obligations expose nonbank servicers to funding pressure \citep{KimEtAl2018}, and servicer frictions shape the delivery of forbearance \citep{KimLeeScharlemannVickery2024,CherryEtAl2021}. The closest comparison is \citet{Aiello2022}, who finds that financially constrained private-label servicers foreclose and modify more often. Those actions can end advances without requiring payment of the full loan balance. A Ginnie Mae buyout instead uses cash to end advances. The contrasting predictions follow from the financing of the action: the present paper studies selection and delay when resolution itself requires a large upfront outlay. The growth and regulation of nonbank intermediation make this distinction consequential \citep{BuchakEtAl2018,JiangEtAl2020,GeteReher2021,KimEtAl2022,HamdiEtAl2023}.

Within the buyout literature, \citet{Ahsin2026} studies the consequences of buyout eligibility for loan outcomes, while \citet{BKS2026} documents issuer selection and information in ordinary times. This paper focuses on how selection changes during funding stress and connects those changes to a theory of information value under a budget. It also relates to intermediary liquidity and internal allocation \citep{HolmstromTirole1997,HolmstromTirole1998,Stein1997,Stein2002,FrootScharfsteinStein1993}, financing-constrained real options \citep{BoyleGuthrie2003}, and information and pledgeability \citep{LibertiPetersen2019,MyersMajluf1984,DeMarzoDuffie1999}. The information results apply established Blackwell and convex-order methods \citep{Blackwell1953,ShakedShanthikumar2007}. The model-evaluation result supplies a funding interpretation of known AUC identities \citep{Hand2009,HernandezOralloFlachFerri2012}, complementing work on prediction and allocation in credit markets \citep{FusterEtAl2022,BergEtAl2020,FusterPlosserSchnablVickery2019,KleinbergEtAl2018}.

The policy implication is conditional. In the model, unpledgeable compensation paid after completion cannot relax a fixed cash budget, whereas cash available at the decision date can. The data support examining that distinction; they do not measure the effect or welfare benefit of a buyout-financing facility. Direct funding records and post-repurchase borrower outcomes would be needed for that evaluation.

Section~\ref{sec:setting} describes the setting. Sections~\ref{sec:model} and \ref{sec:information} develop the allocation and information results. Section~\ref{sec:evidence} presents the evidence and its limits. Section~\ref{sec:welfare} considers borrower surplus and policy, and Section~\ref{sec:conclusion} concludes. Appendix~\ref{app:proofs} contains proofs; the Internet Appendix provides extensions, data details, robustness analyses, and numerical verification.

\section{Institutional Setting}\label{sec:setting}
\subsection{Cash-intensive relief}
A Ginnie Mae issuer pools government-insured mortgages into securities and guarantees timely payment to investors. When a borrower misses a payment, the issuer advances scheduled principal and interest from its own funds for as long as the loan remains delinquent and in the pool. Once the borrower has missed three payments, the issuer may repurchase the loan at par. Repurchase ends the advance obligation on that loan. It also requires the full unpaid balance in cash and places a nonearning asset on the issuer's balance sheet until the loan is cured, modified, or liquidated. A repurchased loan that reperforms can be redelivered into a new pool, which is valuable when market rates are below the loan's rate. The option to repurchase belongs to the issuer of record and cannot be sold separately from the servicing. Depositories fund these outlays from deposits. Nonbanks, which by 2020 serviced most Ginnie Mae collateral, fund them from capped warehouse and servicing-advance facilities, much of it supplied by the banks they compete with \citep{KimEtAl2018,JiangEtAl2020,Jiang2023,KimEtAl2022}.

A repurchase is not itself debt relief. It changes who holds the loan and the holder's incentives without changing the borrower's obligations. It can be the step that makes relief possible: a loan cannot be modified while it is in a pool, and \citet{Ahsin2026} shows, from Ginnie Mae's 2003 suspension of the buyout option for rolling delinquencies, that earlier buyout eligibility lowers foreclosure by about 40 percent and raises cures by about 50 percent, because issuers pursue loss mitigation once they own the loan. Relief comes from forbearance, deferral, or modification, each with its own cash requirement, compensation, and consequences for the borrower. The CARES Act of March 27, 2020, entitled borrowers with federally backed mortgages to forbearance on request, which multiplied delinquencies, and therefore advance obligations, across every issuer at once. Private-label servicers faced no statutory mandate but made discretionary relief decisions under their own contracts and funding \citep{BandyopadhyayRelief2026}. The framework below treats these actions as different uses of the same scarce resource.

\subsection{Two interventions of different kinds}
Two 2020 interventions illustrate the distinction the model draws between liquidity and value. On April 10, 2020, Ginnie Mae made its Pass-Through Assistance Program available to issuers facing pandemic-related shortfalls \citep{GNMA2003}. The program advanced funds to cover principal and interest owed to investors at a fixed interest rate, required full repayment, and was intended as an extraordinary measure for issuers whose other resources were exhausted. It was liquidity at a steep marginal cost, not a transfer. On June 29, 2020, Ginnie Mae restricted the redelivery of reperforming loans \citep{GNMA2007}. A loan that entered forbearance on or after March 1, 2020, and was repurchased on or after July 1, 2020, became ineligible for existing pool types. It could be securitized again only in a new pool type, after six consecutive timely payments and at least 210 days after its last delinquency. The restriction lowered the value of repurchasing a dated subset of loans without supplying or absorbing any cash. In the model's terms, the first intervention relaxes the budget at a cost and the second shifts the value of action. A market-wide shift in value followed in 2022, when mortgage rates more than doubled and the redelivery value of most delinquent loans fell.

\subsection{Four values that must be kept distinct}\label{sec:values}
For each action I distinguish four quantities, all measured relative to the same feasible continuation. \emph{Private incremental value} is the discounted payoff to the intermediary, including contractual transfers and financing at normal terms. \emph{Cash required now} is the net outlay needed to implement the action before its proceeds arrive. \emph{Borrower benefit} is the incremental effect on the borrower's payments, housing retention, and resources. \emph{Total surplus} adds the gains of all parties net of real resource and fiscal costs, counting transfers once. Principal paid by an issuer to bondholders at repurchase is a private cash outflow and a transfer between stakeholders, not a social loss. Conversely, a loan that reperforms after an intervention can be privately valuable without any borrower gain if the borrower would have cured anyway. The model keeps these objects separate because the policy conclusions turn on the differences among them.

\section{Rationing Relief}\label{sec:model}
\subsection{Primitives}
A risk-neutral intermediary holds a unit mass of resolution opportunities indexed by $i$, with distribution $F$. For each opportunity, it chooses between a normalized continuation and an intervention, $x_i\in[0,1]$; interior values represent randomization across otherwise identical opportunities. The intervention has private incremental value $g_i$ relative to continuation and requires net cash $\ell_i>0$ at the decision date. The value $g_i$ already includes the present value of all future receipts and costs. The cash requirement appears separately because cash cannot be moved freely across time. Liquidity $K\ge 0$ is usable cash and committed borrowing capacity net of unavoidable obligations. The intermediary solves
\begin{equation}\label{eq:primal}
V(K)=\sup_{0\le x_i\le 1}\int g_i x_i\,dF(i)\quad\text{subject to}\quad\int \ell_i x_i\,dF(i)\le K.
\end{equation}

\begin{assumption}\label{ass:regular}
The population is nonatomic; $g$ is bounded and measurable; and $0<\underline\ell\le\ell_i\le\bar\ell<\infty$. Liquidity cannot be raised costlessly at the decision date.
\end{assumption}

An aggregate shock can move $F$, $g$, $\ell$, and $K$ together. The comparative statics below vary one primitive at a time; Section~\ref{sec:evidence} discusses how the data can do the same.

\subsection{Rationing by value per dollar}
\begin{proposition}[Rationing]\label{prop:allocation}
Under Assumption~\ref{ass:regular}, there is a multiplier $\lambda\ge 0$ such that an optimal allocation funds every opportunity with $g_i>\lambda\ell_i$, funds none with $g_i<\lambda\ell_i$, and randomizes on equality. The value function satisfies
\begin{equation}\label{eq:dual}
V(K)=\min_{\lambda\ge 0}\Big\{\lambda K+\int\pos{g_i-\lambda\ell_i}dF(i)\Big\},
\end{equation}
is nondecreasing and concave, and $V'(K)=\lambda$ wherever it is differentiable. Opportunities with $0<g_i<\lambda\ell_i$ are profitable but unfunded. The funding threshold $\lambda$ is nonincreasing in $K$, and it is zero once $K$ covers every positive-value opportunity.
\end{proposition}

The multiplier $\lambda$ is the shadow value of a dollar at the decision date. An action worth $g_i$ is forgone when the cash it uses would earn more elsewhere in the portfolio. With heterogeneous cash needs, total value is the wrong ranking: a high-balance or high-coupon loan can be passed over precisely because it absorbs more scarce cash. The structure is a continuous knapsack. The contribution here is what it implies once information, verification, and borrower welfare enter.

\subsection{Selection and the appearance of success}
To connect rationing to information, let $q_i\in[0,1]$ be the intermediary's posterior probability that the intervention succeeds, for instance that the loan reperforms, and specialize to
\begin{equation}\label{eq:linear}
g(q)=Rq-c,\qquad \ell(q)=k-\omega q,\qquad R>c\ge 0,\quad k>\omega\ge 0,\quad Rk>\omega c.
\end{equation}
Success raises the private payoff by $R$ at cost $c$; the parameter $\omega$ captures offsets to the cash requirement that are larger for better loans (for example, a higher advance rate on a loan expected to reperform). Setting $\omega=0$ gives a constant cash need. With slight abuse of notation, $F$ now denotes the distribution of $q$, with density $f$.

\begin{proposition}[Selection]\label{prop:selection}
The optimal policy funds $q>q^*(\lambda)=(c+\lambda k)/(R+\lambda\omega)$, with $dq^*/d\lambda=(Rk-\omega c)/(R+\lambda\omega)^2>0$. If the budget binds and $q$ has a positive density near the cutoff, a reduction in $K$ lowers the funded share and raises $\E[q\mid x=1]$. If $q$ is the true success probability under intervention, the success rate among funded interventions rises as access contracts.
\end{proposition}

\begin{figure}[H]
\centering\includegraphics[width=\textwidth]{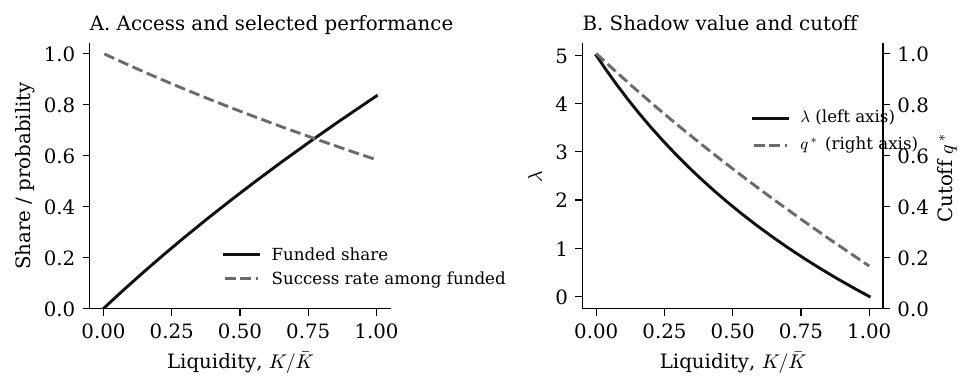}
\caption{Rationing relief under a funding constraint.\label{fig:rationing}}
\begin{minipage}{0.97\textwidth}\footnotesize\setstretch{1.0}
The figure plots the solution of problem (\ref{eq:primal}) with $g(q)=1.2q-0.2$, $\ell(q)=0.3-0.1q$, and posteriors uniform on $[0,1]$. Liquidity is scaled by $\bar K$, the cash needed to fund every positive-value opportunity. Panel A shows the share of opportunities funded and the success rate among funded opportunities. Panel B shows the shadow value of cash $\lambda$ (left axis) and the posterior cutoff $q^*$ (right axis). All quantities are exact model calculations; none is estimated.
\end{minipage}
\end{figure}

Proposition~\ref{prop:selection} is a selection result. It says nothing about the effect of intervention relative to continuation, which requires both potential outcomes. Its empirical content is that a contraction in liquidity improves the measured performance of the relief that is delivered: cure rates among modified or repurchased loans rise just as fewer borrowers are helped. Performance metrics computed on treated loans therefore cannot be read as evidence that a servicer's relief improved. Figure~\ref{fig:rationing} and Table~\ref{tab:uniform} illustrate with a uniform posterior.

\subsection{Value shocks and liquidity shocks}
The rationing structure yields a diagnostic for whether the budget binds. Let the cash need be constant, $\ell_i=k$, so the fundable share is $m=K/k$.

\begin{proposition}[Value shocks versus liquidity shocks]\label{prop:dissociation}
Let $\bar m=\Pr(g>0)$ be the share of positive-value opportunities.
\begin{enumerate}[label=(\alph*)]
\item If the budget binds ($m<\bar m$), the share of opportunities funded equals $m$. It rises by $1/k$ per dollar of liquidity and is unchanged by any shock to values that leaves at least a share $m$ of opportunities with positive value.
\item If the budget is slack ($m>\bar m$), the share funded equals $\bar m$. It does not respond to liquidity and moves one-for-one with the share of positive-value opportunities as long as that share remains below $m$.
\item With heterogeneous cash needs, the cash deployed under a binding budget equals $K$ for every value shock that leaves positive-value cash demand above $K$.
\end{enumerate}
\end{proposition}

Proposition~\ref{prop:dissociation} is a double dissociation: liquidity moves constrained intermediaries, and value moves unconstrained ones. A common shock that moves every opportunity's value in the same direction cannot move two intermediaries' relief in opposite directions. A common drain on cash combined with a common rise in value can: the rise in value raises relief where funding is slack, and the drain lowers it where funding binds. The 2020 policy environment supplies both kinds of variation. Pass-through assistance and facility conditions shifted liquidity, while the June 2020 redelivery restriction shifted value for a dated subset of loans and the 2022 rise in mortgage rates lowered the value of repurchase for most of them.

\subsection{Where scarcity comes from}
Write liquidity as $K=A-M$, where $A$ is usable cash and committed borrowing capacity and $M$ is the cash required by existing obligations over the decision interval, chiefly advances on delinquent loans. A forbearance wave raises $M$ for every issuer. For a depository, $A$ is large and does not contract with delinquency. For a nonbank, $A$ is capped by facility limits that tighten as collateral values fall. The same event that raises the value of repurchase, a surge in delinquencies, also drains the resource needed to exercise it. This is why a regression of relief on contemporaneous delinquency does not isolate the liquidity channel: delinquency is at once a cash demand, a state variable for asset quality, and a determinant of the value of action.

\section{Liquidity and the Value of Information}\label{sec:information}
\subsection{Setup}
Fix a cell of opportunities that are identical on hard information and have success probability $p$ under intervention. Before acting, the intermediary observes a signal with posterior $Q=\Pr(\text{success}\mid\text{signal})$ distributed according to $G$ on $[0,1]$; Bayes plausibility requires $\E[Q]=p$. Each intervention yields $RQ-c$ and requires cash $k$, so the fundable share is $m=K/k\in[0,1]$. I assume $Rp>c$, so that intervention is profitable on hard information alone and the intermediary would act without the signal, and $\Pr(Q=p)=0$, so that every signal is either good or bad news.\footnote{If $\Pr(Q=p)>0$, parts (ii) and (iv) of Proposition~\ref{prop:info} hold with $m^*$ replaced by any point of $[\Pr(Q>p),\Pr(Q\ge p)]$, an interval on which $\Delta V$ is constant and the marginal values of liquidity with and without the signal are equal. Where $V$ has kinks, marginal values refer to one-sided derivatives.} Write $G^{-1}(u)=\inf\{q:G(q)\ge u\}$ for the quantile function and
\begin{equation}\label{eq:T}
T_G(m)=\int_{1-m}^{1}G^{-1}(u)\,du
\end{equation}
for the expected number of successes among the top-ranked share $m$. By Proposition~\ref{prop:allocation}, the intermediary funds the highest posteriors first and stops at zero value, so its value with and without the signal is
\begin{equation}\label{eq:VG}
V_G(m)=\int_{1-m}^{1}\pos{RG^{-1}(u)-c}du,\qquad V_0(m)=m\,(Rp-c).
\end{equation}
The allocation value of the signal is $\Delta V(m)=V_G(m)-V_0(m)$.

\subsection{The main result}
\begin{proposition}[Liquidity and the value of information]\label{prop:info}
Let $m^*=\Pr(Q>p)$. The allocation value of the signal has the following properties.
\begin{enumerate}[label=(\roman*)]
\item \emph{Decomposition.} $\Delta V(m)=\mathcal A(m)+\mathcal S(m)$, where the ranking value $\mathcal A(m)=R\,[T_G(m)-mp]$ and the screening value $\mathcal S(m)=\int_{1-m}^{1}\pos{c-RG^{-1}(u)}du$ are nonnegative. $\mathcal A$ is concave with $\mathcal A(0)=\mathcal A(1)=0$. $\mathcal S$ is convex, nondecreasing, and zero whenever $m\le\Pr(RQ\ge c)$.
\item \emph{Shape.} $\Delta V$ is concave, with $\Delta V(0)=0$ and $\Delta V(1)=\E\pos{c-RQ}\ge 0$. It is nondecreasing on $[0,m^*]$ and nonincreasing on $[m^*,1]$.
\item \emph{Complementarity.} The marginal value of liquidity is higher with the signal than without it if $K<km^*$ and lower if $K>km^*$. Information and liquidity are complements when liquidity is scarce and substitutes when it is abundant.
\item \emph{Peak.} The maximal allocation value is $\Delta V(m^*)=R\,\E\pos{Q-p}=\tfrac{R}{2}\,\E|Q-p|$, and it consists entirely of ranking value.
\end{enumerate}
\end{proposition}

\begin{figure}[H]
\centering\includegraphics[width=\textwidth]{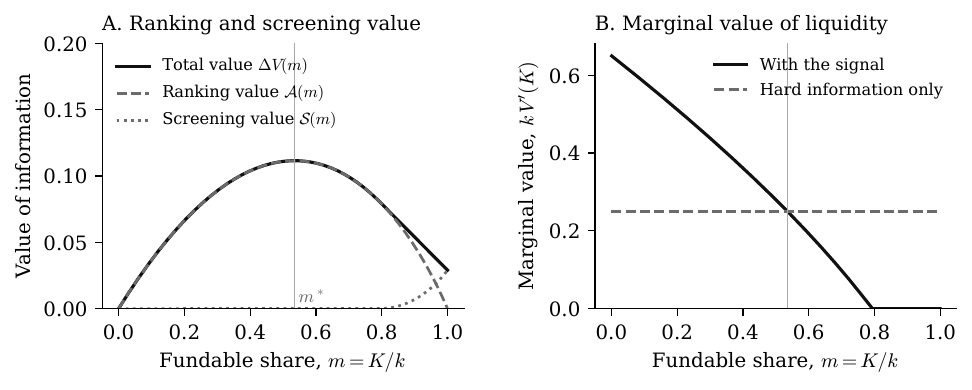}
\caption{The value of information is hump-shaped in liquidity.\label{fig:information}}
\begin{minipage}{0.97\textwidth}\footnotesize\setstretch{1.0}
Posteriors are distributed Beta(1.5,\,1), with mean $p=0.6$; $R=1$ and $c=0.35$, so 20.7\% of borrowers are unprofitable on the finer information. Panel A plots the allocation value of the signal, $\Delta V(m)=V_G(m)-V_0(m)$, and its ranking and screening components (Proposition~\ref{prop:info}). The peak is at $m^*=\Pr(Q>p)=0.535$ with value $R\,\E[Q-p]_+=0.112$. Panel B plots the marginal value of liquidity per unit of fundable share with and without the signal; the curves cross at $m^*$.
\end{minipage}
\end{figure}

Three implications follow. First, the value of the same information differs sharply across balance sheets. An intermediary that can fund almost nothing gains little from knowing whom it would help, and one that can fund every opportunity uses information only to decline. The value is largest in between, at the fundable share equal to the share of borrowers whose signal is good news. Second, the statistic that governs the maximal allocation value is the mean absolute deviation of the posterior. It is not the variance, the coefficient of determination, or classification accuracy, and it measures how far the signal moves beliefs, weighted equally in both directions. Third, the regime determines the kind of value information delivers. When liquidity is scarce, information is used to decide whom to help first. As liquidity approaches full funding, it is used increasingly to decide whom not to help, and its value becomes the option to decline actions that are unprofitable on the finer information; if no such actions exist, the value of information vanishes at full funding. Figure~\ref{fig:information} illustrates with a continuous posterior.

\begin{corollary}[Binary signal]\label{cor:binary}
Let a binary signal have symmetric accuracy $\eta\in(1/2,1)$, meaning that it is good with probability $\eta$ after success and bad with probability $\eta$ after failure, so that $Q\in\{q_L,q_H\}$ with $\Pr(Q=q_H)=\pi_H=p\eta+(1-p)(1-\eta)$, and suppose $Rq_L>c$. Then $\Delta V(m)=R(q_H-p)m$ for $m\le\pi_H$ and $\Delta V(m)=R(p-q_L)(1-m)$ for $m\ge\pi_H$.
\end{corollary}

With $p=1/2$, $\eta=0.8$, $R=1$, and $c=0.1$, the posteriors are 0.8 and 0.2 and the peak value is 0.15 at $m^*=1/2$. Concavity implies that if adopting the signal carries a fixed cost $\psi$ paid from a separate operating budget, the intermediary adopts it on an interval of liquidity levels containing $m^*$ whenever $\psi<\Delta V(m^*)$. In the example, a cost of 0.06 makes adoption worthwhile only for fundable shares between 0.2 and 0.8. If the cost is paid from the same scarce cash, the comparison becomes $V_G(m-\psi/k)$ against $V_0(m)$, and adoption is further deterred at low liquidity. Both intermediaries with ample liquidity and intermediaries with very little have weak incentives to invest in borrower information.

\subsection{Reliance on information}
Proposition~\ref{prop:info} concerns the value of information. A closely related object governs how strongly decisions respond to information, which is what loan-level data reveal. Let $r=h(Q)$ be any ranking variable that is nondecreasing in the posterior, such as a value proxy, with $0<\mathrm{Var}(r)<\infty$. Define the level response $\tau_r(m)=\mathrm{Cov}(x,r)/\mathrm{Var}(r)$, the slope of a linear probability regression of intervention on $r$, and the relative response $\tau_r(m)/\Pr(x=1)$.

\begin{proposition}[Reliance on information]\label{prop:reliance}
Suppose $RQ\ge c$ almost surely, so the intermediary funds the top-ranked share $m$.
\begin{enumerate}[label=(\roman*)]
\item The level response is $\tau_r(m)=\big[\int_{1-m}^1 h(G^{-1}(u))\,du-m\,\E[r]\big]/\mathrm{Var}(r)$. It is concave in $m$, zero at $m=0$ and $m=1$, and maximized where the funded share equals $\Pr(r>\E[r])$. For $r=Q$, $\tau_Q(m)=\mathcal A(m)/(R\,\mathrm{Var}(Q))$: the level response is proportional to the ranking value of information and peaks at $m^*$.
\item The relative response is $\tau_r(m)/m=\big(\E[r\mid x=1]-\E[r]\big)/\mathrm{Var}(r)$, which is nonincreasing in $m$. As liquidity contracts, the response relative to the base rate weakly rises.
\item The difference in intervention rates between borrowers with good and bad news, $D(m)=\Pr(x=1\mid Q>p)-\Pr(x=1\mid Q<p)$, equals $m/m^*$ for $m\le m^*$ and $(1-m)/(1-m^*)$ for $m\ge m^*$.
\end{enumerate}
\end{proposition}

The proof uses only that the intermediary funds a top share of its ranking and that $r$ is monotone in it. The proposition therefore applies to any ranking index, including hard characteristics such as a loan's coupon that enter the ranking through the value of action, with $Q$ replaced by the index. A constrained intermediary keeps ranking and funds less of its ranking. When liquidity contracts below the peak, the level sensitivity of relief to a ranking variable falls toward zero, while the sensitivity relative to the collapsed base rate rises. A flattening of the level relationship between relief and a value proxy is therefore what rationing predicts, not evidence that the intermediary stopped discriminating. It is also not, by itself, a signature of a funding constraint. A value shock that compresses values across the ranking variable can produce the same flattening, and a uniform shift in values leaves an unconstrained intermediary funding the top of the same ranking (Internet Appendix~\ref{ia:slope}). The slopes are diagnostic in combination with the dated shocks of Proposition~\ref{prop:dissociation}.

\subsection{Which model is worth more?}
Servicers rank delinquent borrowers with statistical and machine-learning models, and the standard measure of such a model's quality is its AUC, the probability that it ranks a randomly chosen success above a randomly chosen failure. Proposition~\ref{prop:info} implies that the relevant measure depends on the balance sheet. Define the cumulative accuracy profile $\Gamma_G(m)=T_G(m)/p$, the share of all eventual successes captured by the top-ranked share $m$. Then the ranking value in Proposition~\ref{prop:info} is
\begin{equation}\label{eq:cap}
\mathcal A_G(m)=Rp\,[\Gamma_G(m)-m],
\end{equation}
proportional to the vertical distance between the accuracy profile and the diagonal at the fundable share.

\begin{proposition}[Evaluating scores for a constrained intermediary]\label{prop:auc}
Consider two scores that produce calibrated posteriors $Q_A\sim G_A$ and $Q_B\sim G_B$ with the same mean $p$.
\begin{enumerate}[label=(\roman*)]
\item Score $A$ is worth at least as much as score $B$ at every fundable share $m\in[0,1]$ and every cost $c\ge 0$ (a comparison that does not involve the no-signal benchmark) if and only if $Q_A\cx Q_B$, that is, if and only if $\Gamma_A(m)\ge\Gamma_B(m)$ for all $m$. For a binary outcome, this is Blackwell's ordering of the two experiments.
\item If $RQ\ge c$ almost surely under a score, its allocation value averaged over fundable shares drawn uniformly from $[0,1]$ is
\begin{equation}\label{eq:aucid}
\int_0^1\Delta V_G(m)\,dm=R\,p\,(1-p)\,\Big(\AUC_G-\tfrac12\Big),
\end{equation}
where $\AUC_G$ counts ties as one-half.
\item Under the condition in (ii), if the accuracy profiles cross, there are fundable shares at which each score is strictly more valuable, and AUC ranks the scores by their average value in (ii).
\end{enumerate}
\end{proposition}

Part (i) says that a score is better for every intermediary only if it is more informative in Blackwell's sense. That is a partial order, and most pairs of models that practitioners compare are not ranked by it. Under the profitability condition in part (ii), AUC measures the average allocation value of a score to an intermediary whose liquidity is equally likely to take any value. \citet{HernandezOralloFlachFerri2012} show that AUC is linear in expected loss when the classification threshold is chosen by a uniformly distributed rate. Equation~(\ref{eq:aucid}) is the economic version of that result for a capacity-constrained intermediary, with the payoff scale and the base rate made explicit. A servicer that knows its liquidity should evaluate models at that liquidity.

\begin{figure}[H]
\centering\includegraphics[width=\textwidth]{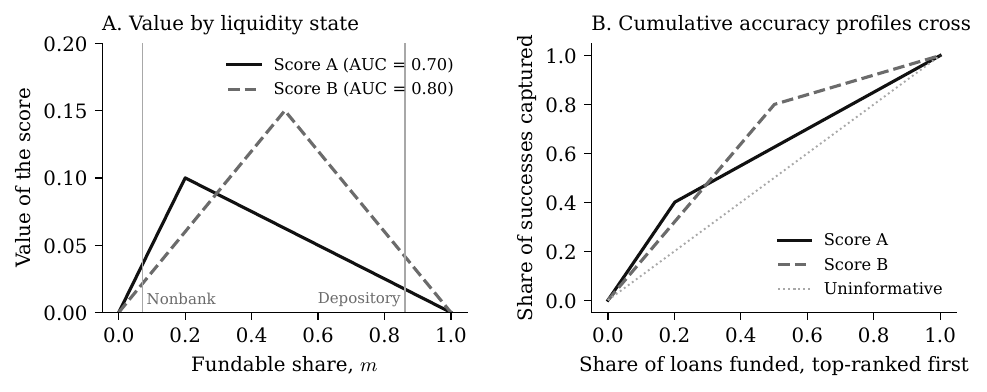}
\caption{Which score is worth more depends on liquidity.\label{fig:scores}}
\begin{minipage}{0.97\textwidth}\footnotesize\setstretch{1.0}
Two calibrated scores with mean $p=0.5$, $R=1$, and $c=0.1$. Score A assigns posterior 1 to 20\% of borrowers and 0.375 to the rest; score B assigns 0.8 and 0.2 to halves. Panel A plots each score's allocation value by fundable share. Vertical lines mark \eNbJunM\% and \eDepJunM\%, the nonbank and depository buyout exercise rates in June 2020 in the sample of Section~\ref{sec:evidence}; the signal structures are hypothetical. Panel B plots the cumulative accuracy profiles; the allocation value in Panel A is $Rp$ times the vertical distance between each profile and the diagonal.
\end{minipage}
\end{figure}

Table~\ref{tab:scores} and Figure~\ref{fig:scores} illustrate with two calibrated scores with the same mean $p=0.5$. Score $A$ is precise at the top: it identifies 20\% of borrowers who succeed with certainty and assigns everyone else 0.375. Score $B$ is broad: it splits borrowers into halves with posteriors 0.8 and 0.2. Score $B$ has the higher AUC, 0.80 against 0.70. At a fundable share of \eNbJunM\%, the nonbank exercise rate in June 2020 in the sample of Section~\ref{sec:evidence}, score $A$ is worth two-thirds more than score $B$. At \eDepJunM\%, the depository rate, score $B$ is worth 2.4 times as much. The two issuer types would choose different models, and both would be right.

\subsection{When information expands funding}
So far, information allocates a fixed budget. It can also change the budget. Suppose each funded loan can be pledged to a lender against a claim that pays $L$ if the intervention succeeds, with $L<R$ and $L<k$, so that each action requires some internal cash. The hard-information facility below has an exogenously fixed advance; it is a benchmark contract, not a competitive equilibrium in which the lender observes and prices selection. Competitive pricing applies to the recognized-information regime. Internal liquidity is $K_0\in(0,k-Lp)$, and $RQ>c$ almost surely, so the budget binds.

\begin{proposition}[When information is financeable]\label{prop:finance}
\begin{enumerate}[label=(\alph*)]
\item If the lender prices the facility on hard information, advancing $Lp$ per funded loan, then whatever the intermediary's signal, the fundable share is $m_H=K_0/(k-Lp)$.
\item If the lender knows the intermediary's information technology $G$ and observes which loans are funded, but not their posteriors, the competitive advance on a pool of the top-ranked share $m$ is $L\,T_G(m)/m$ per loan. The fundable share $m_R$ solves $km-L\,T_G(m)=K_0$, and loan-level disclosure of posteriors yields the same $m_R$.
\item $m_R\ge m_H$, strictly if $Q$ is nondegenerate.
\item The intermediary's private value is $V_G(m_R)$ under (b) and $V_G(m_H)-L\,[T_G(m_H)-m_Hp]$ under (a), and the former is at least as large.
\end{enumerate}
\end{proposition}

Under the stipulated facility pricing, information that the lender does not recognize is valuable for allocation but not for financing. Under (a), the intermediary also cross-subsidizes the lender, because the funded pool is better than the facility's price assumes. Under (b), the lender infers the pool's quality from the intermediary's known selection rule, and the budget expands although no borrower-level information changes hands. What must be verifiable is the technology and its use, not each score. The distinction matters for how intermediaries adopt predictive technology. An internal model that facility lenders cannot validate improves who is helped but not how many can be helped. A model whose existence and use lenders can verify, through validation, audit, or shared data standards, can do both. In the numerical example of Corollary~\ref{cor:binary}, with $k=1$, $L=0.6$, and $K_0=0.2$, recognition raises the fundable share from 28.6\% to 38.5\%. The assumption in (b) that the lender observes which loans are funded is natural for buyout facilities, where the funded loans are the collateral. If the intermediary instead chooses which of its loans to pledge, pooled pricing invites adverse selection in the pledged set \citep{MyersMajluf1984,DeMarzoDuffie1999}, and loan-level verification regains a role.

\section{Evidence}\label{sec:evidence}
This section takes the rationing, reliance, and value-shock predictions to Ginnie Mae buyout decisions from 2019 to 2022, a period with a liquidity shock concentrated on nonbank issuers in 2020, its reversal in 2021, and two changes in the value of repurchase, Ginnie Mae's June 2020 restriction on redelivery and the rise in mortgage rates in 2022. Before turning to the data, I state which statistic each kind of shock should move. Under a pure liquidity shock, the funded share moves at constrained intermediaries. The value-per-dollar ordering remains fixed, while the decision index $g_i-\lambda\ell_i$ increasingly penalizes cash-intensive actions as $\lambda$ rises (Propositions~\ref{prop:allocation} and \ref{prop:dissociation}). Under a value shock, the funded share moves at unconstrained intermediaries, and the ranking itself can change for everyone; if the shock pushes part of the opportunity set below zero, intermediaries that fund every profitable action start to use their ranking to decline (Propositions~\ref{prop:dissociation} and \ref{prop:info}). Section~\ref{sec:data} describes the data and a ranking index. Section~\ref{sec:fixedrank} follows the ranking of constrained issuers through the contraction and recovery of their budgets, and Section~\ref{sec:shape} checks the same benchmark across the whole range of funded shares. Section~\ref{sec:perdollar} tests the rule's other half, that scarce cash tilts the ranking against the actions that use most of it, and Section~\ref{sec:funding} relates exercise to each issuer's own cash drain. Section~\ref{sec:unconstrained} turns to unconstrained issuers and to the value shocks. Section~\ref{sec:queue} shows that the options constrained issuers postponed in 2020 were exercised later according to the same ranking, and Section~\ref{sec:forbearance} documents repurchases that unconstrained issuers made and constrained issuers did not. Section~\ref{sec:alternatives} weighs the alternative explanations against the evidence, Section~\ref{sec:companion} connects the results to companion work, Section~\ref{sec:illustration} puts the estimates together in a quantitative illustration, and Section~\ref{sec:tests} collects the predictions and the tests that remain.

\subsection{Data and a ranking index}\label{sec:data}
The data are Ginnie Mae's public monthly loan-level disclosures for single-family pools from December 2013 to December 2022, in files assembled from the published disclosures for \citet{BKS2026}. I use them in two forms. The first has one observation per buyout option, in the month the option vests because the loan first reaches three missed payments, with the loan's hard characteristics at that date, the month and reason of any removal from its pool over the following twelve months, and its last recorded status through December 2022. The second records, for every month, the stock of loans that are at least three payments delinquent and therefore eligible for repurchase. A repurchase is a removal under reason 2 before any cure. The main outcome is exercise within three months of vesting, which captures the decision at vesting and is observed for every cohort through September 2022; Section~\ref{sec:queue} studies later exercise. The sample contains the \eEmpN{} options that vested from January 2019 to September 2022 at the \eEmpIssuers{} issuers classified as depositories or nonbanks: \eEmpNbN{} at nonbanks and \eEmpDepN{} at depositories. Issuers are classified by charter (Internet Appendix~\ref{ia:sample}). Nonbanks fund advances and repurchases from capped facilities; depositories fund them from deposits \citep{KimEtAl2018,JiangEtAl2020}; Section~\ref{sec:funding} adds a direct measure of each issuer's cash drain.

The model ranks actions by value per dollar of cash, which is not observed. Under a stable ranking and additive cutoff representation, earlier exercise decisions provide a proxy for that ordering. Proposition~\ref{prop:allocation} alone does not identify the ranking from observed covariates. I therefore estimate the ranking from nonbanks' own earlier decisions. The ranking index is the linear predictor of a logit of exercise on hard loan characteristics at vesting, estimated with issuer-month fixed effects on the \eNTrain{} nonbank options that vested from 2016 to 2018 in issuer-months in which some but not all options were exercised. The fixed effects absorb common issuer-month exercise propensities. Interpreting the remaining predictor as a value ranking requires stable omitted determinants of exercise and an additive cutoff representation. The inputs are the loan's note rate minus the 30-year mortgage rate in the vesting month, which governs the premium at which a repurchased loan can be redelivered, the change in the ten-year Treasury rate since origination, loan age, balance, credit score, current loan-to-value ratio, judicial-foreclosure state, insurance program, pool issue type, first-time-buyer status, loan purpose, prior missed payments, and the annual insurance premium. The mortgage spread dominates: a one-point higher spread raises the log-odds of exercise by \eCoefMspread{} within an issuer-month, as the redelivery value of a repurchased loan implies \citep{BKS2026}, and VA loans rank lower. On decisions after the estimation window, the index's size-weighted area under the ROC curve within issuer-month cells is \eAUCMainNbNineteen{} for nonbank decisions in 2019, \eAUCMainNbTwenty{} in 2020, \eAUCMainNbTwentyOne{} in 2021, and \eAUCMainNbTwentyTwo{} in 2022. For depositories, which exercised four in five options and selected little before 2022, it is \eAUCMainDepNineteen{} in 2019. The index is blind to forbearance, which does not occur in the training period, and a repurchase of a loan in forbearance ties up cash for longer (Internet Appendix~\ref{ia:dynamic}). I therefore define decision cells as issuer by month by forbearance status, so that the ranking is compared only among options that use cash in the same way. The disclosures flag forbearance from May 2020; in March and April 2020 the cells are issuer-months. Issuers also rank on information the disclosures do not contain \citep{BKS2026}; the estimation assumes that this part of the ranking is logistic, independent of the index, and stable. I standardize the index over all options in the sample.

\subsection{Rationing along a fixed ranking}\label{sec:fixedrank}
Propositions~\ref{prop:allocation} and \ref{prop:reliance} imply that when a constrained intermediary's budget shrinks, it keeps ranking opportunities the same way and funds a smaller top share of its ranking, and that it moves back down the same ranking when the budget recovers. Four alternatives are natural. Under random rationing, a constrained issuer funds a random subset of the options it would otherwise exercise, for instance in the order in which they are processed; its response to value then falls in proportion to its funded share, and its response relative to the base rate does not rise. If an issuer in distress stops discriminating, both responses fall. Changes in relative loan values can alter the weight on the index. Finally, a uniform decline in the value of every option would lead even an unconstrained issuer to fund a smaller top share of the same ranking, and it produces the same pattern as rationing (Internet Appendix~\ref{ia:slope}). The comparisons below bear on the first three alternatives; Section~\ref{sec:perdollar} examines the additional balance gradient; and the fourth is addressed by the contrast with depositories and by the value shocks in Section~\ref{sec:unconstrained}.

The fixed-ranking prediction is implemented in two steps. Let $x_{ijt}$ indicate exercise of option $i$ in decision cell $j$ (an issuer, month, and forbearance status) and let $r_{ijt}$ be its ranking index. First, I estimate a logit with cell fixed effects,
\begin{equation}\label{eq:felogit}
\Pr(x_{ijt}=1)=\Lambda\big(\alpha_{jt}+\beta\, r_{ijt}\big),
\end{equation}
on nonbank decisions from March 2019 to February 2020. Under the maintained logit specification, the fixed effects absorb common cell-level exercise propensities, and $\beta$ measures the association with the index; the logistic error represents the part of their ranking that the index misses. Second, holding $\beta$ at its pre-shock value, for every later cell I choose $\alpha_{jt}$ to match the observed funded share and compute the implied level response of exercise to the index, the within-cell covariance of exercise with the index divided by its variance, and the implied relative response, the level response divided by the funded share. By Proposition~\ref{prop:reliance}(ii), the relative response is also the selection statistic of Proposition~\ref{prop:selection}: the mean index of exercised options relative to all options, scaled by the variance of the index. The predictions use the pre-shock weight, each cell's funded share, and each cell's distribution of the index, but not which loans were exercised after February 2020. For the random-rationing benchmark, each issuer funds, along the same ranking, the options it would have funded at its own pre-shock funded share, and then exercises a random fraction of them that matches its observed funded share.

\begin{figure}[H]
\centering\includegraphics[width=\textwidth]{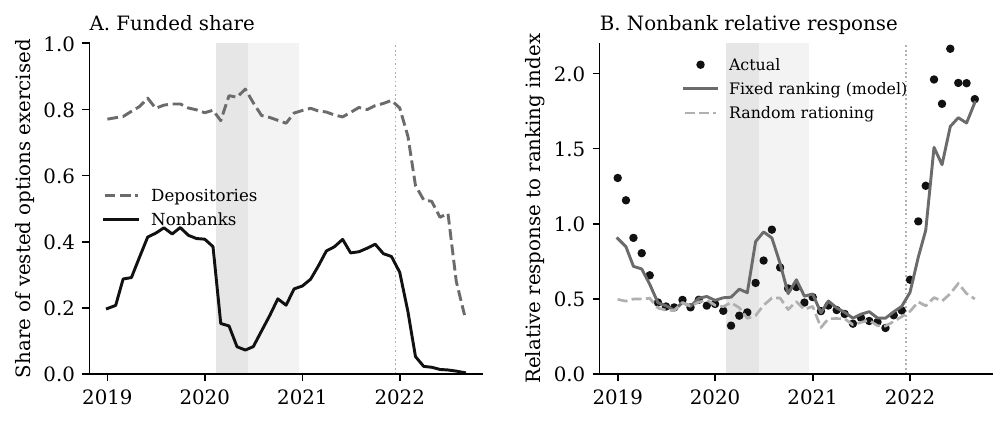}
\caption{Rationing along a fixed ranking: Ginnie Mae buyout decisions, 2019--2022.\label{fig:empirics}}
\begin{minipage}{0.97\textwidth}\footnotesize\setstretch{1.0}
Panel A plots the share of vested buyout options exercised within three months of vesting, by vesting month, at depository and nonbank issuers. Panel B plots, for nonbanks, the relative response of exercise to the ranking index: the within-cell covariance of exercise with the index, divided by its variance and by the funded share, where a cell is an issuer, vesting month, and forbearance status. The fixed-ranking line holds the latent weight on the index at its March 2019--February 2020 estimate from equation (\ref{eq:felogit}) and matches each cell's funded share; it uses no information on which loans were exercised outside that window. The random-rationing benchmark has each issuer fund, along the same ranking, the options it would have funded at its own pre-shock share and exercise a random fraction of them that matches its observed funded share. Shading marks March--June 2020 (freeze) and July--December 2020 (release); the dotted line marks the start of 2022.
\end{minipage}
\end{figure}

Figure~\ref{fig:empirics} and Table~\ref{tab:empirics} report the results. Nonbanks' funded share fell from \eNbPreM\% in the year before the pandemic to \eNbFrzM\% in March through June 2020. Their level response to the index fell from \eNbPreLvl{} to \eNbFrzLvl, as every alternative predicts. The response relative to the base rate, which separates the alternatives, was \eNbPreRel{} before the shock and \eNbFrzRel{} in the freeze. Rationing along the pre-shock ranking, given each cell's funded share, predicts a rise to \eNbFrzRelPred; random rationing predicts a fall to \eNbFrzRelRand; a loss of discrimination predicts a larger fall. The actual response lies between the first two: the difference from the fixed-ranking prediction is \eNbFrzGapFix{} (bootstrap s.e.\ \eNbFrzGapFixSE) and from the random-rationing benchmark \eNbFrzGapRand{} (\eNbFrzGapRandSE), and the latent weight on the index changed by \eNbBetaDFrz{} (s.e.\ \eNbBetaDFrzSE) against a pre-shock weight of \eNbBetaPre{} (\eNbBetaPreSE). The freeze months are therefore consistent with rationing along a ranking that the index captures less well than before the shock, or with a mixture of ranked and random rationing, and they do not by themselves distinguish the two. The point estimate shows no fall in the relative response; the bootstrap does not exclude a fall of up to about \eNbFrzRelFallPI.

The aggregate conceals heterogeneity across issuers. Internet Appendix Table~\ref{tab:issuers} reports the comparison separately for the five nonbanks with the most options before the shock, among issuers that exercised at least 5\% of them (an issuer that exercised almost none has no estimable pre-shock weight), and for the other 78 nonbanks as a group. The five held \eTopFiveSharePre\% of pre-shock options and \eTopFiveShareFrz\% of freeze options, and each row re-estimates the pre-shock weight on that issuer's own decisions. Three of the five, and the group of small issuers, show a freeze response at or above the fixed-ranking prediction. For the issuer with the most pre-shock options, whose funded share fell from \eIssAPreM\% to \eIssAFrzM\%, the response rose from \eIssAPreRel{} to \eIssAFrzRel, between the fixed-ranking prediction of \eIssAFrzRelPred{} and the random-rationing benchmark of \eIssAFrzRelRand{} and closer to the former in the change from its pre-shock level. The exception is the second-largest issuer, whose funded share fell to \eIssBFrzM\% and whose response fell to \eIssBFrzRel, below both benchmarks (\eIssBFrzRelPred{} and \eIssBFrzRelRand); it accounts for the aggregate's shortfall from the fixed ranking. Month by month, the nonbank response was below both benchmarks in March and April, when the cells cannot separate loans in forbearance from the rest, and moved toward the fixed ranking in May and June (\eNbMayRel{} against \eNbMayRelPred{} and \eNbMayRelRand; \eNbJunRel{} against \eNbJunRelPred{} and \eNbJunRelRand). Internet Appendix Table~\ref{tab:robust} varies the ranking variable, the decision cell, the exercise window, and the sample, and Internet Appendix Table~\ref{tab:stock} repeats the test on the monthly stock of eligible loans.

The cycle then reversed. In the second half of 2020 the nonbank funded share rose to \eNbRelM\% and the relative response to \eNbRelRel, close to the fixed-ranking prediction of \eNbRelRelPred{} (a difference of \eNbRelGapFix, s.e.\ \eNbRelGapFixSE) and \eNbRelGapRand{} above the random-rationing benchmark (s.e.\ \eNbRelGapRandSE; 95\% percentile interval \eNbRelGapRandPLo{} to \eNbRelGapRandPHi). This is the period in which the comparison is most informative, with the two benchmarks well separated and the narrowest bootstrap interval, and it favors the fixed ranking. In the first half of 2021 the funded share recovered to \eNbRecOneM\% and the response fell to \eNbRecOneRel; the fixed ranking, with the pre-shock weight, each cell's funded share, and no information on which loans were exercised after February 2020, predicts \eNbRecOneRelPred, a difference of \eNbRecOneGapFix{} (s.e.\ \eNbRecOneGapFixSE). In the second half of 2021 the response was \eNbRecTwoRel{} against a prediction of \eNbRecTwoRelPred. Once funded shares are back near their pre-shock levels the two benchmarks nearly coincide, so 2021 is consistent with both; what it shows is that the response moved back to where the pre-shock weight and the recovered funded share put it. The latent weight on the index changed by \eNbBetaDRel{} (s.e.\ \eNbBetaDRelSE), \eNbBetaDRecOne{} (\eNbBetaDRecOneSE), and \eNbBetaDRecTwo{} (\eNbBetaDRecTwoSE) in these three periods. The fixed-ranking computation also holds each period's distribution of the index fixed; within-cell dispersion of the index was similar before the shock and during the freeze (\eWVarNbPre{} and \eWVarNbFrz) and larger in 2021 (\eWVarNbRecOne{} and \eWVarNbRecTwo), so the predictions in Table~\ref{tab:empirics} should be read as reflecting the funded share and the index distribution together.

\subsection{Reliance across funded shares}\label{sec:shape}
Proposition~\ref{prop:reliance}(i) says that the level response of exercise to a ranking is concave in the funded share and vanishes as the funded share approaches zero and one. That shape is not by itself evidence for the model. Within a cell, the covariance of a yes-or-no decision with any ranking must vanish when almost no options or almost all of them are exercised, and every binary-choice model with a constant weight on the ranking produces a humped profile. The informative question is whether the observed profile lies on the one that the issuers' own pre-shock weight implies, or below it, as a loss of discrimination or partly random rationing would imply.

\begin{figure}[H]
\centering\includegraphics[width=\textwidth]{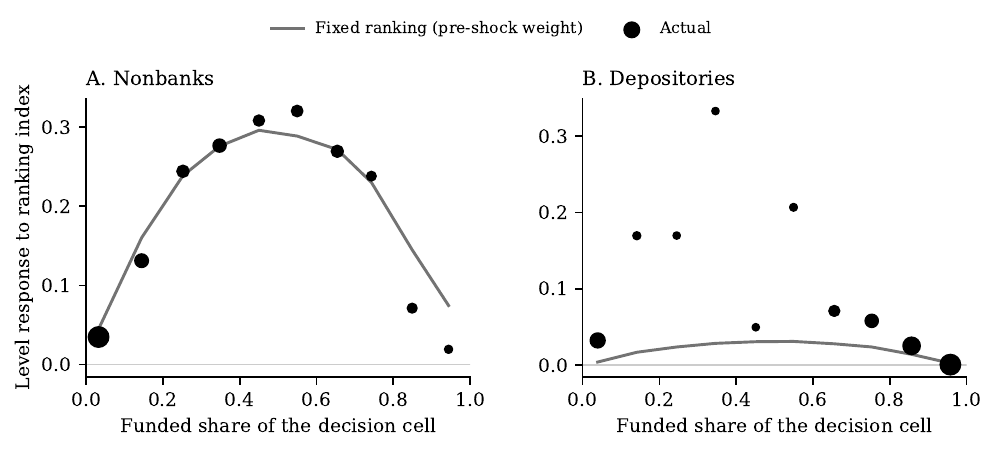}
\caption{Reliance on the ranking across funded shares.\label{fig:shape}}
\begin{minipage}{0.97\textwidth}\footnotesize\setstretch{1.0}
Decision cells (issuer by vesting month by forbearance status) with at least \eShapeMinN{} options and a funded share strictly between zero and one, January 2019--September 2022, grouped into tenths of the funded share. Dots plot the level response within each bin, the sum of within-cell covariances of exercise with the ranking index divided by the sum of within-cell variances, with size proportional to the number of cells; the line plots the same statistic under the fixed ranking, which holds the latent weight at its pre-shock estimate and matches each cell's funded share (Section~\ref{sec:fixedrank}). A humped profile is implied by any binary-choice model with a constant weight on the ranking; the comparison of interest is between the dots and the line.
\end{minipage}
\end{figure}

Figure~\ref{fig:shape} answers it across the whole range of funded shares, not only in the freeze. For bins of the cell funded share it plots the level response, the sum of within-cell covariances of exercise with the index divided by the sum of within-cell variances, together with the same statistic under the fixed ranking of Section~\ref{sec:fixedrank}. The sample is the \eShapeNbCells{} nonbank cells from \eShapeNbIssuers{} issuers with at least \eShapeMinN{} options and a funded share strictly between zero and one, from January 2019 to September 2022. Nonbank cells lie on the fixed-ranking profile for funded shares up to about eight-tenths; above that, in the \eShapeNbTopCells{} cells where nonbanks exercised almost everything, their response falls below it; these cells carry the least within-cell variation in exercise, and the departure is the one part of the range where the fixed ranking overstates reliance. Summarizing each profile by the curvature $\gamma$ in $\tau(m)=\gamma\,m(1-m)$, fitted by least squares through the origin on the cell covariances, the actual curvature is \eShapeNbGammaAct{} and the fixed-ranking curvature \eShapeNbGammaFix, a ratio of \eShapeNbRatio{} (issuer bootstrap s.e.\ \eShapeNbRatioSE; 95\% interval \eShapeNbRatioLo{} to \eShapeNbRatioHi). The pre-shock cells match the fixed ranking by construction, because the weight is estimated on them; on the \eShapeNbCellsExPre{} cells from March 2020 onward the ratio is \eShapeNbRatioExPre. The one-parameter form describes both profiles well, the fixed-ranking profile almost exactly (an $R^2$ of \eShapeNbRtwoFix{} on cell covariances, against \eShapeNbRtwoAct{} for the actual), which is what one should expect of a binary-choice model and why the shape alone proves little. What the figure shows is that constrained issuers used their ranking with the intensity their pre-shock weight implies at nearly every funded share the cycle produced, so the freeze months are the exception in Table~\ref{tab:empirics} rather than the rule. Depositories' cells sit mostly near full funding, where both profiles are flat, and their few interior cells are the screening months of 2022; for them the comparison has limited power (Panel B). A depository-trained index could alter measured fit, but would not remove the limited decision variation near full exercise.

\subsection{The price of cash in the ranking}\label{sec:perdollar}
Proposition~\ref{prop:allocation} ranks actions by value per dollar of cash, not by value. With heterogeneous cash needs, the cutoff $g_i>\lambda\ell_i$ tilts against actions that absorb more cash as the shadow price $\lambda$ rises (Internet Appendix~\ref{ia:hetero} extends the information results to this case), so a larger loan should become less likely to be repurchased than a smaller loan of the same value exactly when the constrained issuer's cash is scarcest. The gross cash a repurchase requires is the loan's unpaid balance; the disclosures do not identify how much can be financed immediately. Random rationing and a uniform value shift with a stable balance gradient do not predict this change. A value shock that varies with balance can do so. A general loss of discrimination would attenuate every weight toward zero, the weight on balance among them, but it cannot push the weight below zero or move it more than proportionately. Depositories, whose $\lambda$ is zero, should show no tilt.

Table~\ref{tab:balance} adds log balance to the cell-fixed-effects logit of equation (\ref{eq:felogit}) with period interactions for both regressors. The index already carries the 2016--2018 weight on log balance, so the coefficient on balance measures the departure from that weight. Before the shock, nonbanks' weight on balance was \eBalNbPre{} (s.e.\ \eBalNbPreSE): holding the index fixed, larger loans were if anything slightly more likely to be repurchased. In the freeze the weight fell by \eBalNbDFrzAbs{} (s.e.\ \eBalNbDFrzSE), five standard errors from zero and the sharpest of the changes the paper reports, to a level of \eBalNbFrzLevel; the weight on the index fell by \eIdxBNbFrzShrink\% in the same regression, and attenuation of both weights by that proportion would have left the weight on balance at \eBalNbAttenuated, not below zero. The weight returned to or above its pre-shock level in the release (a change of \eBalNbDRel, s.e.\ \eBalNbDRelSE). Depositories' weight on balance was \eBalDepPre{} (\eBalDepPreSE) before the shock and changed by \eBalDepDFrz{} (\eBalDepDFrzSE) in the freeze. The raw numbers point the same way: before the shock nonbanks repurchased \eDolNbPreMthreeBal\% of the balance that vested against \eNbPreM\% of the options, so the options they exercised were about \eDolNbPreSizeGap\% smaller than the average vesting option; in the freeze they repurchased \eDolNbFrzMthreeBal\% of the balance against \eNbFrzM\% of the options, and the options they exercised were about \eDolNbFrzSizeGap\% smaller than average. The regression conditions on the exercise index and observed forbearance status; it does not hold unobserved private value fixed. The tilt is not an artifact of forbearance. In March and April the cells cannot separate loans in forbearance from the rest, and take-up may have varied with balance; in May and June, when forbearance status enters the cell definition, the change in nonbanks' weight was \eBalNbDFrzLate{} (s.e.\ \eBalNbDFrzLateSE), against \eBalNbDFrzEarly{} (\eBalNbDFrzEarlySE) in March and April, and depositories' was \eBalDepDFrzLate{} (\eBalDepDFrzLateSE). The temporary balance penalty is consistent with a higher shadow value of cash. It rejects simple random thinning and proportional attenuation, but not every change in loan value or funding eligibility. Its interpretation differs from the fixed-ranking comparison: balance enters the model directly as the cash requirement, and the test asks only whether its weight, conditional on the index, moved with the budget. Figure~\ref{fig:tilt} traces the same regression quarter by quarter. The weight was positive in each of the five quarters through 2020Q1, turned negative in 2020Q2, the only quarter in which it did, and was positive again in 2020Q3. The funded share stayed low in the third quarter, but after June it also reflects the redelivery restriction of June 29, which lowered the value of repurchasing loans in forbearance (Section~\ref{sec:unconstrained}); Proposition~\ref{prop:dissociation} allows a low funded share to reflect lower value even when funding is slack. The reversal of the tilt is consistent with an easing constraint, but does not establish that the budget became slack. The weight then rose above its pre-shock range in 2020Q4 and 2021Q1, which the static model does not predict and which I do not interpret, before settling back through 2022. Depositories' weight was $-$0.02 in 2020Q2 and within \eTiltQDepMaxAbs{} of zero in every plotted quarter. A value shock that lowered the value of large loans more than small ones in exactly these four months, and reversed thereafter, could also move the weight conditional on the index; issuer-specific valuation, collateral eligibility, or borrower composition could make such a shock differ across types. The stable depository coefficient is informative but does not eliminate those possibilities.

\subsection{Funding pressure within issuer types}\label{sec:funding}
Issuer type classifies funding conditions but does not measure them. The disclosures allow one measure of the cash that existing obligations absorb, the $M$ of Section~\ref{sec:model}: the principal and interest an issuer has advanced on its delinquent loans and not yet recovered, relative to the balance of its portfolio. \citet{Aiello2022} measures servicer constraint by the one-year change in this ratio; the specifications here include issuer effects, so they also use changes within issuer. For nonbanks, advances outstanding averaged \eAdvNbPre\% of balance before the pandemic, \eAdvNbFrz\% in the freeze, and \eAdvNbRecOne\% in the first half of 2021; for depositories, \eAdvDepPre\%, \eAdvDepFrz\%, and \eAdvDepRecOne\%. Within issuer and month, a rise in a nonbank's advances lowers its funded share at vesting by \eAdvNbMAbs{} percentage points per log point (coefficient \eAdvNbM, s.e.\ \eAdvNbMSE), or \eAdvNbMperSD{} points per within-issuer standard deviation of the measure, and its monthly exercise rate on the stock of eligible loans by \eAdvNbStockAbs{} points per log point (s.e.\ \eAdvNbStockSE). The corresponding depository coefficients are \eAdvDepM{} (\eAdvDepMSE) and \eAdvDepStock{} (\eAdvDepStockSE). The difference between types in the funded-share response, \eAdvDiff{} (s.e.\ \eAdvDiffSE), is not statistically significant, and the nonbank estimate is larger in magnitude with the issuer's serious-delinquency rate as a control (\eAdvNbMCtrl, s.e.\ \eAdvNbMCtrlSE). Within issuer, the weight on the index also falls with a nonbank's advance burden (\eAdvNbBeta{} per log point, s.e.\ \eAdvNbBetaSE) and not significantly with a depository's (\eAdvDepBeta, \eAdvDepBetaSE), the within-type counterpart of the decline in the latent weight during the freeze. Internet Appendix Table~\ref{tab:funding} reports alternative measures.

These within-type estimates are small next to the 2020 collapse: a within-issuer standard deviation of the advance burden (\eAdvNbLogSD{} log points) moves the funded share by a few percentage points, whereas the freeze lowered the funded shares of the five largest nonbanks by between twenty and sixty points at once (Internet Appendix Table~\ref{tab:issuers}). The advance burden is the $M$ of Section~\ref{sec:model}, the cash that existing obligations absorb. The budget is $K=A-M$, and nothing so far measures $A$.

For a nonbank, the main source of cash in the spring of 2020 was origination. Mortgage rates fell to record lows, refinancing volume rose, and the margin between primary and secondary rates widened \citep{FusterEtAl2021}. The liquidity crisis among nonbank servicers that was feared in March 2020 did not arrive \citep{Pence2022}, and \citet{Loewenstein2021} attributes that in part to origination income; what follows is the issuer-level counterpart of that account. The same fall in rates, together with forbearance, lowered the value of servicing rights, which are the collateral for servicing-secured facilities. An issuer that originates heavily relative to the book it services has income that moves against its servicing obligations; an issuer that mostly holds servicing has none. I measure this with public data fixed before the shock: the dollar volume of mortgages the issuer originated in 2019, from the Home Mortgage Disclosure Act files, divided by the unpaid balance of its Ginnie Mae servicing portfolio in December 2019. Of the \fOrNb{} nonbanks, \fOrMatched{} are matched to a HMDA reporter; the three that are not are servicers that do not originate, and their intensity is zero. The measure varies widely. Among the \fOrActN{} nonbanks that exercised at least 5\% of their options before the shock and had at least 50 options in both periods, which are the issuers with a funded share that can fall and which held \fOrShPre\% of nonbank options before the shock and \fOrShFrz\% in the freeze, its quartiles are \fOrQA, \fOrQB, and \fOrQC.

\begin{figure}[H]\centering
\includegraphics[width=\textwidth]{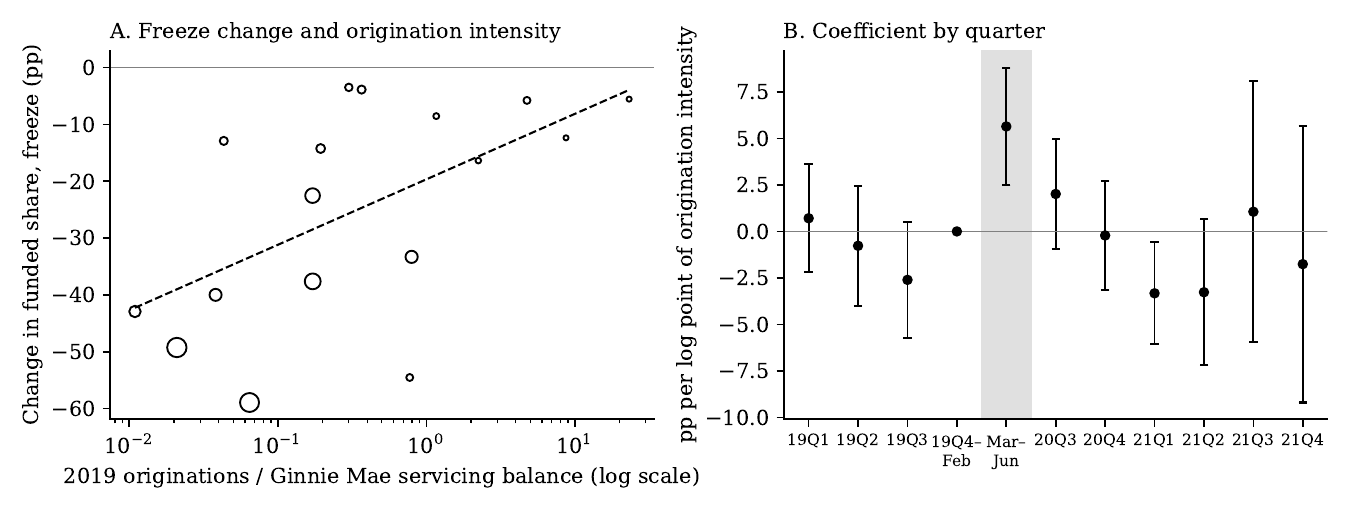}
\caption{Origination intensity and nonbank exercise.\label{fig:origination}}
\begin{minipage}{0.97\textwidth}\footnotesize\setstretch{1.0}\smallskip
Panel A plots, for the 17 nonbanks of Table~\ref{tab:origination}, the change in the funded share from March 2019--February 2020 to March--June 2020 against 2019 originations per dollar of Ginnie Mae servicing balance; marker area is proportional to pre-shock options. Panel B plots the coefficient on log origination intensity by quarter from an issuer-month regression with issuer and month fixed effects, relative to October 2019--February 2020, with 95\% confidence intervals from issuer-clustered standard errors; the shaded quarter is March--June 2020.
\end{minipage}
\end{figure}

Table~\ref{tab:origination} relates the monthly funded share of these issuers to origination intensity interacted with the periods of Table~\ref{tab:empirics}, with issuer and month fixed effects. In the freeze, a log point of origination intensity is associated with a funded share \fOrAFrz{} percentage points higher relative to the pre-shock year (s.e.\ \fOrAFrzSE). With \fOrActN{} clusters, conventional standard errors are unreliable, so the table reports wild cluster bootstrap $p$-values with the null imposed \citep{CameronGelbachMiller2008}; for the freeze coefficient $p=\fOrAP$. Across the interquartile range of intensity the implied difference is \fOrIQR{} percentage points, against a mean fall of \fOrMeanFall. The estimate lies between \fOrLooLo{} and \fOrLooHi{} when any one issuer is dropped, is \fOrAdv{} (s.e.\ \fOrAdvSE) holding the issuer's advance burden fixed, and is \fOrCFrz{} (\fOrCFrzSE; $p=\fOrCP$) with 2018 originations and the December 2018 balance. In the index metric of Section~\ref{sec:alternatives}, a log point of intensity lowers the fall in an issuer's exercise index by \fOrIdx{} (s.e.\ \fOrIdxSE; rank correlation $-\fOrIdxRho$, permutation $p=\fOrIdxP$). Figure~\ref{fig:origination} plots the coefficient by quarter. It is within two standard errors of zero in each of the first three quarters of 2019, \fOrEvFrz{} points (s.e.\ \fOrEvFrzSE) in March through June 2020, \fOrEvQThree{} (\fOrEvQThreeSE) in the third quarter, and zero in the fourth. The five largest nonbanks of Internet Appendix Table~\ref{tab:issuers} line up in the same order: the two with intensity of 0.02 and 0.06 took their funded shares from 53.4\% and 61.3\% to 4.2\% and 2.4\%; the two at 0.17 fell to 25.4\% and 48.9\%; and the one at 0.80 fell to 25.7\%.

Three qualifications belong beside the estimate. First, origination intensity falls with size among these issuers (rank correlation $-\fOrRhoSize$), because the largest Ginnie Mae servicers are aggregators. With log portfolio balance and the advance burden both interacted with the freeze, the coefficient on intensity is \fOrSize{} (s.e.\ \fOrSizeSE; $p=\fOrSizeP$) and that on size is $-\fOrSizeB$ (\fOrSizeBSE); seventeen issuers do not separate the two cleanly. Second, the choice of numerator matters at the margin: adding purchased loans, which matter most for correspondent aggregators, gives \fOrDFrz{} (\fOrDFrzSE; $p=\fOrDP$) unweighted and \fOrPurW{} (\fOrPurWSE; $p=\fOrPurWP$) weighting issuers by options. Third, the relation belongs to the freeze. A break placed one year earlier gives $-\fOrPlac$ (\fOrPlacSE; $p=\fOrPlacP$), of the opposite sign. In the second half of 2020 and the first half of 2021 the percentage-point coefficients are \fOrARel{} (\fOrARelSE) and \fOrARecOne{} (\fOrARecOneSE), and in the index metric the sign reverses (0.41, s.e.\ 0.18, and 0.40, s.e.\ 0.15): servicing-heavy issuers worked down the backlog of Section~\ref{sec:queue}, while several origination-heavy issuers stopped exercising in the months when warehouse capacity earned most in origination. I do not have a test that separates that from the redelivery restriction of June 29, and I do not interpret it further. The advance-based exposures, an analogue of the distant-exposure instrument of \citet{Aiello2022}, a synthetic control built from depositories, and exposure to the FHFA's April 2020 cap on GSE advances reproduce the type contrast but do not separate nonbanks from one another (Internet Appendix~\ref{ia:funding}, \ref{ia:synth}, and \ref{ia:design}). Taken together, these patterns support a funding interpretation. Neither issuer type nor origination intensity supplies an independently exogenous liquidity shock, so they do not identify a causal funding elasticity.

\subsection{Unconstrained issuers and value shocks}\label{sec:unconstrained}
Depositories raised their exercise during the freeze, from \eDepPreM\% to \eDepFrzM\%. For an unconstrained intermediary, the funded share is the share of options with positive value (Proposition~\ref{prop:dissociation}(b)), so a rising funded share indicates that the value of exercise rose, as falling rates would imply. The inference requires two assumptions: that depositories' funding did not bind, which does not hold for every one of them (Section~\ref{sec:companion} describes a depository whose exercise collapsed), and that the change in value was common to both types' loans, which is not guaranteed by common pool membership. Pool-month effects absorb shared security-level conditions, but not borrower risk or issuer-specific continuation values; Section~\ref{sec:companion} reports the within-pool divergence. A uniform decline in values, the alternative that the nonbank comparisons cannot rule out, would have lowered depositories' exercise instead. By Proposition~\ref{prop:dissociation}, a common shock that moved every loan's value in the same direction cannot move the two types' exercise in opposite directions, while a common drain on cash that bound for one type and not the other, combined with rising values, does. Depositories' weight on the index, \eDepBetaPre{} before the shock, fell by \eDepBetaDFrzAbs{} (s.e.\ \eDepBetaDFrzSE) in the freeze, to a point estimate below zero; with \eDepFrzM\% of options exercised, the weight rests on the few options they declined, and their level response stayed within two hundredths of zero from the freeze through 2021.

Two dated shocks to value allow the converse comparison. The first is the redelivery restriction of June 29, 2020, which lowered the value of repurchasing loans that had entered forbearance on or after March 1 once repurchased on or after July 1. In the second half of 2020, depositories' funded share fell from \eDepFrzM\% to \eDepRelM\%, as Proposition~\ref{prop:dissociation}(b) implies when the value of part of the opportunity set falls. The restriction targets loans in forbearance, which is also the variable that defines the decision cells, so its effect on the ranking within cells is not identified separately from the return of funding and the record-low rates of the same months.

\begin{figure}[H]
\centering\includegraphics[width=\textwidth]{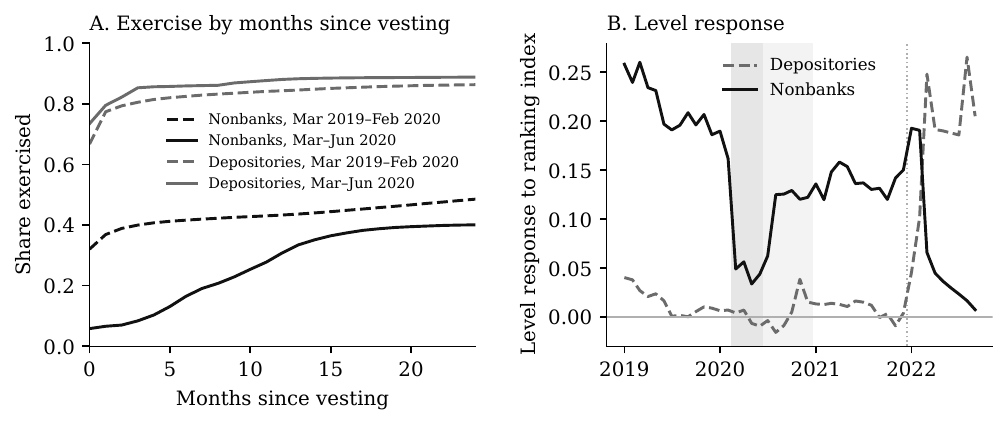}
\caption{Postponed exercise and the switch to screening.\label{fig:queue}}
\begin{minipage}{0.97\textwidth}\footnotesize\setstretch{1.0}
Panel A plots the share of options exercised by each number of months after vesting, observed through December 2022, for options that vested at nonbanks and depositories in March 2019--February 2020 (dashed) and March--June 2020 (solid). Panel B plots the level response of exercise within three months to the ranking index, the within-cell covariance of exercise with the index divided by its variance, by vesting month. Shading marks March--June 2020 (freeze) and July--December 2020 (release); the dotted line marks the start of 2022.
\end{minipage}
\end{figure}

The second value shock is the rise in mortgage rates in 2022. The 30-year rate averaged 3.1\% in December 2021 and 6.1\% in September 2022,\footnote{Monthly means of Freddie Mac's Primary Mortgage Market Survey, retrieved from FRED (series MORTGAGE30US).} so that \eNegSpNbRate\% of the options that vested at nonbanks in 2022 and \eNegSpDepRate\% of those at depositories were on loans whose note rate was below the current mortgage rate, against about \eNegSpNbFrz\% in the freeze. A repurchased loan is valuable largely because it can be redelivered once it reperforms, and a loan with a below-market rate cannot be redelivered at a premium, so much of the eligible stock lost its value. Depositories' funded share fell from \eDepRecTwoM\% in the second half of 2021 to \eDepRateM\% in 2022, as Proposition~\ref{prop:dissociation}(b) predicts for an unconstrained intermediary when the share of positive-value options falls. Proposition~\ref{prop:info} predicts something sharper. Once part of the opportunity set has negative value, information acquires screening value, and an intermediary that can fund every profitable action starts to use its ranking to decide which actions to decline. Depositories' level response to the index, \eDepPreLvl{} before the pandemic and within a few hundredths of zero through 2021, rose to \eDepRateLvl{} in 2022 (Figure~\ref{fig:queue}, Panel B), and their relative response rose to \eDepRateRel{} against \eDepRateRelPred{} under the pre-shock weight (a difference of \eDepRateGapFix, s.e.\ \eDepRateGapFixSE; percentile interval \eDepRateGapFixPLo{} to \eDepRateGapFixPHi). The ranking they began to apply is the mortgage spread itself: in 2022 the spread alone ranks their decisions as well as the index (within-cell AUCs in 2022 of \eAUCSPDepTwentyTwo{} for the spread and \eAUCMainDepTwentyTwo{} for the index at depositories, and \eAUCSPNbTwentyTwo{} and \eAUCMainNbTwentyTwo{} at nonbanks). Nonbanks' weight on the index also rose at vesting, by \eNbBetaDRate{} (s.e.\ \eNbBetaDRateSE), as a value shock that reorders opportunities implies for every intermediary; on the monthly stock of eligible loans, whose index is measured at vesting and is stale for loans that vested years earlier, the nonbank weight falls in 2022 (Internet Appendix Table~\ref{tab:stock}). Within cells, the index in 2022 is essentially the coupon, so the data cannot distinguish declining negative-value loans from re-ranking on the coupon; both are what a value shock does. In 2020, when the budget moved, depositories' level response to the index stayed near zero; in 2022, when value moved, they began to select, in the way that screening on a ranking implies. Nonbanks' funded share also fell, to \eNbRateM\%. For them the value shock cannot be separated from funding costs: an issuer that finances repurchased loans at market rates also faced higher carrying costs in 2022, which lower the private value of repurchase more for nonbanks than for depositories funded by deposits.

\subsection{Postponed exercise}\label{sec:queue}
The freeze delayed exercise more than it prevented it. Of the options that vested at nonbanks from March to June 2020, \eNbFrzM\% were exercised within three months of vesting, \eNbFrzHtwelve\% within twelve months, and \eNbFrzEver\% by December 2022. Before the pandemic the corresponding shares were \eNbPreM\%, \eNbPreHtwelve\%, and \eNbPreEver\% (Figure~\ref{fig:queue}, Panel A). Among the freeze options not exercised at vesting and still in their pools, \eNbQFrzRate\% were exercised in the fourth through twelfth months and \eNbQFrzEver\% by December 2022, against \eNbQPreRate\% and \eNbQPreEver\% for the corresponding pre-shock options. About \eQueueOctDec\% of the delayed exercises took place between October and December 2020; almost none took place in July through September.

The model says in what order a backlog is worked down. If the ranking is fixed and the budget expands, the options funded next are the highest-ranked ones not yet funded; under random rationing, the queue also retains top-ranked options, so later exercise still loads on the index; only a loss of discrimination makes later exercise unrelated to it. Table~\ref{tab:queue} reports, for options not exercised within three months of vesting and still in their pools, a logit of exercise in the following nine months on the index, with cell fixed effects. For nonbank options that vested in the freeze, the weight on the index is \eNbQFrzBeta{} (s.e.\ \eNbQFrzSE), about twice the weight in the pre-shock queue, \eNbQPreBeta{} (\eNbQPreSE), and about seven-tenths of the pre-shock weight on exercise at vesting, \eNbBetaPre. At depositories it is \eDepQFrzBeta{} (\eDepQFrzSE) for the freeze cohort. The options that constrained issuers postponed were exercised later with a weight on the ranking close to the one their decisions had carried before the shock; the issuers had not stopped discriminating. Liquidity changed the timing of repurchases more than the ranking behind them, as the dynamic version of the model implies when the budget recovers (Internet Appendix~\ref{ia:dynamic}). Postponement had a cash price of its own: the issuer kept advancing scheduled payments on loans it had not repurchased, and at a common horizon of 24 months the freeze cohort's exercise, \eNbFrzHtwentyfour\%, still fell \eQueueForgoneHtwentyfour{} percentage points short of the pre-shock cohort's \eNbPreHtwentyfour\%. Of the \eQueueGapHthree-point gap between the two cohorts at three months, about \eQueueMadeUpShare\% had closed within two years. In dollars, options on \$\eDolNbFrzUpb{} billion of unpaid balance vested at nonbanks in the freeze. They repurchased \$\eDolNbFrzXthree{} billion within three months; at the pre-shock balance-weighted rate they would have repurchased \$\eDolNbShortThree{} billion more. By 24 months they had repurchased \$\eDolNbFrzXtwentyfour{} billion, and the shortfall against the pre-shock rate had fallen to \$\eDolNbShortTwentyfour{} billion. The balance repurchased between the fourth and the twenty-fourth month, \$\eDolNbFrzLateBn{} billion (by coincidence the same figure as the three-month shortfall), waited \eDolNbFrzLateDelay{} months beyond the three-month window on average across loans, and in the meantime the issuers advanced roughly \$\eDolNbFrzLateAdv{} billion of scheduled principal and interest on those loans (the level payment implied by each loan's note rate and remaining term; escrow advances are not observed, so this is a lower bound on the cash the delay absorbed). Advances are recoverable, so this is cash tied up rather than value lost; the value lost to delay depends on the redelivery premium, which the disclosures do not observe.

\subsection{Repurchases depositories made and nonbanks did not}\label{sec:forbearance}
Forbearance gives a view of rationing when the budget binds over several periods. The disclosures flag loans in forbearance from May 2020. In May and June, \eNbFbShare\% of the options that vested at nonbanks were on loans in forbearance, and \eDepFbShare\% at depositories. Depositories exercised \eDepFbRate\% of them within three months, more than the \eDepNoFbRate\% of options on loans not in forbearance. Nonbanks exercised \eNbFbRate\%, against \eNbNoFbRate\% of options on loans not in forbearance (Table~\ref{tab:forbearance}). The second figure is close to nonbanks' pre-shock funded share of \eNbPreM\%: in these two months the collapse of the nonbank funded share was the non-repurchase of loans in forbearance, which were \eNbFbShare\% of their options. Within issuer-months, forbearance is associated with a probability of exercise \eNbFbCoefAbs{} percentage points lower at nonbanks (s.e.\ \eNbFbSE) and \eDepFbCoef{} points higher at depositories (\eDepFbSE); conditioning also on deciles of the ranking index leaves both differences essentially unchanged. Because the redelivery restriction lowered the value of repurchasing these loans from July 1, the cleaner comparison is exercise by June 30: depositories had repurchased \eDepFbByJune\% of the May--June options on loans in forbearance by then, and nonbanks \eNbFbByJune\%, against \eDepNoFbByJune\% and \eNbNoFbByJune\% for options on loans not in forbearance. By December 2022, nonbanks had exercised \eNbFbEver\% of the options on loans in forbearance and \eNbNoFbEver\% of the others.

The repurchases that depositories made most often were the ones nonbanks made least often. Depositories fund repurchased loans on their own balance sheets, so the value of these repurchases to them need not equal their value to a nonbank, and their profitability is not observed. What the contrast shows is that a repurchase attractive enough for unconstrained issuers to make more often than any other was one that constrained issuers left aside, and then made later. The cash required at repurchase is the loan's balance in either case, so a static budget does not explain the difference. A multi-period budget does. A loan in forbearance cannot be modified and redelivered until its forbearance ends, a constraint one issuer cited in its filings as its reason not to repurchase. Its repurchase therefore ties up cash longer, and in the dynamic version of the model (Internet Appendix~\ref{ia:dynamic}) an action's cost in scarce cash grows with how long it holds funds. On that interpretation, the pattern is the multi-period analogue of Proposition~\ref{prop:allocation}: actions with positive value and low value per dollar-month of cash are the ones a constrained intermediary leaves unfunded. Differences in the eligibility of loans in forbearance as facility collateral would work in the same direction; the disclosures do not allow the two to be separated.

\subsection{Alternative explanations}\label{sec:alternatives}
The comparisons above were designed so that each alternative to rationing along a fixed ranking makes at least one prediction that the data can check. Table~\ref{tab:alternatives} collects them. Random rationing fails the release comparison, where the response exceeds the benchmark by \eNbRelGapRand{} (s.e.\ \eNbRelGapRandSE; percentile interval \eNbRelGapRandPLo{} to \eNbRelGapRandPHi), and the tilt. A loss of discrimination cannot be excluded for the freeze months themselves, but it would have attenuated the weight on balance rather than pushed it below zero, the weight on the index was within one to two standard errors of its pre-shock level in 2021, and the postponed options were exercised on the same ranking. A uniform fall in value with unchanged funding and opportunity sets cannot explain depositories' higher exercise. A more specific version of that alternative deserves its own answer: that the pandemic lowered the chance that a delinquent borrower would cure, and with it the value of repurchase, so that nonbanks cut exercise because the trade was worth less to them and not because they could not fund it. The two explanations can be told apart in the units of the exercise decision. Write the probability that an option with mortgage spread $c$ is exercised as $\Lambda(v(c)-h)$, where $v$ is the value index and $h$ the hurdle the issuer applies. A rise in the shadow price of cash raises $h$ and lowers the index by the same amount at every spread. A fall in the probability of cure lowers $v(c)$ in proportion to the redelivery premium, which rises with the spread, so the index should fall most where the spread is highest. Table~\ref{tab:spreadindex} reports the fall in the log-odds of exercise from the pre-shock year to the freeze by quintile of the mortgage spread. At nonbanks it is \fSpNbOne, \fSpNbTwo, \fSpNbThree, and \fSpNbFour{} in the bottom four quintiles and \fSpNbFive{} in the top one. The top quintile's fall is \fSpTopMid{} smaller than the average of the middle three (issuer bootstrap s.e.\ \fSpTopMidSE; 95\% interval $-\fSpTopMidLo$ to $-\fSpTopMidHi$), and the slope of the fall in the spread is $-\fSpSlope$ per percentage point ($-\fSpSlopeLo$ to $\fSpSlopeHi$). A uniform decline in cure probabilities under this value specification predicts the opposite sign. This does not exclude heterogeneous changes in borrower prospects or issuer-specific continuation values. In proportional terms the same pattern is a decline of \eSpNbTopFall\% in the top quintile against \eSpNbOtherFallLo\% to \eSpNbOtherFallHi\% in the others. Depositories' index rose in every quintile, by \fSpDepLo{} to \fSpDepHi. The balance reversal is an additional restriction for that simple alternative, although the index need not absorb all balance-related changes in cure prospects. Composition accounts for the level of the nonbank collapse in May and June, when exercise on loans not in forbearance was near its pre-shock rate, but not for the gap between \eDepFbRate\% at depositories and \eNbFbRate\% at nonbanks on the loans that were, nor for the tilt within forbearance status. Aggregation is addressed issuer by issuer. What the evidence does not separate is stated as well: the 2022 fall in nonbank exercise mixes a value shock with higher funding costs, and the 2020 freeze alone does not separate a fixed ranking from a mixture of ranked and random rationing.

\subsection{The type contrast, and evidence from companion work}\label{sec:companion}
The contrast between issuer types survives controls for several dimensions of observable composition. With issuer and state-by-month fixed effects on the \fDidN{} options that vested from March 2019, nonbank exercise fell relative to depository exercise by \fDidFrz{} percentage points in the freeze (s.e.\ \fDidFrzSE), \fDidRel{} in the release (\fDidRelSE), \fDidRecOne{} in the first half of 2021 (\fDidRecOneSE), \fDidRecTwo{} in the second half (\fDidRecTwoSE), and \fDidRate{} in 2022 (\fDidRateSE) (Internet Appendix Table~\ref{tab:typecontrast}). Most Ginnie Mae loans sit in multi-issuer pools, in which many issuers contribute loans to one security and each services its own. Of the options that vested from March 2019 to June 2020, \fPoolShare\% were in pool-months containing loans of both types, and within pool and month a nonbank-serviced loan was \fPool{} points less likely to be repurchased in the freeze than before it (s.e.\ \fPoolSE). Five nonbanks hold 59\% of pre-shock options, so clustered standard errors overstate precision; the wild cluster bootstrap $p$-values are \fDidP{} for the freeze contrast and \fPoolP{} within pools, against $t$-statistics of \fDidT{} and \fPoolT. The two types were not on parallel paths before the shock: a break placed at March 2019 gives \fPlac{} points (s.e.\ \fPlacSE; $p=\fPlacP$), of the opposite sign, and on the window from July 2019, in which the pre-period is flat, the freeze contrast is \fFlat{} (\fFlatSE; $p=\fFlatP$). I therefore treat the contrast as a description and rest the interpretation on the statistics that do not use parallel trends: the tilt, the fall in the index by spread, and the origination gradient within nonbanks. At the issuer level, the index fell in the freeze at \fIssNbPos{} of the \fIssNbN{} nonbanks with at least 50 options in both periods and at \fIssDepPos{} of \fIssDepN{} depositories (rank-sum $p=\fIssRankP$); the mean difference is \fIssDiff{} log-odds (s.e.\ \fIssDiffSE; permutation $p=\fIssPermP$), and \fIssAdv{} (\fIssAdvSE) holding the issuer's pre-shock advance burden fixed, on which the coefficient is $-\fIssAdvB$ (\fIssAdvBSE).

The unfunded options stayed where they were. In the monthly histories, a change in the issuer of record between two consecutive months is a servicing transfer. Seriously delinquent loans left nonbank issuers, including moves between issuer numbers of the same firm, at a monthly rate of \fTrPre\% in the year before the shock, \fTrFrz\% in the freeze, \fTrRel\% in the release, and \fTrTwoOne\% in 2021, and of the \fTrN{} such transfers from March 2020 to December 2021, none went to a depository. Servicing did not move back toward the institutions with deposit funding, whereas \citet{HamdiEtAl2023} document transfers from banks to nonbanks following Basel III. Because the option cannot be sold apart from the servicing, transfers did not move these unexercised options toward deposit-funded holders. Their profitability to the original issuer is not directly observed. The same set appears on audited balance sheets. At the four nonbanks that report the balance of vested and unexercised options under ASC 860-50, that balance grew \fWdBalLo{} to \fWdBalHi{} times from 2019 to 2020, against \fWdFlowLo{} to \fWdFlowHi{} times for the options newly vesting at the same firms (Internet Appendix~\ref{ia:funding} and Table~\ref{tab:wedge}).

The information results also require that intermediaries hold decision-relevant borrower information and exercise discretion over relief, and both premises are documented. \citet{BKS2026} show that before the pandemic, a one percentage point higher interest rate spread raised the probability of buyout by seven to nine percentage points, and that issuers acquired soft information during delinquent servicing that predicted reperformance and used it in their buyout decisions. Using one servicer's call-center transcripts for 19,418 borrowers, \citet{BandyopadhyayDelegated2026} shows that servicer discretion over pandemic forbearance operated through documentation, verification, and referral rather than outright denial. \citet{BandyopadhyayRelief2026} follows 12,691 loans at the same servicer and finds that an April 2020 forbearance conversation predicts a 9 to 15 percentage point higher probability of subsequent modification, concentrated among borrowers already close to foreclosure and only where the servicer retained foreclosure discretion. \citet{BandyopadhyayInside2026} shows that documentation frictions in loss-mitigation applications raise foreclosure hazard.

\subsection{A quantitative illustration}\label{sec:illustration}
For illustration only, suppose funding capacities equal the two issuer types' observed exercise shares in the freeze, \eNbFrzM\% and \eDepFrzM\%, lie on opposite sides of $m^*$ for any signal whose share of good news lies between them. Observed exercise is not a measure of funding capacity when the budget is slack. Under the illustrative capacity assignment, Proposition~\ref{prop:info} implies that introducing such a signal would raise the marginal value of liquidity for a nonbank and lower it for a depository. Table~\ref{tab:scores} makes the implication concrete for model choice. The signal structures in the table are hypothetical, and the fundable shares are set near the observed exercise rates only to fix ideas. For a constrained servicer, AUC is not a close proxy for the value of a model; it is an average over balance sheets the servicer does not have. The 2022 episode is consistent with the screening margin: when a value shock pushed part of the opportunity set below zero, the unconstrained issuers' use of information shifted from nothing to screening.

\subsection{What the evidence establishes, and what it does not}\label{sec:tests}
Table~\ref{tab:predictions} separates model implications from their observable counterparts. The temporary balance penalty, subsequent exercise of the backlog, and greater selection in the release period are consistent with cash-constrained allocation. The depository response in 2022 is consistent with value-based screening. The freeze comparison alone does not distinguish fixed-ranking rationing from a mixture of ranked and random rationing.

Three boundaries are essential. First, the index predicts exercise rather than recovery under intervention, so selection on that index is not a test of improved cure rates or the monetary value of information. Second, funding is not observed directly. Pre-trends limit the issuer-type contrast, and the origination proxy is correlated with issuer size in a small, selected sample. The evidence supports a mechanism interpretation rather than a causal estimate of liquidity's effect. Third, post-repurchase outcomes are unavailable, so borrower benefits and welfare remain theoretical. Direct facility records, such as the nonpublic Mortgage Call Report data used by \citet{KimLeeScharlemannVickery2024} and \citet{BegleySrinivasan2023}, and linked borrower outcomes would strengthen those tests. Internet Appendix~\ref{ia:design} describes possible extensions; none is presented as a completed result.

\section{Borrowers, Stakeholders, and Policy}\label{sec:welfare}
\subsection{Private selection and borrower surplus}
Let $b_i$ be the borrower's incremental surplus from intervention, $v_i$ the gain to other private claimants, and $t_i$ the net gain to taxpayers, all relative to the same continuation. Total incremental surplus is
\begin{equation}\label{eq:welfare}
w_i=g_i+b_i+v_i+t_i,
\end{equation}
with real resource costs counted once and transfers netting out. A planner facing the same funding technology ranks by $w_i/\ell_i$; the intermediary ranks by $g_i/\ell_i$. In the information setting of Section~\ref{sec:information}, write $\beta(q)=\E[b\mid Q=q]$ and $\bar w(q)=\E[w\mid Q=q]$.

\begin{proposition}[Private selection and borrower surplus]\label{prop:welfare}
Suppose $RQ\ge c$ almost surely, so the intermediary funds the top-ranked share $m$.
\begin{enumerate}[label=(\roman*)]
\item If $\beta$ is nonincreasing, the signal weakly lowers borrower surplus at every fundable share, strictly for $m\in(0,1)$ if $\beta(Q)$ is nondegenerate.
\item If $\bar w$ is nonincreasing, the signal weakly lowers total surplus at every fundable share while raising private value.
\item A planner with the same signal and the same budget weakly raises total surplus relative to allocation without the signal.
\end{enumerate}
\end{proposition}

The condition in (i) is the self-cure logic of \citet{AdelinoGerardiWillen2013}. Borrowers with the best prospects under intervention are often those most likely to recover without it, so their incremental gain is smallest. When that ordering holds, better private selection systematically steers relief away from the borrowers who gain most. For example, let half the loans have posterior 0.8 and half 0.2, with $R=1$, $c=0.1$, borrower gains of 0.1 and 0.9, and capacity for half the loans. Without the signal, private value is 0.20 and borrower surplus 0.25 per loan. With it, private value is 0.35 and borrower surplus 0.05; total surplus falls from 0.45 to 0.40. Part (iii) locates the problem. The welfare loss arises from using information to maximize private recovery value, not from the information itself, so improvements in servicing technology should be evaluated with borrower outcomes rather than with investor returns or the performance of treated loans.

The public data allow a partial description of the borrowers who were affected in 2020. For loans originated from 2018 onward I attach borrower income, race and ethnicity, and census-tract characteristics from Home Mortgage Disclosure Act records, matching on state, origination year, loan type, loan amount, note rate, term, loan-to-value ratio, and debt-to-income ratio (Internet Appendix~\ref{ia:hmda}). The public fields do not identify loans uniquely, so each attribute is an average over the HMDA records that are consistent with the loan, and the merge covers 11\% of nonbank options before the shock and 15\% in the freeze. Within that sample, nonbanks cut exercise in about the same proportion for every group of borrowers. Exercise fell from 42.6\% to 5.4\% for borrowers in the lowest third of income relative to the area median and from 37.2\% to 4.6\% in the highest third. Relative to depositories, the change in the freeze does not differ with borrower income ($-0.3$ percentage points per standard deviation, s.e.\ 0.6) or for Black borrowers ($-0.4$, s.e.\ 1.2), in specifications with issuer-by-month effects. Before the shock nonbanks exercised more often on loans to Hispanic borrowers and on loans in majority-minority tracts, by 12.6 (s.e.\ 3.3) and 5.7 (s.e.\ 2.3) percentage points relative to depositories, and these differences narrowed in the freeze by 10.6 (s.e.\ 3.3) and 5.2 (s.e.\ 2.3). The decline in repurchase rates in percentage points was therefore larger for Hispanic borrowers, 40.6 against 34.0 for other borrowers, although the proportional fall was the same. These comparisons describe heterogeneity in repurchase rates within the matched sample. They do not identify discriminatory intent, establish the absence of differential treatment, or measure borrower welfare. The merge is partial and is limited to recently originated loans, and the results should be read with that in mind.

Two consequences of a binding budget are worth recording before turning to policy. Relief to one loan is funded by relief denied to another, so a loan-level estimate that ignores this interference can overstate the effect of an institution-wide change and a credible evaluation must measure effects across the intermediary's whole book (Internet Appendix~\ref{ia:displacement}). And cash is not the only constraint: staff capacity limits relief in the same way, and the two are observationally equivalent when staff and cash requirements are proportional across actions, which is why documentation and queue data are part of the mechanism rather than controls \citep{Agarwal2017,BandyopadhyayInside2026} (Internet Appendix~\ref{ia:menus}).

\subsection{Liquidity support versus ex post compensation}
Policy can support relief with liquidity, which supplies cash at the decision date, or with compensation paid after an action is completed, such as the incentive payments per completed modification that servicers received under the Home Affordable Modification Program \citep{Agarwal2017}. The two are often treated as substitutes. They are not when funding binds.

\begin{proposition}[Liquidity versus compensation]\label{prop:policy}
Consider an ex post payment $s\ge0$ per completed action and a liquidity injection $\Delta K\ge0$. The payment is valued at the decision date but cannot be pledged or advanced against; it does not change the funding technology.
\begin{enumerate}[label=(\alph*)]
\item \emph{Binding budget, constant cash need.} Suppose values have a positive density at the margin. The payment leaves the set of funded actions unchanged and raises the shadow value of cash by $s/k$. The injection raises the funded share by $\Delta K/k$ as long as the budget still binds.
\item \emph{Binding budget, heterogeneous cash needs.} A flat payment leaves the cash deployed equal to $K$ and weakly raises the number of actions only by tilting relief toward actions with lower cash needs. A payment proportional to the cash need leaves the allocation unchanged.
\item \emph{Elastic funding.} Suppose external funds $B$ cost $\Phi(B)$ with $\Phi$ increasing and strictly convex, cash needs are constant, values are $g=RQ-c$ with $Q$ continuously distributed, and at the optimum the intermediary borrows and funds an interior share. For a payment $s>0$ whose optimal funded share remains interior and has nonnegative marginal value before the payment, there is an injection that produces the same funded share. The payment raises the intermediary's external borrowing and the injection lowers it.
\end{enumerate}
\end{proposition}

Under a hard budget and the no-pledging assumption, compensation paid after completion is inframarginal: it raises the value of each action but not the cash available to take one, so the budget spends it on the same actions or on cheaper ones. Liquidity support expands relief directly. Relief that is mandated without funding is the limiting case, and its cost need not stay in the relief program: the banks that bore the Hungarian early-repayment scheme lent less afterward \citep{SchindeleLoranthOlah2026}. When borrowing has a rising marginal cost, compensation can expand relief. A liquidity injection can match this expansion only while the marginal action remains nonnegative in value before compensation; a subsidy that induces negative-private-value actions cannot in general be replicated by liquidity alone. Where the equivalence holds, facility draws distinguish the instruments. A payment made at the time of the action combines the two: it is both liquidity and compensation. The Pass-Through Assistance Program was liquidity priced as a last resort, which places it in case (c) with a steep $\Phi$. At a continuous interior cutoff, the social return to a marginal dollar of liquidity is $\bar w(q^*)/\ell(q^*)$. That is a local expression; a full benefit-cost calculation also requires fiscal costs, default risk, and the counterfactual private financing.

\section{Conclusion}\label{sec:conclusion}
The financing of an intervention can shape which loans receive it and when. In Ginnie Mae disclosures, nonbanks temporarily shifted repurchases away from large balances during the 2020 disruption, continued to select on an index of earlier decisions, and subsequently repurchased many of the options initially left unexercised. Depositories followed a different path. These patterns are consistent with a cash constraint, but unobserved funding conditions, pre-trends, and issuer-specific valuations limit a causal interpretation.

The model explains why information and funding capacity must be evaluated jointly. With common cash needs, the allocation value of information peaks at an intermediate capacity, while full funding leaves only screening value. Under the stated profitability conditions, AUC averages ranking value across capacities and need not select the best model for a given intermediary. These are theoretical results, not estimates of the value of the empirical exercise index.

The distinction between repurchase and borrower relief also matters for policy. The benchmark repurchase shortfall describes the timing of an intermediary action; it is not a welfare loss. Evaluating liquidity support requires evidence on funding, subsequent resolution, and borrower outcomes. The framework identifies the relevant margins and clarifies why compensation that cannot be pledged need not substitute for cash available when the action is taken.

\section*{Declarations}
\ifdefined\ANON\else
\paragraph{Declaration of competing interest.} From 2019 to 2021 the author was Principal Data Scientist at Rocktop Partners, a mortgage credit investor active in the Ginnie Mae early-buyout market. The firm is not a data source for this paper and had no role in its design, analysis, or writing. The author has no other competing interests to declare.

\fi
\paragraph{Data availability.} The empirical analysis uses Ginnie Mae's public loan-level MBS disclosures, together with Freddie Mac's Primary Mortgage Market Survey (FRED) and state unemployment rates (BLS), both bundled with the code, the public Home Mortgage Disclosure Act loan application registers for 2018 through 2021 (FFIEC), from which lender-level origination totals and borrower characteristics are taken, and Ginnie Mae's monthly new-issue loan-level files for January 2018 through March 2022, which supply the debt-to-income ratio at origination. The loan-level analysis files were assembled from the published disclosures for \citet{BKS2026}, and issuers are classified by charter from public records; the accompanying analysis scripts depend on those assembled loan-level files and additional intermediate inputs. The package includes model code, empirical scripts, and aggregate outputs, but is not a self-contained raw-data replication archive. Under Ginnie Mae's terms of use the loan-level files are not redistributed; the package contains the aggregate outputs. The paper uses no proprietary data.

\paragraph{Funding.} This research did not receive any specific grant from funding agencies in the public, commercial, or not-for-profit sectors.

\paragraph{Declaration of generative AI and AI-assisted technologies in the writing process.} During the preparation of this work the author used AI language models (Claude, Anthropic, and ChatGPT, OpenAI) to assist with drafting, editing, and code for the numerical examples and the empirical analysis. The author is responsible for reviewing and verifying the final text, code, references, and results before submission and for the content of the publication.

\clearpage
\input{references.tex}

\clearpage
\appendix
\renewcommand{\theequation}{A\arabic{equation}}
\setcounter{equation}{0}
\section{Proofs}\label{app:proofs}

\paragraph{Proof of Proposition~\ref{prop:allocation}.} For any $\lambda\ge 0$ and feasible $x$,
\begin{equation*}
\int g_ix_i\,dF=\lambda\int\ell_ix_i\,dF+\int(g_i-\lambda\ell_i)x_i\,dF\le\lambda K+\int\pos{g_i-\lambda\ell_i}dF,
\end{equation*}
which gives weak duality. Let $C(\lambda)=\int\ell_i\1\{g_i>\lambda\ell_i\}dF$ be the cash needed to fund every opportunity whose value per dollar exceeds $\lambda$. $C$ is nonincreasing and right-continuous, and it is zero for $\lambda\ge\sup g/\underline\ell$. If $C(0)\le K$, set $\lambda=0$ and fund every positive-value opportunity; the bound holds with equality. Otherwise let $\lambda=\inf\{\lambda':C(\lambda')\le K\}>0$. Fund all opportunities with $g_i>\lambda\ell_i$, and fund those with $g_i=\lambda\ell_i$ with the probability that exhausts $K$; nonatomicity makes this feasible. Then $\int\ell_ix_i\,dF=K$, and $x_i=1$ exactly where $g_i-\lambda\ell_i>0$, so both inequalities hold with equality and (\ref{eq:dual}) follows. Monotonicity is immediate because a larger budget enlarges the feasible set. For concavity, if $x^1$ and $x^2$ are feasible at $K_1$ and $K_2$, their convex combination is feasible at the same combination of budgets and attains the same combination of values. By (\ref{eq:dual}), $V$ is the lower envelope of affine functions of $K$ with slopes $\lambda$, so its derivative, where it exists, equals the minimizing $\lambda$, and the minimizers are nonincreasing in $K$. \qed

\paragraph{Proof of Proposition~\ref{prop:selection}.} Under (\ref{eq:linear}), $g/\ell$ is strictly increasing in $q$, since its derivative is $(Rk-\omega c)/(k-\omega q)^2>0$, so the threshold rule of Proposition~\ref{prop:allocation} is a cutoff in $q$: $Rq-c>\lambda(k-\omega q)$ if and only if $q>(c+\lambda k)/(R+\lambda\omega)$. Differentiating gives $dq^*/d\lambda$. The cash used at cutoff $u$ is $\int_u^1(k-\omega q)dF(q)$, with derivative $-(k-\omega u)f(u)<0$, so a lower $K$ raises the cutoff. The funded share $1-F(u)$ falls. With $\bar q(u)=\E[q\mid q\ge u]$, $\bar q'(u)=f(u)[\bar q(u)-u]/(1-F(u))\ge 0$, strictly when there is mass above $u$. If $q$ is the success probability under intervention, the law of iterated expectations makes $\bar q(u)$ the success rate among funded interventions. \qed

\paragraph{Proof of Proposition~\ref{prop:dissociation}.} With $\ell_i=k$, the threshold rule funds opportunities in decreasing order of $g$ until either cash runs out or value reaches zero, so the funded share is $\min\{m,\Pr(g>0)\}$ (with randomization at atoms). Parts (a) and (b) follow directly. For (c), Proposition~\ref{prop:allocation} implies that cash deployed equals $\min\{K,\int\ell_i\1\{g_i>0\}dF\}$, which equals $K$ whenever positive-value cash demand exceeds $K$. \qed

\paragraph{Proof of Proposition~\ref{prop:info}.} \emph{Value formula.} Couple $Q=G^{-1}(U)$ with $U$ uniform on $[0,1]$. For any policy $x(\cdot)\in[0,1]$ with $\E[x]\le m$, $\E[(RQ-c)x]\le\E[\pos{RQ-c}x]\le\int_{1-m}^1\pos{RG^{-1}(u)-c}du$, because $\pos{RG^{-1}(u)-c}$ is nondecreasing in $u$ and the top-$m$ integral maximizes $\E[h(U)x]$ over such policies for any nonnegative nondecreasing $h$. The policy $x=\1\{U\ge 1-m\}\1\{RQ\ge c\}$ attains the bound, which gives (\ref{eq:VG}). Without the signal, all opportunities are identical with value $Rp-c>0$, so $V_0(m)=m(Rp-c)$.

(i) Writing $\pos{z}=z+\pos{-z}$ inside (\ref{eq:VG}) gives $V_G(m)=RT_G(m)-cm+\mathcal S(m)$, so $\Delta V=R[T_G(m)-mp]+\mathcal S(m)$. Since $\E[Q]=p$, $T_G(1)=p$ and $\mathcal A(1)=0$. $\mathcal A'(m)=R[G^{-1}(1-m)-p]$ is nonincreasing, so $\mathcal A$ is concave, and a concave function that vanishes at both endpoints is nonnegative. $\mathcal S'(m)=\pos{c-RG^{-1}(1-m)}$ is nonnegative and nondecreasing in $m$, and it is zero whenever $RG^{-1}(1-m)\ge c$, which holds for all $m\le\Pr(RQ\ge c)$.

(ii) $V_G'(m)=\pos{RG^{-1}(1-m)-c}$ is nonincreasing, so $V_G$ is concave; $V_0$ is linear; hence $\Delta V$ is concave. $\Delta V(1)=\E\pos{RQ-c}-(Rp-c)=\E\pos{c-RQ}$. For almost every $m$, $\Delta V'(m)=\pos{RG^{-1}(1-m)-c}-(Rp-c)$. If $m<m^*$, then $1-m>G(p)$, so $G^{-1}(1-m)>p$ and $\Delta V'(m)>0$. If $m>m^*$, then $1-m<G(p)$, and because $\Pr(Q=p)=0$, $G^{-1}(1-m)<p$. Since $Rp-c>0$, $\pos{RG^{-1}(1-m)-c}<Rp-c$, so $\Delta V'(m)<0$.

(iii) Since $V'(K)=V_G'(m)/k$ with the signal and $V_0'(m)/k$ without it, the cross-difference in marginal values equals $\Delta V'(m)/k$, whose sign is given in (ii).

(iv) For $u>G(p)=1-m^*$, $G^{-1}(u)>p$, so $RG^{-1}(u)-c>0$. Hence $\Delta V(m^*)=R\int_{1-m^*}^1[G^{-1}(u)-p]du=R\,\E[(Q-p)\1\{Q>p\}]=R\,\E\pos{Q-p}$, and $\mathcal S(m^*)=0$. Because $\E[Q-p]=0$, $\E\pos{Q-p}=\E\pos{p-Q}=\tfrac12\E|Q-p|$. \qed

\paragraph{Proof of Corollary~\ref{cor:binary}.} Bayes' rule gives $q_H$, $q_L$, and $\pi_Hq_H+(1-\pi_H)q_L=p$. With $Rq_L>c$, $\mathcal S\equiv 0$ and $\Delta V=R[T_G(m)-mp]$. For $m\le\pi_H$, $T_G(m)=q_Hm$. For $m\ge\pi_H$, $T_G(m)=\pi_Hq_H+(m-\pi_H)q_L=p-(1-m)q_L$, so $\Delta V=R(p-q_L)(1-m)$. Here $m^*=\pi_H$. \qed

\paragraph{Proof of Proposition~\ref{prop:reliance}.} With $RQ\ge c$, the intermediary funds the top share $m$ of posteriors, so $x=\1\{U\ge 1-m\}$ in the coupling $Q=G^{-1}(U)$, and $r=h(G^{-1}(U))$ is nondecreasing in $U$. (i) $\mathrm{Cov}(x,r)=\E[xr]-\E[x]\E[r]=\int_{1-m}^1h(G^{-1}(u))\,du-m\,\E[r]$. Its derivative in $m$ is $h(G^{-1}(1-m))-\E[r]$, which is nonincreasing, so the covariance is concave; it is zero at $m=0$ and, since the integral equals $\E[r]$ at $m=1$, at $m=1$; and its derivative changes sign where $h(G^{-1}(1-m))$ crosses $\E[r]$, that is, at $m=\Pr(r>\E[r])$. For $h$ the identity, the covariance is $T_G(m)-mp=\mathcal A(m)/R$, and the crossing is at $m^*$. (ii) Dividing the covariance by $\Pr(x=1)=m$ gives $\E[r\mid x=1]-\E[r]$, and $\E[r\mid x=1]=m^{-1}\int_0^mh(G^{-1}(1-v))\,dv$ is the average of a nonincreasing function over $[0,m]$, hence nonincreasing in $m$. (iii) If $m\le m^*$, every funded loan has $Q>p$, so $\Pr(x=1\mid Q>p)=m/m^*$ and $\Pr(x=1\mid Q<p)=0$. If $m\ge m^*$, every loan with $Q>p$ is funded and the remaining $m-m^*$ is drawn from the $1-m^*$ with $Q<p$, so $D=1-(m-m^*)/(1-m^*)=(1-m)/(1-m^*)$. \qed

\paragraph{Proof of Proposition~\ref{prop:auc}.} Equation (\ref{eq:cap}) follows from the definitions. (i) Let $t=c/R$ and $h_G(m')=T_G(m')-tm'$. Since $h_G'(m')=G^{-1}(1-m')-t$ is nonincreasing, $h_G$ is concave and maximized over $[0,m]$ at $\min\{m,\Pr(Q>t)\}$. The top-$m$ integral in (\ref{eq:VG}) keeps exactly the quantiles with $G^{-1}(u)>t$, so $V_G(m)=R\max_{m'\le m}h_G(m')$. If $T_A\ge T_B$ pointwise, then $h_A\ge h_B$ pointwise and $V_A(m)\ge V_B(m)$ for every $m$ and $c$. Conversely, with $c=0$, $V_G(m)=RT_G(m)$, so dominance at every $m$ requires $T_A\ge T_B$. For distributions with equal means, $T_A(m)\ge T_B(m)$ for all $m$ is equivalent to $Q_A\cx Q_B$ \citep[][Theorem 3.A.5]{ShakedShanthikumar2007}. For a binary state, one experiment is Blackwell more informative than another if and only if its distribution of posteriors is a mean-preserving spread of the other's \citep{Blackwell1953}. Since $\Gamma_G=T_G/p$, the condition is equivalent to $\Gamma_A\ge\Gamma_B$.

(ii) With $RQ\ge c$ almost surely, $\mathcal S\equiv 0$ and $\Delta V_G=\mathcal A_G$. Exchanging the order of integration,
\begin{equation*}
\int_0^1\mathcal A_G(m)\,dm=R\int_0^1[G^{-1}(u)-p]\,u\,du=R\,\mathrm{Cov}(Q,U),
\end{equation*}
where $U$ is the uniform rank in the coupling $Q=G^{-1}(U)$, which randomizes within ties. Let $Y$ be the success indicator, with $\E[Y\mid Q,U]=Q$, so $\mathrm{Cov}(Q,U)=\mathrm{Cov}(Y,U)=p\,(\E[U\mid Y=1]-\tfrac12)$. Let $U'$ be the rank of an independent draw. Then $\E[U\mid Y=1]=\Pr(U'<U\mid Y=1)=p\cdot\tfrac12+(1-p)\Pr(U_0<U_1)$, where $U_1$ and $U_0$ are the ranks of an independent success and failure. With randomization within ties, $\Pr(U_0<U_1)=\Pr(Q_0<Q_1)+\tfrac12\Pr(Q_0=Q_1)=\AUC_G$. Substituting gives $\mathrm{Cov}(Y,U)=p(1-p)(\AUC_G-\tfrac12)$.

(iii) If $\Gamma_A(m_1)>\Gamma_B(m_1)$ and $\Gamma_A(m_2)<\Gamma_B(m_2)$, then with $c$ small enough that $RQ\ge c$ under both scores, (\ref{eq:cap}) gives strict rankings in opposite directions at $m_1$ and $m_2$, and (ii) gives the AUC ranking of the averages. \qed

\paragraph{Proof of Proposition~\ref{prop:finance}.} (a) Each funded loan requires internal cash $k-Lp>0$ whatever its posterior, so $m\le K_0/(k-Lp)$. Since $RQ>c$, the budget binds.

(b) Given a per-loan advance $\alpha$, the intermediary receives $\alpha$ now and repays $L$ upon success, so a funded loan contributes $(R-L)Q-c+\alpha$ and uses internal cash $k-\alpha$. Because $R>L$, the intermediary funds the highest posteriors and, since each marginal loan has positive value, exhausts the budget: $m(k-\alpha)=K_0$. A competitive lender that knows $G$ and observes the funded set breaks even at $\alpha=L\,\E[Q\mid\text{top }m]=L\,T_G(m)/m$. Combining gives $km-LT_G(m)=K_0$. The left side has derivative $k-LG^{-1}(1-m)\ge k-L>0$, so the solution is unique. At the solution, the marginal loan has $Q_m\le\alpha/L$, so $(R-L)Q_m-c+\alpha\ge RQ_m-c>0$, which confirms that the intermediary funds up to the budget. Under loan-level disclosure, the advance is $LQ_i$ and each loan uses internal cash $k-LQ_i$, which is lowest for the highest posteriors; funding the top share $m$ uses $km-LT_G(m)$, which gives the same $m_R$.

(c) $T_G(m)\ge mp$, with strict inequality for $m\in(0,1)$ when $Q$ is nondegenerate (Proposition~\ref{prop:info}(i)). Hence $km-LT_G(m)\le m(k-Lp)$, so any $m$ feasible under (a) is feasible under (b), and $m_H$ is strictly slack under (b). Because $K_0<k-Lp$, both shares are below one.

(d) Financing is fair at the pool level under (b), so private value is $V_G(m_R)$. Under (a), the intermediary receives $Lp$ per loan but repays $L$ with probability equal to its own posterior, a transfer of $L[T_G(m_H)-m_Hp]\ge 0$ to the lender. Since $V_G$ is nondecreasing and $m_R\ge m_H$, $V_G(m_R)\ge V_G(m_H)\ge V_G(m_H)-L[T_G(m_H)-m_Hp]$. \qed

\paragraph{Proof of Proposition~\ref{prop:welfare}.} Without the signal, the intermediary cannot distinguish loans within the cell, so any funded share $m$ is independent of $Q$ and borrower surplus is $m\,\E[\beta(Q)]$. With it, the intermediary funds the top share $m$, and borrower surplus is $\int_{1-m}^1\beta(G^{-1}(u))du$. If $\beta$ is nonincreasing, $\beta\circ G^{-1}$ is nonincreasing in $u$, and its average over $[1-m,1]$ is at most its average over $[0,1]$, which gives (i); the inequality is strict for $m\in(0,1)$ unless $\beta\circ G^{-1}$ is almost everywhere constant. The same argument applied to $\bar w$ gives (ii), and private value rises by Proposition~\ref{prop:info}. For (iii), the planner can replicate random allocation, so its optimum with the signal is at least $m\,\E[\bar w(Q)]$. \qed

\paragraph{Proof of Proposition~\ref{prop:policy}.} (a) A payment $s$ raises every $g_i$ by $s$, which preserves the ranking and weakly enlarges the positive-value set, so the budget still binds and the funded set is the same top share $K/k$. With a positive density of values at the margin, the multiplier is unique, and the marginal opportunity's value $g^*$ satisfies $g^*+s=\lambda'k$ against $g^*=\lambda k$ before, so $\lambda'=\lambda+s/k$. An injection raises $K/k$ one-for-one while the budget binds.

(b) The budget binds before the payment and the positive-value set weakly expands, so cash deployed remains $K$. Let $x$ and $x'$ be optimal at payments $s<s'$, with values $\Gamma(x)=\int g_ix_i\,dF$ and counts $N(x)=\int x_i\,dF$. Optimality gives $\Gamma(x')+s'N(x')\ge \Gamma(x)+s'N(x)$ and $\Gamma(x)+sN(x)\ge \Gamma(x')+sN(x')$. Adding the two yields $(s'-s)[N(x')-N(x)]\ge 0$, so the count weakly rises. It rises only if the composition of funded actions changes, and the cash constraint then requires that actions with larger cash needs be replaced by actions with smaller ones. A payment $\sigma\ell_i$ changes the ranking ratio from $g_i/\ell_i$ to $g_i/\ell_i+\sigma$, which preserves the order, so the allocation is unchanged.

(c) With funded share $m$ and marginal posterior $q(m)=G^{-1}(1-m)$, external borrowing is $B=\pos{mk-K}$ and the first-order condition at an interior optimum with $B>0$ is $Rq(m)-c+s=k\,\Phi'(mk-K)$. The left side is decreasing in $m$ and the right side is increasing, so $m$ rises with $s$ and, holding $s=0$, with $K$. For a payment $s$ that yields share $m_s$, the injection $\Delta K=m_sk-K-(\Phi')^{-1}\big((Rq(m_s)-c)/k\big)$ yields the same share, provided $(Rq(m_s)-c)/k\ge\Phi'(0)$; if instead $0\le(Rq(m_s)-c)/k<\Phi'(0)$, the injection optimum is a corner with $B=0$ and $\Delta K=m_sk-K$. If $Rq(m_s)-c<0$, liquidity alone cannot implement the subsidized share: the intermediary would not fund a negative-value marginal action even with internal cash. Under the payment, $B=m_sk-K$ exceeds its initial value because $m_s$ rises. Under the injection, $\Phi'(B)=(Rq(m_s)-c)/k$ is below its initial value because $q(m_s)$ falls, so $B$ falls, to zero in the corner case. \qed

\clearpage
\section*{Figures and Tables}

\begin{table}[H]\centering
\caption{Rationing as Liquidity Increases\label{tab:uniform}}
\begin{tabular}{rrrrrrr}\toprule
 & & Cutoff & Shadow value & Funded & Private & Success\\
$K/\bar K$ & $K$ & $q^*$ & $\lambda$ & share & value & among funded\\\midrule
\UniformRows
\bottomrule\end{tabular}
\begin{minipage}{0.97\textwidth}\footnotesize\setstretch{1.0}\smallskip
The table reports the model of Figure~\ref{fig:rationing} at four liquidity levels. Private value and liquidity are per unit mass of opportunities. Success is the success probability among funded interventions. An independent linear program with 4,000 posterior types reproduces private value to within $1.3\times10^{-8}$ at 17 liquidity levels.
\end{minipage}
\end{table}

\begin{table}[H]\centering
\caption{Value of Two Scores at Different Fundable Shares\label{tab:scores}}
\begin{tabular}{rrrr}\toprule
Fundable share $m$ & Score A (AUC = \AUCA) & Score B (AUC = \AUCB) & Ratio A/B\\\midrule
\ScoreRows
\bottomrule\end{tabular}
\begin{minipage}{0.97\textwidth}\footnotesize\setstretch{1.0}\smallskip
Allocation value per loan in the model of Proposition~\ref{prop:auc}, with the scores of Figure~\ref{fig:scores}. The first and last rows set the fundable share to the nonbank and depository buyout exercise rates in June 2020 in the sample of Section~\ref{sec:evidence}. Averaged over fundable shares drawn uniformly from $[0,1]$, the values are $0.05$ and $0.075$, equal to $Rp(1-p)(\AUC-\tfrac12)$ as in equation~(\ref{eq:aucid}).
\end{minipage}
\end{table}

\begin{table}[H]\centering
\caption{Rationing along a Fixed Ranking, 2019--2022\label{tab:empirics}}
\scriptsize
\setlength{\tabcolsep}{2.6pt}
\begin{tabular}{lcccccc}\toprule
 & Pre-shock & Freeze & Release & \multicolumn{2}{c}{Recovery} & Rate rise\\\cmidrule(lr){5-6}
 & 3/19--2/20 & 3/20--6/20 & 7/20--12/20 & 1/21--6/21 & 7/21--12/21 & 1/22--9/22\\\midrule
\multicolumn{7}{l}{\emph{Panel A. Nonbank issuers}}\\
\NbRows\addlinespace
\multicolumn{7}{l}{\emph{Panel B. Depository issuers}}\\
\DepRows\bottomrule
\end{tabular}
\begin{minipage}{0.97\textwidth}\footnotesize\setstretch{1.0}\smallskip
The sample is \eEmpN{} Ginnie Mae buyout options that vested from January 2019 to September 2022 at \eEmpIssuers{} depository and nonbank issuers; exercise is repurchase within three months of vesting. The six periods begin in March 2019, so options that vested in January and February 2019 enter the sample and the index diagnostics but no column. The ranking index is the linear predictor of a logit of exercise on hard loan characteristics with issuer-month fixed effects, estimated on nonbank decisions in 2016--2018 and standardized over the sample. Cells are issuer by vesting month by forbearance status. $^{a}$The pre-shock column reports the level; later columns report the change from it. The latent weight is the coefficient on the index in a logit with cell fixed effects and period interactions; standard errors are clustered by issuer. The level response is the within-cell covariance of exercise with the index divided by its variance; the relative response divides it by the funded share. Fixed-ranking predictions hold the latent weight at its pre-shock estimate and match each cell's funded share; in the pre-shock period they equal the actual responses by construction, because the logit's score equations match the covariance of exercise with the index. The random-rationing benchmark has each issuer fund, along the same ranking, the options it would have funded at its own pre-shock share and exercise a random fraction of them that matches its observed funded share; it lies below the actual response in the pre-shock period because months in which an issuer's share falls below its own average are scaled down, and the last row therefore compares changes from the pre-shock period. Standard errors for differences are from \eNbBootB{} bootstrap draws over issuers that re-estimate the pre-shock weight.
\end{minipage}
\end{table}

\begin{table}[H]\centering
\caption{The Price of Cash in the Ranking: The Weight on Loan Balance\label{tab:balance}}
\scriptsize
\setlength{\tabcolsep}{3pt}
\begin{tabular}{lcccccc}\toprule
 & Pre-shock & Freeze & Release & \multicolumn{2}{c}{Recovery} & Rate rise\\\cmidrule(lr){5-6}
 & 3/19--2/20 & 3/20--6/20 & 7/20--12/20 & 1/21--6/21 & 7/21--12/21 & 1/22--9/22\\\midrule
\BalanceRows\bottomrule
\end{tabular}
\begin{minipage}{0.97\textwidth}\footnotesize\setstretch{1.0}\smallskip
Coefficients from a logit of exercise within three months of vesting on the standardized ranking index and standardized log unpaid balance, with cell fixed effects (issuer by vesting month by forbearance status) and interactions of both regressors with period indicators. The pre-shock column reports levels; later columns report changes from the pre-shock level. Splitting the freeze into March--April, when the forbearance flag is not yet observed, and May--June, when it enters the cell definition, gives changes in the weight on balance of \eBalNbDFrzEarly{} (\eBalNbDFrzEarlySE) and \eBalNbDFrzLate{} (\eBalNbDFrzLateSE) for nonbanks and \eBalDepDFrzEarly{} (\eBalDepDFrzEarlySE) and \eBalDepDFrzLate{} (\eBalDepDFrzLateSE) for depositories. Standard errors are clustered by issuer. The sample is as in Table~\ref{tab:empirics}.
\end{minipage}
\end{table}

\begin{table}[H]\centering
\caption{Postponed Options Were Exercised Later on the Same Ranking\label{tab:queue}}
\footnotesize
\begin{tabular}{lcccccc}\toprule
 & \multicolumn{3}{c}{Nonbanks} & \multicolumn{3}{c}{Depositories}\\\cmidrule(lr){2-4}\cmidrule(lr){5-7}
Vesting period & Options & Later (\%) & Weight & Options & Later (\%) & Weight\\\midrule
\QueueRows\bottomrule
\end{tabular}
\begin{minipage}{0.97\textwidth}\footnotesize\setstretch{1.0}\smallskip
The sample is options not exercised within three months of vesting and still in their pools at the end of the third month. Later is the share exercised in the fourth through twelfth months after vesting. Weight is the coefficient on the ranking index in a logit of later exercise with cell fixed effects (issuer by vesting month by forbearance status); standard errors clustered by issuer are in parentheses. For comparison, the pre-shock weight on exercise within three months is \eNbBetaPre{} for nonbanks (Table~\ref{tab:empirics}).
\end{minipage}
\end{table}

\begin{table}[H]\centering
\caption{Repurchases Depositories Made and Nonbanks Did Not: Forbearance, May--June 2020\label{tab:forbearance}}
\begin{tabular}{lcc}\toprule
 & Nonbanks & Depositories\\\midrule
Share of options in forbearance (\%) & \eNbFbShare & \eDepFbShare\\
Exercised within three months, in forbearance (\%) & \eNbFbRate & \eDepFbRate\\
Exercised within three months, not in forbearance (\%) & \eNbNoFbRate & \eDepNoFbRate\\
Difference within issuer-month (pp) & \eNbFbCoef{} (\eNbFbSE) & \eDepFbCoef{} (\eDepFbSE)\\
Same, within index decile (pp) & \eNbFbCoefCtrl{} (\eNbFbSECtrl) & \eDepFbCoefCtrl{} (\eDepFbSECtrl)\\
Exercised by June 30, 2020, in forbearance (\%) & \eNbFbByJune & \eDepFbByJune\\
Exercised by June 30, 2020, not in forbearance (\%) & \eNbNoFbByJune & \eDepNoFbByJune\\
Exercised by December 2022, in forbearance (\%) & \eNbFbEver & \eDepFbEver\\
Issuers & \eNbFbIssuers & \eDepFbIssuers\\\bottomrule
\end{tabular}
\begin{minipage}{0.97\textwidth}\footnotesize\setstretch{1.0}\smallskip
Buyout options that vested in May and June 2020, the first months in which the disclosures flag forbearance. The within-issuer-month difference is the coefficient on a forbearance indicator in a linear probability model of exercise with issuer-month fixed effects; the next row adds indicators for deciles of the ranking index. Standard errors are clustered by issuer.
\end{minipage}
\end{table}

\begin{table}[H]\centering\small
\caption{Origination Intensity and the Funded Share of Nonbank Issuers\label{tab:origination}}
\setstretch{1.0}
\setlength{\tabcolsep}{4pt}\begin{tabular}{lccccc}\toprule
 & (1) & (2) & (3) & (4) & (5)\\
 & Baseline & Weighted & 2018 & Purchases & All\\\midrule
Intensity $\times$ freeze (Mar--Jun 2020) & \fOrAFrz & \fOrBFrz & \fOrCFrz & \fOrDFrz & \fOrEFrz\\
 & (\fOrAFrzSE) & (\fOrBFrzSE) & (\fOrCFrzSE) & (\fOrDFrzSE) & (\fOrEFrzSE)\\
 & [\fOrAP] & [\fOrBP] & [\fOrCP] & [\fOrDP] & [\fOrEP]\\
Intensity $\times$ release (Jul--Dec 2020) & \fOrARel & \fOrBRel & \fOrCRel & \fOrDRel & \fOrERel\\
 & (\fOrARelSE) & (\fOrBRelSE) & (\fOrCRelSE) & (\fOrDRelSE) & (\fOrERelSE)\\
Intensity $\times$ Jan--Jun 2021 & \fOrARecOne & \fOrBRecOne & \fOrCRecOne & \fOrDRecOne & \fOrERecOne\\
 & (\fOrARecOneSE) & (\fOrBRecOneSE) & (\fOrCRecOneSE) & (\fOrDRecOneSE) & (\fOrERecOneSE)\\
Intensity $\times$ Jul--Dec 2021 & \fOrARecTwo & \fOrBRecTwo & \fOrCRecTwo & \fOrDRecTwo & \fOrERecTwo\\
 & (\fOrARecTwoSE) & (\fOrBRecTwoSE) & (\fOrCRecTwoSE) & (\fOrDRecTwoSE) & (\fOrERecTwoSE)\\
Intensity $\times$ 2022 & \fOrARate & \fOrBRate & \fOrCRate & \fOrDRate & \fOrERate\\
 & (\fOrARateSE) & (\fOrBRateSE) & (\fOrCRateSE) & (\fOrDRateSE) & (\fOrERateSE)\\
\midrule
Issuers & \fOrAG & \fOrBG & \fOrCG & \fOrDG & \fOrEG\\
Issuer-months & \fOrAN & \fOrBN & \fOrCN & \fOrDN & \fOrEN\\\bottomrule
\end{tabular}
\begin{minipage}{0.97\textwidth}\footnotesize\setstretch{1.0}\smallskip
Issuer-month regressions of the funded share (options exercised within three months of vesting, in percentage points) on the log of origination intensity interacted with the periods of Table~\ref{tab:empirics}, with issuer and month fixed effects, March 2019 to September 2022; the omitted period is March 2019 to February 2020. Origination intensity is the dollar volume of mortgages the issuer originated in 2019 (HMDA, all loan types) divided by the unpaid balance of its Ginnie Mae servicing portfolio in December 2019, floored at 0.01. Columns 1 to 4 use the nonbanks that exercised at least 5\% of their pre-shock options and had at least 50 options before the shock and in the freeze. Column 2 weights issuer-months by options. Column 3 uses 2018 originations and the December 2018 balance. Column 4 adds loans purchased in 2019 to the numerator. Column 5 uses every nonbank with a HMDA match or no originations; issuers that exercised almost nothing before the shock have no funded share to lose, which biases its freeze coefficient upward. Issuer-months with fewer than ten options are dropped. Standard errors clustered by issuer in parentheses; wild cluster bootstrap $p$-values (Rademacher weights, null imposed, 1,999 draws) in brackets for the freeze coefficient. The 2022 coefficient in the unweighted columns reflects the pre-shock level: issuers that exercised more before the shock had further to fall when rates rose, and it is $-0.07$ (s.e.\ 0.31) in the index metric. Dropping the five issuers matched to a HMDA reporter through a trade name that addresses cannot confirm leaves the column 5 freeze coefficient at \fOrDrop{} (\fOrDropSE; [\fOrDropP]; \fOrDropG{} issuers); none of them is among the issuers of columns 1 to 4.
\end{minipage}
\end{table}

\begin{table}[H]\centering\small
\caption{The Fall in the Exercise Index by Mortgage Spread\label{tab:spreadindex}}
\setstretch{1.0}
\begin{tabular}{lccccc}\toprule
Quintile of mortgage spread & 1 & 2 & 3 & 4 & 5\\
Mean spread (percentage points) & \fSpMsOne & \fSpMsTwo & \fSpMsThree & \fSpMsFour & \fSpMsFive\\\midrule
\multicolumn{6}{l}{\emph{Nonbanks}}\\
Exercised, Mar 2019--Feb 2020 (\%) & \fSpRNbPreOne & \fSpRNbPreTwo & \fSpRNbPreThree & \fSpRNbPreFour & \fSpRNbPreFive\\
Exercised, Mar--Jun 2020 (\%) & \fSpRNbFrzOne & \fSpRNbFrzTwo & \fSpRNbFrzThree & \fSpRNbFrzFour & \fSpRNbFrzFive\\
Fall in log-odds & \fSpNbOne & \fSpNbTwo & \fSpNbThree & \fSpNbFour & \fSpNbFive\\
 & (\fSpNbSEOne) & (\fSpNbSETwo) & (\fSpNbSEThree) & (\fSpNbSEFour) & (\fSpNbSEFive)\\
\multicolumn{6}{l}{\emph{Depositories}}\\
Exercised, Mar 2019--Feb 2020 (\%) & \fSpRDepPreOne & \fSpRDepPreTwo & \fSpRDepPreThree & \fSpRDepPreFour & \fSpRDepPreFive\\
Exercised, Mar--Jun 2020 (\%) & \fSpRDepFrzOne & \fSpRDepFrzTwo & \fSpRDepFrzThree & \fSpRDepFrzFour & \fSpRDepFrzFive\\
Fall in log-odds & $-$\fSpDepOne & $-$\fSpDepTwo & $-$\fSpDepThree & $-$\fSpDepFour & $-$\fSpDepFive\\\midrule
Nonbank less depository & \fSpNetOne & \fSpNetTwo & \fSpNetThree & \fSpNetFour & \fSpNetFive\\
 & (\fSpNetSEOne) & (\fSpNetSETwo) & (\fSpNetSEThree) & (\fSpNetSEFour) & (\fSpNetSEFive)\\\bottomrule
\end{tabular}
\begin{minipage}{0.97\textwidth}\footnotesize\setstretch{1.0}\smallskip
Options that vested from March 2019 to June 2020, by quintile of the note rate minus the 30-year mortgage rate in the vesting month (quintiles of the pooled sample). The fall in log-odds is the logit of the pre-shock exercise rate minus the logit of the freeze rate. Standard errors in parentheses from 999 bootstrap draws that resample issuers within type. Top quintile less the mean of quintiles 2 to 4, nonbanks: $-\fSpTopMid$ (s.e.\ \fSpTopMidSE).
\end{minipage}
\end{table}

\begin{table}[H]\centering
\caption{Alternative Explanations and the Evidence\label{tab:alternatives}}
\scriptsize\setstretch{1.0}
\begin{tabularx}{\textwidth}{>{\raggedright\arraybackslash}p{.20\textwidth}>{\raggedright\arraybackslash}p{.34\textwidth}Y}\toprule
Alternative & Distinguishing prediction & Evidence\\\midrule
Random rationing: exercise cut without regard to the ranking & Relative response stays near its pre-shock level as the funded share falls (\eNbFrzRelRand{} in the freeze, \eNbRelRelRand{} in the release); no change in the weight on balance & Release response \eNbRelRel{} against \eNbRelRelRand{} (difference \eNbRelGapRand, s.e.\ \eNbRelGapRandSE; Table~\ref{tab:empirics}); weight on balance fell by \eBalNbDFrzAbs{} (\eBalNbDFrzSE) and recovered (Table~\ref{tab:balance})\\\addlinespace[3pt]
Loss of discrimination: the index describes the ranking less well in the crisis & Relative response falls; latent weight on the index falls and stays low; postponed options exercised without regard to the index & Freeze change in the latent weight \eNbBetaDFrz{} (\eNbBetaDFrzSE) does not exclude a partial loss; 2021 changes \eNbBetaDRecOne{} and \eNbBetaDRecTwo{} (Table~\ref{tab:empirics}); postponed options exercised with weight \eNbQFrzBeta{} (\eNbQFrzSE) on the index (Table~\ref{tab:queue}); attenuation cannot push the weight on balance below zero (Section~\ref{sec:perdollar}); outside the freeze, reliance tracks the pre-shock weight across funded shares (Figure~\ref{fig:shape})\\\addlinespace[3pt]
Uniform value decline under stable funding and composition & Both issuer types weakly cut exercise; a uniform cure-probability decline in the stated specification affects high-spread loans most & The fall in the exercise index is \fSpTopMid{} smaller in the top spread quintile than in the middle three (s.e.\ \fSpTopMidSE; Table~\ref{tab:spreadindex}); nonbank exercise fell \eSpNbTopFall\% in the top spread quintile against \eSpNbOtherFallLo--\eSpNbOtherFallHi\% in the others, and depositories' rose in every quintile (Section~\ref{sec:alternatives}); Depositories' funded share rose from \eDepPreM\% to \eDepFrzM\% (Table~\ref{tab:empirics}); within the same pool and month, nonbank-serviced loans were far less likely to be repurchased (Section~\ref{sec:companion}); depositories' weight on balance unchanged (Table~\ref{tab:balance})\\\addlinespace[3pt]
Composition: loans in forbearance arrived in large numbers from March 2020 & The collapse reflects a kind of loan no issuer would repurchase; differences disappear once forbearance status is held fixed & In May--June the nonbank collapse is concentrated on forbearance loans, but depositories repurchased \eDepFbRate\% of them and nonbanks \eNbFbRate\%, a gap that remains within issuer-months and index deciles (Table~\ref{tab:forbearance}); May--June change in the weight on balance, within forbearance status, \eBalNbDFrzLate{} (\eBalNbDFrzLateSE) (Table~\ref{tab:balance})\\\addlinespace[3pt]
Aggregation: one large issuer drives the aggregate & The aggregate response is not shared by most issuers & Four of the five largest nonbanks with an estimable pre-shock weight and the small issuers as a group kept responding, in point estimates; one large issuer did not (Internet Appendix Table~\ref{tab:issuers})\\\addlinespace[3pt]
Local loan quality: issuers with portfolios where the pandemic hit hardest held worse loans, not tighter cash & An issuer's collapse scales with the shock in the states where its loans sit & Within type, the size of an issuer's collapse is not significantly related to its exposure to state unemployment shocks or to its forbearance share, with imprecise estimates (Internet Appendix~\ref{ia:funding}); depositories in the same pools did not cut (Section~\ref{sec:companion}). An analogue of the distant-exposure design of \citet{Aiello2022} has no first stage in these data, so it cannot identify the funding channel across issuers here (Internet Appendix~\ref{ia:funding})\\\addlinespace[3pt]
Not separated by the data & & The 2022 fall in nonbank exercise mixes a value shock with higher funding costs (Section~\ref{sec:unconstrained}); the freeze months alone do not separate a fixed ranking from a mixture of ranked and random rationing (Section~\ref{sec:fixedrank})\\\bottomrule
\end{tabularx}
\begin{minipage}{0.97\textwidth}\footnotesize\setstretch{1.0}\smallskip
Each row states an alternative to rationing along a fixed ranking, the prediction that distinguishes it, and the evidence bearing on that prediction; the last row records what the evidence does not separate. Predictions are conditional on the stated specifications; heterogeneous value shocks and issuer-specific collateral constraints are not ruled out. Numbers are from the tables cited.
\end{minipage}
\end{table}

\begin{table}[H]\centering
\caption{Predictions and Evidence\label{tab:predictions}}
\scriptsize\setstretch{1.0}
\begin{tabularx}{\textwidth}{>{\raggedright\arraybackslash}p{.23\textwidth}>{\raggedright\arraybackslash}p{.32\textwidth}Y}\toprule
Prediction & Observable implication & Current evidence\\\midrule
1. Under binding funding, profitable actions can go unfunded, ranked by value per dollar (Prop.~\ref{prop:allocation}) & Constrained intermediaries forgo or postpone actions whose value has not fallen, tilt against cash-intensive ones when cash is scarce, and cut most where inflows are smallest & Nonbanks' weight on balance fell by \eBalNbDFrzAbs{} in the freeze and recovered; depositories' unchanged (Table~\ref{tab:balance}); by June 30, 2020, depositories had repurchased \eDepFbByJune\% of May--June options in forbearance, nonbanks \eNbFbByJune\% (Table~\ref{tab:forbearance})\; among \fOrActN{} nonbanks the freeze fall was \fOrAFrz{} points smaller per log point of origination intensity (wild cluster bootstrap $p=\fOrAP$; Table~\ref{tab:origination})\\addlinespace[3pt]
2. Tighter liquidity improves measured performance (Prop.~\ref{prop:selection}) & Cure rate among funded actions rises as the funded share falls, holding the pre-event distribution of borrower states fixed & Selection on the index did not fall as nonbank funded shares fell almost fivefold, and rose, in issuer-level point estimates, at four of the five largest nonbanks and at the small issuers as a group (Table~\ref{tab:empirics}; Internet Appendix Table~\ref{tab:issuers}); cure rates require outcomes after repurchase\\\addlinespace[3pt]
3. Liquidity moves constrained intermediaries; value moves unconstrained ones (Prop.~\ref{prop:dissociation}) & Opposite responses when a common cash drain coincides with a common rise in value; asymmetric responses to a dated value shock & 2020: nonbank funded share fell almost fivefold, depository share rose; 2022 value shock: depository share fell from \eDepRecTwoM\% to \eDepRateM\% and the ranking weight rose at both types (Table~\ref{tab:empirics})\\\addlinespace[3pt]
4. Level reliance on information is hump-shaped; relative reliance rises as liquidity contracts (Prop.~\ref{prop:reliance}) & Level slope of relief in a ranking variable falls as the funded share falls below the peak; slope relative to the base rate rises & Nonbank relative response was \eNbFrzRel{} in the freeze, between random rationing (\eNbFrzRelRand) and the fixed ranking (\eNbFrzRelPred); \eNbRelRel{} in the release and \eNbRecOneRel{} and \eNbRecTwoRel{} in 2021 against fixed-ranking values of \eNbRelRelPred, \eNbRecOneRelPred, and \eNbRecTwoRelPred; postponed options exercised on the same ranking (Table~\ref{tab:queue}); across funded shares, nonbank reliance tracks the fixed-ranking profile (curvature ratio \eShapeNbRatio, s.e.\ \eShapeNbRatioSE; Figure~\ref{fig:shape})\\\addlinespace[3pt]
5. Information and liquidity switch from complements to substitutes at $m^*$ (Props.~\ref{prop:info} and \ref{prop:reliance}) & Funding shock $\times$ prior signal interaction positive below and negative above $\hat m^*$ & Indirect screening evidence: depositories' level response rose from \eDepPreLvl{} to \eDepRateLvl{} during the 2022 rate increase (Figure~\ref{fig:queue}); funding interaction untested (Internet Appendix~\ref{ia:design})\\\bottomrule
\end{tabularx}
\begin{minipage}{0.97\textwidth}\footnotesize\setstretch{1.0}\smallskip
Proposition numbers refer to Sections~\ref{sec:model} and \ref{sec:information}. The four implications the disclosures cannot bear on are in Internet Appendix Table~\ref{tab:predictions_untested}. Numbers are from the tables cited.
\end{minipage}
\end{table}

\ifdefined\NOIA\else
\clearpage
\begin{center}{\Large Internet Appendix for}\\[4pt]{\large ``The Financing of Debt Relief: Intermediary Liquidity and the Value of Borrower Information''}\end{center}
\setcounter{section}{0}
\renewcommand{\thesection}{IA.\arabic{section}}
\renewcommand{\theHsection}{IA.\arabic{section}}
\setcounter{table}{0}\renewcommand{\thetable}{IA.\arabic{table}}\renewcommand{\theHtable}{IA.\arabic{table}}
\setcounter{figure}{0}\renewcommand{\thefigure}{IA.\arabic{figure}}\renewcommand{\theHfigure}{IA.\arabic{figure}}
\renewcommand{\theequation}{IA\arabic{equation}}
\setcounter{equation}{0}
\input{internet_appendix.tex}

\fi
\end{document}

%% file: preamble.tex
\documentclass[12pt]{article}
\usepackage[letterpaper,margin=1in]{geometry}
\usepackage[T1]{fontenc}
\usepackage{lmodern,microtype,setspace,amsmath,amssymb,amsthm,booktabs,tabularx,longtable,graphicx,xcolor}
\usepackage[authoryear,round]{natbib}
\usepackage{enumitem,caption,float,array,footmisc}
\usepackage[colorlinks=true,linkcolor=black,citecolor=black,urlcolor=black]{hyperref}
\newcolumntype{Y}{>{\raggedright\arraybackslash}X}
\hypersetup{pdftitle={The Financing of Debt Relief: Intermediary Liquidity and the Value of Borrower Information},pdfauthor={Arka P. Bandyopadhyay}}
\ifdefined\ANON\hypersetup{pdfauthor={},pdfsubject={}}\fi
\setlist{nosep,leftmargin=*}
\theoremstyle{plain}
\newtheorem{proposition}{Proposition}
\newtheorem{corollary}{Corollary}

\newtheorem{assumption}{Assumption}
\theoremstyle{remark}

\newcommand{\E}{\mathbb{E}}
\newcommand{\1}{\mathbf{1}}

\newcommand{\pos}[1]{\left[#1\right]_{+}}
\newcommand{\AUC}{\mathrm{AUC}}
\newcommand{\cx}{\ge_{\mathrm{cx}}}
\DeclareMathOperator*{\argmax}{arg\,max}
\input{tables/uniform_rows.tex}
\input{tables/score_rows.tex}
\input{tables/emp_macros.tex}

\input{tables/robust_rows.tex}
\input{tables/empirics_rows.tex}
\input{tables/queue_rows.tex}
\input{tables/stock_rows.tex}
\input{tables/issuer_rows.tex}
\input{tables/balance_rows.tex}
\input{tables/funding_rows.tex}
\input{tables/shape_macros.tex}

\input{tables/synth_macros.tex}
\input{tables/fhfa_macros.tex}
\input{tables/tilt_macros.tex}
\input{tables/spread_macros.tex}

\input{tables/fold_macros.tex}
\ifdefined\NOIA\InputIfFileExists{tables/ia_labels.tex}{}{}\fi

%% file: tables/uniform_rows.tex
\newcommand{\UniformRows}{%
0.100 & 0.020 & 0.902 & 4.204 & 0.098 & 0.092 & 0.951 \\
0.300 & 0.060 & 0.718 & 2.898 & 0.282 & 0.234 & 0.859 \\
0.600 & 0.121 & 0.467 & 1.422 & 0.533 & 0.363 & 0.733 \\
1.000 & 0.201 & 0.167 & 0.000 & 0.833 & 0.417 & 0.583 \\
}

%% file: tables/score_rows.tex
\newcommand{\ScoreRows}{%
0.072 & 0.0360 & 0.0216 & 1.67 \\
0.250 & 0.0938 & 0.0750 & 1.25 \\
0.500 & 0.0625 & 0.1500 & 0.42 \\
0.750 & 0.0313 & 0.0750 & 0.42 \\
0.862 & 0.0173 & 0.0414 & 0.42 \\
}
\newcommand{\AUCA}{0.70}
\newcommand{\AUCB}{0.80}

%% file: tables/emp_macros.tex
\newcommand{\eEmpN}{1,922,668}
\newcommand{\eEmpIssuers}{120}
\newcommand{\eEmpNbN}{1,298,771}
\newcommand{\eEmpDepN}{623,897}
\newcommand{\eNbFrzDrop}{32}

\newcommand{\eNbPreM}{40.0}
\newcommand{\eNbPreLvl}{0.201}

\newcommand{\eNbPreRel}{0.50}

\newcommand{\eNbFrzM}{8.3}
\newcommand{\eNbFrzLvl}{0.043}

\newcommand{\eNbFrzRel}{0.52}
\newcommand{\eNbFrzRelPred}{0.75}
\newcommand{\eNbFrzRelRand}{0.40}
\newcommand{\eNbFrzGapFix}{$-$0.23}
\newcommand{\eNbFrzGapFixSE}{0.25}
\newcommand{\eNbFrzGapRand}{0.11}
\newcommand{\eNbFrzGapRandSE}{0.16}

\newcommand{\eNbBetaDFrz}{$-$0.49}
\newcommand{\eNbBetaDFrzSE}{0.31}

\newcommand{\eNbRelM}{15.9}

\newcommand{\eNbRelRel}{0.67}
\newcommand{\eNbRelRelPred}{0.71}
\newcommand{\eNbRelRelRand}{0.47}
\newcommand{\eNbRelGapFix}{$-$0.03}
\newcommand{\eNbRelGapFixSE}{0.11}
\newcommand{\eNbRelGapRand}{0.20}
\newcommand{\eNbRelGapRandSE}{0.08}

\newcommand{\eNbRelGapRandPLo}{0.00}
\newcommand{\eNbRelGapRandPHi}{0.31}

\newcommand{\eNbBetaDRel}{$-$0.09}
\newcommand{\eNbBetaDRelSE}{0.24}

\newcommand{\eNbRecOneM}{32.4}

\newcommand{\eNbRecOneRel}{0.43}
\newcommand{\eNbRecOneRelPred}{0.45}

\newcommand{\eNbRecOneGapFix}{$-$0.02}
\newcommand{\eNbRecOneGapFixSE}{0.05}

\newcommand{\eNbBetaDRecOne}{$-$0.07}
\newcommand{\eNbBetaDRecOneSE}{0.21}

\newcommand{\eNbRecTwoRel}{0.37}
\newcommand{\eNbRecTwoRelPred}{0.40}

\newcommand{\eNbBetaDRecTwo}{$-$0.19}
\newcommand{\eNbBetaDRecTwoSE}{0.13}

\newcommand{\eNbRateM}{7.2}

\newcommand{\eNbBetaDRate}{0.60}
\newcommand{\eNbBetaDRateSE}{0.20}

\newcommand{\eNbBetaPre}{1.45}
\newcommand{\eNbBetaPreSE}{0.29}

\newcommand{\eNbFbRate}{3.9}
\newcommand{\eNbNoFbRate}{38.6}
\newcommand{\eNbFbCoef}{$-$27.2}
\newcommand{\eNbFbSE}{10.3}
\newcommand{\eNbFbCoefAbs}{27.2}
\newcommand{\eNbFbShare}{90}
\newcommand{\eNbFbIssuers}{77}
\newcommand{\eNbFbCoefCtrl}{$-$26.6}
\newcommand{\eNbFbSECtrl}{10.0}

\newcommand{\eNbFbEver}{38.5}
\newcommand{\eNbNoFbEver}{52.0}

\newcommand{\eNbQPreRate}{5.2}
\newcommand{\eNbQPreBeta}{0.56}
\newcommand{\eNbQPreSE}{0.15}
\newcommand{\eNbQPreEver}{21.4}

\newcommand{\eNbQFrzRate}{26.4}
\newcommand{\eNbQFrzBeta}{1.05}
\newcommand{\eNbQFrzSE}{0.25}
\newcommand{\eNbQFrzEver}{37.5}

\newcommand{\eNbMayRel}{0.41}
\newcommand{\eNbMayRelPred}{0.54}
\newcommand{\eNbMayRelRand}{0.37}
\newcommand{\eNbJunM}{7.2}
\newcommand{\eNbJunRel}{0.61}
\newcommand{\eNbJunRelPred}{0.88}
\newcommand{\eNbJunRelRand}{0.39}
\newcommand{\eNbFbByJune}{1.1}
\newcommand{\eNbNoFbByJune}{37.1}
\newcommand{\eNbBootB}{200}
\newcommand{\eNbBootIssuers}{83}
\newcommand{\eDepPreM}{80.5}
\newcommand{\eDepPreLvl}{0.010}

\newcommand{\eDepFrzM}{85.3}

\newcommand{\eDepBetaDFrzSE}{0.16}
\newcommand{\eDepBetaDFrzAbs}{0.35}

\newcommand{\eDepRelM}{79.2}

\newcommand{\eDepRecTwoM}{81.0}

\newcommand{\eDepRateM}{51.9}
\newcommand{\eDepRateLvl}{0.177}

\newcommand{\eDepRateRel}{0.34}
\newcommand{\eDepRateRelPred}{0.03}

\newcommand{\eDepRateGapFix}{0.31}
\newcommand{\eDepRateGapFixSE}{0.12}

\newcommand{\eDepRateGapFixPLo}{0.10}
\newcommand{\eDepRateGapFixPHi}{0.59}

\newcommand{\eDepBetaPre}{0.13}

\newcommand{\eDepFbRate}{87.6}
\newcommand{\eDepNoFbRate}{72.1}
\newcommand{\eDepFbCoef}{13.2}
\newcommand{\eDepFbSE}{3.0}

\newcommand{\eDepFbShare}{88}
\newcommand{\eDepFbIssuers}{36}
\newcommand{\eDepFbCoefCtrl}{13.2}
\newcommand{\eDepFbSECtrl}{3.0}

\newcommand{\eDepFbEver}{91.0}

\newcommand{\eDepQFrzBeta}{0.39}
\newcommand{\eDepQFrzSE}{0.24}

\newcommand{\eDepJunM}{86.2}

\newcommand{\eDepFbByJune}{76.4}
\newcommand{\eDepNoFbByJune}{66.8}

\newcommand{\eDepBootIssuers}{37}
\newcommand{\eNbFrzTwelve}{30.7}
\newcommand{\eNbFrzEver}{40.2}

\newcommand{\eNbPreEver}{51.3}

\newcommand{\eNbFrzRelFallPI}{0.17}

\newcommand{\eQueueOctDec}{40}
\newcommand{\eTopFiveSharePre}{59}
\newcommand{\eTopFiveShareFrz}{67}
\newcommand{\eIssAFrzM}{4.2}
\newcommand{\eIssAPreM}{53.4}
\newcommand{\eIssAPreRel}{0.69}
\newcommand{\eIssAFrzRel}{1.53}
\newcommand{\eIssAFrzRelPred}{2.06}
\newcommand{\eIssAFrzRelRand}{0.74}

\newcommand{\eIssBFrzM}{2.4}

\newcommand{\eIssBFrzRel}{0.38}
\newcommand{\eIssBFrzRelPred}{2.11}
\newcommand{\eIssBFrzRelRand}{0.61}

\newcommand{\eNbPreHtwelve}{43.3}
\newcommand{\eNbPreHtwentyfour}{48.5}

\newcommand{\eNbFrzHtwelve}{30.8}
\newcommand{\eNbFrzHtwentyfour}{40.1}

\newcommand{\eQueueForgoneHtwentyfour}{8.4}
\newcommand{\eQueueGapHthree}{31.7}
\newcommand{\eQueueMadeUpShare}{73}
\newcommand{\eAdvNbPre}{0.22}
\newcommand{\eAdvNbFrz}{0.31}
\newcommand{\eAdvNbRecOne}{0.55}

\newcommand{\eAdvNbM}{$-$9.2}
\newcommand{\eAdvNbMSE}{3.1}
\newcommand{\eAdvNbMAbs}{9.2}
\newcommand{\eAdvNbMperSD}{4.1}
\newcommand{\eAdvNbLogSD}{0.44}
\newcommand{\eAdvNbMCtrl}{$-$15.1}
\newcommand{\eAdvNbMCtrlSE}{5.4}

\newcommand{\eAdvNbStockSE}{1.6}
\newcommand{\eAdvNbStockAbs}{4.1}

\newcommand{\eAdvNbN}{2,358}
\newcommand{\eAdvNbBeta}{$-$0.97}
\newcommand{\eAdvNbBetaSE}{0.47}
\newcommand{\eSSNbIssuers}{58}
\newcommand{\eSSNbLab}{16.1}
\newcommand{\eSSNbLabSE}{10.7}
\newcommand{\eSSNbFb}{$-$3.0}
\newcommand{\eSSNbFbSE}{2.3}
\newcommand{\eSSNbLabSD}{1.2}
\newcommand{\eSSNbFbSD}{4.3}
\newcommand{\eBalNbPre}{0.19}
\newcommand{\eBalNbPreSE}{0.11}

\newcommand{\eBalNbDFrzSE}{0.06}
\newcommand{\eBalNbDFrzAbs}{0.32}

\newcommand{\eBalNbDRel}{0.07}
\newcommand{\eBalNbDRelSE}{0.06}

\newcommand{\eBalNbAdv}{$-$0.06}
\newcommand{\eBalNbAdvSE}{0.12}
\newcommand{\eIdxBNbAdv}{$-$1.11}
\newcommand{\eIdxBNbAdvSE}{0.51}
\newcommand{\eSSNbLabPerSD}{19.9}
\newcommand{\eSSNbLabPerSDSE}{13.2}
\newcommand{\eSSNbFbPerSD}{$-$12.8}
\newcommand{\eSSNbFbPerSDSE}{9.9}
\newcommand{\eBalNbDFrzEarly}{$-$0.31}
\newcommand{\eBalNbDFrzEarlySE}{0.15}

\newcommand{\eBalNbDFrzLate}{$-$0.32}
\newcommand{\eBalNbDFrzLateSE}{0.08}

\newcommand{\eAdvDepPre}{0.06}
\newcommand{\eAdvDepFrz}{0.17}
\newcommand{\eAdvDepRecOne}{0.14}

\newcommand{\eAdvDepM}{$-$3.2}
\newcommand{\eAdvDepMSE}{3.5}

\newcommand{\eAdvDepStock}{$-$1.5}
\newcommand{\eAdvDepStockSE}{3.8}

\newcommand{\eAdvDepN}{1,229}
\newcommand{\eAdvDepBeta}{$-$0.10}
\newcommand{\eAdvDepBetaSE}{0.17}

\newcommand{\eSSDepLab}{$-$1.2}
\newcommand{\eSSDepLabSE}{8.4}
\newcommand{\eSSDepFb}{$-$1.8}
\newcommand{\eSSDepFbSE}{2.0}

\newcommand{\eBalDepPre}{$-$0.01}
\newcommand{\eBalDepPreSE}{0.01}
\newcommand{\eBalDepDFrz}{$-$0.02}
\newcommand{\eBalDepDFrzSE}{0.03}

\newcommand{\eBalDepAdv}{$-$0.14}
\newcommand{\eBalDepAdvSE}{0.15}
\newcommand{\eIdxBDepAdv}{$-$0.65}
\newcommand{\eIdxBDepAdvSE}{0.39}

\newcommand{\eBalDepDFrzEarly}{$-$0.04}
\newcommand{\eBalDepDFrzEarlySE}{0.05}

\newcommand{\eBalDepDFrzLate}{$-$0.02}
\newcommand{\eBalDepDFrzLateSE}{0.04}

\newcommand{\eAdvDiff}{$-$6.0}
\newcommand{\eAdvDiffSE}{4.6}
\newcommand{\eNTrain}{298,723}
\newcommand{\eNTrainCells}{649}
\newcommand{\eCoefMspread}{1.24}

\newcommand{\eNegSpDepRate}{74}

\newcommand{\eWVarNbPre}{0.38}

\newcommand{\eWVarNbFrz}{0.37}
\newcommand{\eNegSpNbFrz}{1}

\newcommand{\eWVarNbRecOne}{0.52}

\newcommand{\eWVarNbRecTwo}{0.60}

\newcommand{\eNegSpNbRate}{86}

\newcommand{\eAUCMainDepNineteen}{0.54}
\newcommand{\eAUCMainDepTwenty}{0.50}
\newcommand{\eAUCMainDepTwentyOne}{0.53}
\newcommand{\eAUCMainDepTwentyTwo}{0.77}
\newcommand{\eAUCMainNbNineteen}{0.73}
\newcommand{\eAUCMainNbTwenty}{0.68}
\newcommand{\eAUCMainNbTwentyOne}{0.67}
\newcommand{\eAUCMainNbTwentyTwo}{0.78}

\newcommand{\eAUCPoolNbNineteen}{0.54}

\newcommand{\eAUCSPDepTwentyTwo}{0.79}
\newcommand{\eAUCSPNbNineteen}{0.70}
\newcommand{\eAUCSPNbTwenty}{0.65}
\newcommand{\eAUCSPNbTwentyOne}{0.64}
\newcommand{\eAUCSPNbTwentyTwo}{0.79}
\newcommand{\eStockN}{12,456,455}

\newcommand{\eDolNbPreMthreeBal}{38.2}

\newcommand{\eDolNbFrzUpb}{59.7}
\newcommand{\eDolNbFrzXthree}{4.0}
\newcommand{\eDolNbFrzXtwentyfour}{22.8}
\newcommand{\eDolNbFrzMthreeBal}{6.7}

\newcommand{\eDolNbFrzLateBn}{18.8}
\newcommand{\eDolNbFrzLateDelay}{7.2}
\newcommand{\eDolNbFrzLateAdv}{0.7}
\newcommand{\eDolNbShortThree}{18.8}
\newcommand{\eDolNbShortTwentyfour}{5.1}

\newcommand{\eDolNbPreSizeGap}{4}
\newcommand{\eDolNbFrzSizeGap}{20}

\newcommand{\eBalNbFrzLevel}{$-$0.13}
\newcommand{\eBalNbAttenuated}{0.11}
\newcommand{\eIdxBNbFrzShrink}{41}
\newcommand{\eDistNbEsd}{1.3}
\newcommand{\eDistNbPreLevel}{$-$13.7}
\newcommand{\eDistNbPreLevelSE}{7.2}

\newcommand{\eDistNbFrzLevel}{1.9}
\newcommand{\eDistNbFrzLevelSE}{2.8}
\newcommand{\eDistNbFrzFsF}{0.5}

\newcommand{\eDistNbRelLevel}{$-$3.9}
\newcommand{\eDistNbRelLevelSE}{2.0}
\newcommand{\eDistNbRelFsF}{0.1}

\newcommand{\eDistDepFrzLevel}{$-$3.1}
\newcommand{\eDistDepFrzLevelSE}{4.6}

%% file: tables/robust_rows.tex
\newcommand{\RobustRows}{%
Baseline & 1.45 & $-$0.49 (0.31) & 8.3 & 0.50 & 0.52 & $-$0.23 (0.26) & 0.11 (0.13)\\
Issuer-month cells & 1.45 & $-$0.50 (0.33) & 8.3 & 0.50 & 0.63 & $-$0.28 (0.28) & 0.16 (0.17)\\
Treasury-spread index, piecewise & 2.48 & 0.03 (1.08) & 8.3 & 0.74 & 1.14 & 0.01 (0.47) & 0.34 (0.42)\\
Boosted ranking index & 0.99 & 0.01 (0.23) & 8.3 & 0.34 & 0.47 & 0.00 (0.14) & 0.18 (0.08)\\
Index from depository decisions & 0.81 & $-$0.49 (0.12) & 8.3 & 0.30 & 0.17 & $-$0.24 (0.12) & $-$0.07 (0.06)\\
Depository exercise classifier & 0.07 & 0.07 (0.13) & 8.3 & 0.03 & 0.07 & 0.03 (0.08) & 0.05 (0.07)\\
Mortgage spread only & 1.20 & $-$0.41 (0.39) & 8.3 & 0.44 & 0.50 & $-$0.25 (0.43) & 0.12 (0.22)\\
Exercise within 12 months & 1.42 & $-$0.35 (0.23) & 30.7 & 0.47 & 0.49 & $-$0.13 (0.11) & 0.03 (0.05)\\
Excluding the largest nonbank & 1.28 & $-$0.22 (0.29) & 10.6 & 0.48 & 0.51 & $-$0.09 (0.16) & 0.16 (0.17)\\
Nonbanks and technology-based lenders & 1.41 & $-$0.45 (0.29) & 8.3 & 0.51 & 0.53 & $-$0.21 (0.17) & 0.11 (0.14)\\
Loans not in forbearance & 1.45 & $-$0.34 (0.36) & 25.1 & 0.50 & 0.34 & $-$0.08 (0.17) & 0.01 (0.10)\\}

%% file: tables/empirics_rows.tex
\newcommand{\NbRows}{%
Options & 175,716 & 283,048 & 357,497 & 142,549 & 118,289 & 193,853\\
Funded share (\%) & 40.0 & 8.3 & 15.9 & 32.4 & 37.1 & 7.2\\
Latent weight on index$^{a}$ & 1.45 (0.29) & $-$0.49 (0.31) & $-$0.09 (0.24) & $-$0.07 (0.21) & $-$0.19 (0.13) & 0.60 (0.20)\\
Level response & 0.201 & 0.043 & 0.107 & 0.140 & 0.136 & 0.072\\
Relative response & 0.50 & 0.52 & 0.67 & 0.43 & 0.37 & 0.99\\
\quad Fixed ranking & 0.50 & 0.75 & 0.71 & 0.45 & 0.40 & 0.80\\
\quad Random rationing & 0.46 & 0.40 & 0.47 & 0.37 & 0.35 & 0.48\\
\quad Actual minus fixed & -- & $-$0.23 (0.25) & $-$0.03 (0.11) & $-$0.02 (0.05) & $-$0.04 (0.04) & 0.19 (0.11)\\
\quad Actual minus random & -- & 0.11 (0.16) & 0.20 (0.08) & 0.06 (0.05) & 0.02 (0.03) & 0.52 (0.11)\\
\quad Same, change from pre & -- & 0.07 (0.15) & 0.16 (0.08) & 0.02 (0.05) & $-$0.02 (0.03) & 0.48 (0.11)\\}
\newcommand{\DepRows}{%
Options & 95,934 & 189,061 & 183,529 & 52,483 & 35,943 & 47,870\\
Funded share (\%) & 80.5 & 85.3 & 79.2 & 79.4 & 81.0 & 51.9\\
Latent weight on index$^{a}$ & 0.13 (0.03) & $-$0.35 (0.16) & $-$0.12 (0.14) & 0.11 (0.17) & $-$0.06 (0.12) & 3.89 (1.86)\\
Level response & 0.010 & $-$0.007 & 0.000 & 0.013 & 0.004 & 0.177\\
Relative response & 0.01 & $-$0.01 & 0.00 & 0.02 & 0.00 & 0.34\\
\quad Fixed ranking & 0.01 & 0.00 & 0.01 & 0.01 & 0.01 & 0.03\\
\quad Random rationing & 0.01 & 0.00 & 0.00 & 0.01 & 0.01 & 0.01\\
\quad Actual minus fixed & -- & $-$0.01 (0.02) & $-$0.01 (0.02) & 0.01 (0.02) & 0.00 (0.01) & 0.31 (0.12)\\
\quad Actual minus random & -- & $-$0.01 (0.02) & 0.00 (0.02) & 0.01 (0.02) & 0.00 (0.01) & 0.33 (0.13)\\
\quad Same, change from pre & -- & $-$0.01 (0.02) & 0.00 (0.02) & 0.01 (0.02) & 0.00 (0.01) & 0.33 (0.13)\\}

%% file: tables/queue_rows.tex
\newcommand{\QueueRows}{%
3/2019--2/2020 & 92,921 & 5.2 & 0.56 (0.15) & 13,334 & 23.6 & 0.11 (0.03)\\
3/2020--6/2020 & 240,382 & 26.4 & 1.05 (0.25) & 21,004 & 24.5 & 0.39 (0.24)\\
7/2020--12/2020 & 275,410 & 27.0 & 0.80 (0.20) & 32,859 & 23.4 & 0.21 (0.16)\\
1/2021--6/2021 & 80,916 & 20.2 & 0.46 (0.14) & 8,272 & 17.6 & $-$0.07 (0.16)\\}

%% file: tables/stock_rows.tex
\newcommand{\StockRows}{%
\multicolumn{7}{l}{\emph{Nonbanks}}\\
Monthly exercise rate (\%) & 7.6 & 2.9 & 3.3 & 6.2 & 7.0 & 2.3\\
Relative response & 0.80 & 0.57 & 1.01 & 0.71 & 0.57 & 0.28\\
\quad Fixed ranking & 0.80 & 0.62 & 0.76 & 0.61 & 0.61 & 0.35\\
\quad Random rationing & 0.79 & 0.61 & 0.76 & 0.60 & 0.60 & 0.34\\
Change in latent weight & -- & $-$0.20 (0.52) & 1.27 (0.50) & 0.48 (0.38) & $-$0.12 (0.12) & $-$0.49 (0.12)\\
\multicolumn{7}{l}{\emph{Depositories}}\\
Monthly exercise rate (\%) & 39.1 & 62.3 & 36.9 & 20.0 & 20.9 & 16.0\\
Relative response & 0.08 & 0.00 & 0.01 & 0.16 & 0.12 & 0.37\\
\quad Fixed ranking & 0.08 & 0.02 & 0.03 & 0.06 & 0.06 & 0.13\\
\quad Random rationing & 0.07 & 0.02 & 0.03 & 0.05 & 0.06 & 0.07\\
Change in latent weight & -- & $-$0.36 (0.32) & $-$0.22 (0.36) & 0.65 (0.47) & 0.35 (0.49) & 0.94 (0.30)\\}

%% file: tables/issuer_rows.tex
\newcommand{\IssuerRows}{%
A & 36,265 & 2.19 & 53.4 & 0.69 & 4.2 & 1.53 & 2.06 & 0.74 & 54.6 & 0.70 & 0.63 & 0.58\\
B & 33,343 & 2.04 & 61.3 & 0.58 & 2.4 & 0.38 & 2.11 & 0.61 & 44.1 & 0.65 & 0.78 & 0.57\\
C & 16,256 & 0.87 & 63.1 & 0.29 & 25.4 & 0.50 & 0.38 & 0.27 & 17.0 & 0.58 & 0.61 & 0.31\\
D & 12,054 & 0.42 & 71.4 & 0.10 & 48.9 & 0.14 & 0.14 & 0.07 & 88.6 & 0.04 & 0.05 & 0.05\\
E & 5,747 & 0.52 & 59.0 & 0.20 & 25.7 & 0.62 & 0.25 & 0.17 & 28.5 & 0.45 & 0.26 & 0.21\\
Other 78 nonbanks & 72,051 & 0.95 & 11.3 & 0.50 & 0.7 & 1.10 & 0.90 & 0.71 & 15.9 & 0.12 & 0.25 & 0.24\\}

%% file: tables/balance_rows.tex
\newcommand{\BalanceRows}{%
\multicolumn{7}{l}{\emph{Nonbanks}}\\
Weight on log balance & 0.19 (0.11) & $-$0.32 (0.06) & 0.07 (0.06) & 0.08 (0.11) & $-$0.10 (0.14) & $-$0.06 (0.13)\\
Weight on ranking index & 1.52 (0.30) & $-$0.62 (0.33) & $-$0.03 (0.25) & 0.00 (0.26) & $-$0.21 (0.17) & 0.60 (0.19)\\
\multicolumn{7}{l}{\emph{Depositories}}\\
Weight on log balance & $-$0.01 (0.01) & $-$0.02 (0.03) & $-$0.05 (0.06) & $-$0.06 (0.06) & $-$0.06 (0.04) & $-$0.15 (0.16)\\
Weight on ranking index & 0.12 (0.03) & $-$0.36 (0.16) & $-$0.15 (0.11) & 0.08 (0.15) & $-$0.09 (0.11) & 3.95 (2.33)\\}

%% file: tables/funding_rows.tex
\newcommand{\FundingRows}{%
Log advances outstanding (\% of balance) & $-$9.2 (3.1) & $-$4.1 (1.6) & $-$0.01 (0.29) & $-$3.2 (3.5) & $-$1.5 (3.8) & 0.09 (0.06)\\
Advances outstanding (\% of balance) & $-$13.6 (13.4) & $-$6.1 (5.0) & 0.74 (1.02) & $-$13.2 (12.3) & $-$4.7 (11.4) & 0.12 (0.22)\\
Advances outstanding (months of P\&I) & $-$6.8 (5.3) & $-$3.2 (2.2) & 0.40 (0.53) & $-$7.0 (6.3) & $-$1.2 (5.6) & 0.08 (0.11)\\
Share of portfolio in forbearance (\%) & $-$1.5 (0.8) & $-$0.7 (0.2) & $-$0.03 (0.05) & $-$1.7 (1.2) & $-$2.1 (0.9) & 0.01 (0.01)\\}

%% file: tables/shape_macros.tex
\newcommand{\eShapeNbCells}{1,030}
\newcommand{\eShapeNbIssuers}{52}
\newcommand{\eShapeNbGammaAct}{1.39}
\newcommand{\eShapeNbGammaFix}{1.24}
\newcommand{\eShapeNbRtwoAct}{0.90}
\newcommand{\eShapeNbRtwoFix}{0.99}
\newcommand{\eShapeNbRatio}{1.12}
\newcommand{\eShapeNbRatioSE}{0.20}
\newcommand{\eShapeNbRatioLo}{0.57}
\newcommand{\eShapeNbRatioHi}{1.29}

\newcommand{\eShapeNbRatioExPre}{1.11}
\newcommand{\eShapeNbCellsExPre}{798}
\newcommand{\eShapeNbTopCells}{94}
\newcommand{\eShapeMinN}{25}

%% file: tables/synth_macros.tex
\newcommand{\eSynDonorsM}{19}

\newcommand{\eSynMSecPreAct}{36.5}
\newcommand{\eSynMSecPreSyn}{36.7}
\newcommand{\eSynMSecRmspe}{5.9}
\newcommand{\eSynMSecGapFrz}{$-$31.7}
\newcommand{\eSynMSecGapRel}{$-$28.4}
\newcommand{\eSynMSecGapYone}{$-$9.0}
\newcommand{\eSynMSecGapYtwo}{$-$12.4}
\newcommand{\eSynMSecP}{0.20}

\newcommand{\eSynMAGapFrz}{$-$48.9}

\newcommand{\eSynMBGapFrz}{$-$65.0}

\newcommand{\eSynMCGapFrz}{$-$24.4}

\newcommand{\eSynMDGapFrz}{$-$32.9}

\newcommand{\eSynMEGapFrz}{$-$26.7}

\newcommand{\eSynMPlacebos}{19}

\newcommand{\eTwoMomNb}{$-$0.012}
\newcommand{\eTwoMomNbSE}{0.023}
\newcommand{\eTwoMomDep}{0.014}
\newcommand{\eTwoMomDepSE}{0.005}
\newcommand{\eTwoMomNbIssuers}{78}

%% file: tables/fhfa_macros.tex
\newcommand{\eFhfaNbIssuers}{75}
\newcommand{\eFhfaZmean}{0.43}
\newcommand{\eFhfaZsd}{0.30}
\newcommand{\eFhfaNbB}{$-$14.1}
\newcommand{\eFhfaNbSE}{16.2}
\newcommand{\eFhfaNbP}{0.41}
\newcommand{\eFhfaDepB}{18.6}
\newcommand{\eFhfaDepSE}{14.4}

\newcommand{\eFhfaEvQtwo}{16.3}
\newcommand{\eFhfaEvQtwoSE}{17.2}
\newcommand{\eFhfaEvQthree}{$-$4.4}
\newcommand{\eFhfaEvQthreeSE}{18.8}
\newcommand{\eFhfaEvQfour}{$-$36.2}
\newcommand{\eFhfaEvQfourSE}{22.7}
\newcommand{\eFhfaEvTwoOneLo}{$-$57.0}
\newcommand{\eFhfaEvTwoOneHi}{$-$40.3}
\newcommand{\eFhfaTiltB}{$-$0.26}
\newcommand{\eFhfaTiltP}{0.17}
\newcommand{\eFhfaTiltN}{13}

%% file: tables/tilt_macros.tex
\newcommand{\eTiltQDepMaxAbs}{0.17}

%% file: tables/spread_macros.tex
\newcommand{\eSpNbTopFall}{71}
\newcommand{\eSpNbOtherFallLo}{84}
\newcommand{\eSpNbOtherFallHi}{89}

%% file: tables/fold_macros.tex
\newcommand{\fDidFrz}{35.7}
\newcommand{\fDidFrzSE}{8.9}
\newcommand{\fDidRel}{24.8}
\newcommand{\fDidRelSE}{8.3}
\newcommand{\fDidRecOne}{10.4}
\newcommand{\fDidRecOneSE}{7.7}
\newcommand{\fDidRecTwo}{5.6}
\newcommand{\fDidRecTwoSE}{6.2}
\newcommand{\fDidRate}{4.7}
\newcommand{\fDidRateSE}{9.2}
\newcommand{\fDidP}{0.027}
\newcommand{\fDidT}{4.0}
\newcommand{\fDidN}{1,875,772}
\newcommand{\fDidG}{120}
\newcommand{\fPool}{30.1}
\newcommand{\fPoolSE}{9.3}
\newcommand{\fPoolP}{0.046}
\newcommand{\fPoolT}{3.2}
\newcommand{\fPoolN}{553,364}
\newcommand{\fPoolG}{116}
\newcommand{\fPoolShare}{74}
\newcommand{\fPlac}{9.9}
\newcommand{\fPlacSE}{4.1}
\newcommand{\fPlacP}{0.048}
\newcommand{\fPlacN}{325,611}
\newcommand{\fPlacG}{119}
\newcommand{\fFlat}{36.6}
\newcommand{\fFlatSE}{9.5}
\newcommand{\fFlatP}{0.034}
\newcommand{\fFlatN}{671,036}
\newcommand{\fFlatG}{117}
\newcommand{\fSpNbOne}{2.24}
\newcommand{\fSpNbSEOne}{0.45}
\newcommand{\fSpNetOne}{2.67}
\newcommand{\fSpNetSEOne}{0.57}
\newcommand{\fSpDepOne}{0.43}
\newcommand{\fSpMsOne}{0.04}
\newcommand{\fSpRNbPreOne}{24.3}
\newcommand{\fSpRNbFrzOne}{3.3}
\newcommand{\fSpRDepPreOne}{78.1}
\newcommand{\fSpRDepFrzOne}{84.5}
\newcommand{\fSpNbTwo}{2.62}
\newcommand{\fSpNbSETwo}{0.64}
\newcommand{\fSpNetTwo}{2.91}
\newcommand{\fSpNetSETwo}{0.78}
\newcommand{\fSpDepTwo}{0.28}
\newcommand{\fSpMsTwo}{0.54}
\newcommand{\fSpRNbPreTwo}{41.1}
\newcommand{\fSpRNbFrzTwo}{4.8}
\newcommand{\fSpRDepPreTwo}{81.2}
\newcommand{\fSpRDepFrzTwo}{85.2}
\newcommand{\fSpNbThree}{2.45}
\newcommand{\fSpNbSEThree}{0.70}
\newcommand{\fSpNetThree}{2.68}
\newcommand{\fSpNetSEThree}{0.83}
\newcommand{\fSpDepThree}{0.23}
\newcommand{\fSpMsThree}{0.86}
\newcommand{\fSpRNbPreThree}{45.4}
\newcommand{\fSpRNbFrzThree}{6.7}
\newcommand{\fSpRDepPreThree}{80.8}
\newcommand{\fSpRDepFrzThree}{84.2}
\newcommand{\fSpNbFour}{2.46}
\newcommand{\fSpNbSEFour}{0.68}
\newcommand{\fSpNetFour}{2.65}
\newcommand{\fSpNetSEFour}{0.86}
\newcommand{\fSpDepFour}{0.19}
\newcommand{\fSpMsFour}{1.26}
\newcommand{\fSpRNbPreFour}{51.3}
\newcommand{\fSpRNbFrzFour}{8.2}
\newcommand{\fSpRDepPreFour}{82.5}
\newcommand{\fSpRDepFrzFour}{85.0}
\newcommand{\fSpNbFive}{1.83}
\newcommand{\fSpNbSEFive}{0.62}
\newcommand{\fSpNetFive}{2.00}
\newcommand{\fSpNetSEFive}{0.86}
\newcommand{\fSpDepFive}{0.16}
\newcommand{\fSpMsFive}{2.05}
\newcommand{\fSpRNbPreFive}{53.6}
\newcommand{\fSpRNbFrzFive}{15.6}
\newcommand{\fSpRDepPreFive}{85.2}
\newcommand{\fSpRDepFrzFive}{87.1}
\newcommand{\fSpTopMid}{0.68}
\newcommand{\fSpTopMidSE}{0.24}
\newcommand{\fSpTopMidLo}{1.14}
\newcommand{\fSpTopMidHi}{0.22}
\newcommand{\fSpSlope}{0.24}
\newcommand{\fSpSlopeLo}{0.58}
\newcommand{\fSpSlopeHi}{0.08}
\newcommand{\fSpDepLo}{0.16}
\newcommand{\fSpDepHi}{0.43}
\newcommand{\fIssNbPos}{54}
\newcommand{\fIssNbN}{62}
\newcommand{\fIssDepPos}{19}
\newcommand{\fIssDepN}{35}
\newcommand{\fIssRankP}{0.029}
\newcommand{\fIssDiff}{1.07}
\newcommand{\fIssDiffSE}{0.47}
\newcommand{\fIssPermP}{0.013}
\newcommand{\fIssAdv}{1.18}
\newcommand{\fIssAdvSE}{0.61}
\newcommand{\fIssAdvB}{0.10}
\newcommand{\fIssAdvBSE}{0.34}
\newcommand{\fOrAFrz}{6.6}
\newcommand{\fOrAFrzSE}{1.2}
\newcommand{\fOrARel}{1.9}
\newcommand{\fOrARelSE}{1.2}
\newcommand{\fOrARecOne}{$-$2.3}
\newcommand{\fOrARecOneSE}{1.4}
\newcommand{\fOrARecTwo}{0.6}
\newcommand{\fOrARecTwoSE}{3.4}
\newcommand{\fOrARate}{5.4}
\newcommand{\fOrARateSE}{1.8}
\newcommand{\fOrAP}{0.008}
\newcommand{\fOrAG}{17}
\newcommand{\fOrAN}{694}
\newcommand{\fOrBFrz}{7.6}
\newcommand{\fOrBFrzSE}{1.8}
\newcommand{\fOrBRel}{$-$2.9}
\newcommand{\fOrBRelSE}{3.9}
\newcommand{\fOrBRecOne}{$-$7.7}
\newcommand{\fOrBRecOneSE}{3.8}
\newcommand{\fOrBRecTwo}{$-$8.2}
\newcommand{\fOrBRecTwoSE}{3.8}
\newcommand{\fOrBRate}{$-$0.8}
\newcommand{\fOrBRateSE}{2.0}
\newcommand{\fOrBP}{$<$0.001}
\newcommand{\fOrBG}{17}
\newcommand{\fOrBN}{694}
\newcommand{\fOrCFrz}{6.5}
\newcommand{\fOrCFrzSE}{1.8}
\newcommand{\fOrCRel}{1.2}
\newcommand{\fOrCRelSE}{1.9}
\newcommand{\fOrCRecOne}{$-$2.1}
\newcommand{\fOrCRecOneSE}{1.8}
\newcommand{\fOrCRecTwo}{0.8}
\newcommand{\fOrCRecTwoSE}{3.5}
\newcommand{\fOrCRate}{5.6}
\newcommand{\fOrCRateSE}{2.0}
\newcommand{\fOrCP}{0.018}
\newcommand{\fOrCG}{17}
\newcommand{\fOrCN}{694}
\newcommand{\fOrDFrz}{6.9}
\newcommand{\fOrDFrzSE}{1.9}
\newcommand{\fOrDRel}{1.8}
\newcommand{\fOrDRelSE}{1.8}
\newcommand{\fOrDRecOne}{$-$2.2}
\newcommand{\fOrDRecOneSE}{2.4}
\newcommand{\fOrDRecTwo}{4.4}
\newcommand{\fOrDRecTwoSE}{4.8}
\newcommand{\fOrDRate}{8.5}
\newcommand{\fOrDRateSE}{3.2}
\newcommand{\fOrDP}{0.031}
\newcommand{\fOrDG}{17}
\newcommand{\fOrDN}{694}
\newcommand{\fOrEFrz}{3.5}
\newcommand{\fOrEFrzSE}{1.2}
\newcommand{\fOrERel}{2.1}
\newcommand{\fOrERelSE}{1.0}
\newcommand{\fOrERecOne}{$-$0.1}
\newcommand{\fOrERecOneSE}{1.1}
\newcommand{\fOrERecTwo}{0.5}
\newcommand{\fOrERecTwoSE}{1.1}
\newcommand{\fOrERate}{3.9}
\newcommand{\fOrERateSE}{1.1}
\newcommand{\fOrEP}{0.012}
\newcommand{\fOrEG}{79}
\newcommand{\fOrEN}{2,683}
\newcommand{\fOrGov}{6.8}
\newcommand{\fOrGovSE}{1.6}
\newcommand{\fOrGovP}{0.010}
\newcommand{\fOrPurW}{6.1}
\newcommand{\fOrPurWSE}{4.3}
\newcommand{\fOrPurWP}{0.18}
\newcommand{\fOrLooLo}{5.8}
\newcommand{\fOrLooHi}{7.1}
\newcommand{\fOrAdv}{6.1}
\newcommand{\fOrAdvSE}{1.1}
\newcommand{\fOrSize}{4.8}
\newcommand{\fOrSizeSE}{2.0}
\newcommand{\fOrSizeP}{0.081}
\newcommand{\fOrSizeB}{2.4}
\newcommand{\fOrSizeBSE}{3.4}
\newcommand{\fOrAll}{5.8}
\newcommand{\fOrAllSE}{1.8}
\newcommand{\fOrAllP}{0.059}
\newcommand{\fOrPlac}{2.2}
\newcommand{\fOrPlacSE}{0.8}
\newcommand{\fOrPlacP}{0.062}

\newcommand{\fOrEvFrz}{5.6}
\newcommand{\fOrEvFrzSE}{1.6}
\newcommand{\fOrEvQThree}{2.0}
\newcommand{\fOrEvQThreeSE}{1.5}

\newcommand{\fOrIdx}{0.51}
\newcommand{\fOrIdxSE}{0.21}
\newcommand{\fOrIdxRho}{0.52}
\newcommand{\fOrIdxP}{0.034}
\newcommand{\fOrActN}{17}
\newcommand{\fOrShPre}{69}
\newcommand{\fOrShFrz}{74}
\newcommand{\fOrQA}{0.06}
\newcommand{\fOrQB}{0.30}
\newcommand{\fOrQC}{1.17}
\newcommand{\fOrRhoSize}{0.75}
\newcommand{\fOrMatched}{80}
\newcommand{\fOrNb}{83}
\newcommand{\fOrIQR}{19}
\newcommand{\fOrMeanFall}{25}
\newcommand{\fOrDrop}{3.5}
\newcommand{\fOrDropSE}{1.2}
\newcommand{\fOrDropP}{0.018}
\newcommand{\fOrDropG}{74}

\newcommand{\fTrPre}{1.09}
\newcommand{\fTrFrz}{0.30}
\newcommand{\fTrRel}{0.82}
\newcommand{\fTrTwoOne}{0.44}
\newcommand{\fTrN}{47,562}

\newcommand{\fWdBalLo}{6.3}
\newcommand{\fWdBalHi}{19.3}
\newcommand{\fWdFlowLo}{4.8}
\newcommand{\fWdFlowHi}{6.9}
\newcommand{\fWdCorr}{0.95}
\newcommand{\fWdFlagBal}{26.4}
\newcommand{\fWdFlagPre}{97.2}
\newcommand{\fWdFlagPost}{10.5}
\newcommand{\fPmPre}{1.0}
\newcommand{\fPmPeak}{17.2}
\newcommand{\fPmPost}{4.6}
\newcommand{\fSpillN}{58}

\newcommand{\fSpillB}{0.11}
\newcommand{\fSpillBSE}{0.12}
\newcommand{\fSpillPl}{0.07}
\newcommand{\fSpillPlSE}{0.07}
\newcommand{\WedgeRows}{%
PennyMac Financial & Nonbank & 30 Sep & 0.89 & 17.18 & 19.3 & 6.9 & 2.8 & 54.7 & 7.5\\
Mr. Cooper Group & Nonbank & 31 Dec & 0.56 & 6.16 & 11.0 & 5.1 & 2.2 & 54.1 & 19.8\\
Caliber Home Loans & Nonbank & 30 Sep & 0.20 & 1.92 & 9.8 & 5.0 & 2.0 & 53.2 & 29.8\\
AmeriHome & Nonbank & 30 Sep & 0.38 & 2.43 & 6.3 & 4.8 & 1.3 & 35.1 & 10.4\\
Flagstar Bank & Depository & 31 Dec & 0.07 & 1.85 & 26.4 & 5.1 & 5.2 & 97.2 & 10.5\\
}

%% file: references.tex

%% file: internet_appendix.tex

\section{Auxiliary Results}

\subsection{Blackwell improvements under a fixed funding technology}\label{ia:blackwell}
If one experiment is a garbling of another and the intermediary can ignore the finer signal, every policy feasible under the coarser experiment remains feasible under the finer one. With the funding technology held fixed, private value before information costs cannot fall. The statement concerns the intermediary's own information. A mandatory disclosure policy can change outside financing terms and equilibrium beliefs, and Proposition~\ref{prop:finance} shows that what lenders know about the technology, not the technology alone, determines borrowing capacity. The claim that better information always expands funding conflates these cases.

\subsection{Why a flatter decision slope is not a liquidity signature}\label{ia:slope}
The action rule in Proposition~\ref{prop:allocation} is a threshold, not a linear response, so the slope of relief in a ranking variable depends on where the threshold sits. Three cases illustrate. First, a binding funding constraint preserves the ranking and funds less of it: the level slope falls toward zero as liquidity contracts below the peak, and the slope relative to the base rate rises (Proposition~\ref{prop:reliance}). Second, a value shock that shifts every opportunity's value by the same amount also preserves the ranking. An unconstrained intermediary then funds the top of the same ranking, and at the same funded share it produces the same level and relative slopes as rationing. Third, a value shock that compresses values across the ranking variable, such as a restriction on redelivery that bears most heavily on the loans with the largest redelivery gains, can flatten both slopes without any binding cash constraint. Slopes alone therefore separate funding from value only against the third alternative. Against the second, what separates them is the response to dated shocks of each kind (Proposition~\ref{prop:dissociation}): the funded share of a constrained intermediary moves with liquidity and not with value, and that of an unconstrained one does the reverse.

\subsection{Staff capacity and funding}\label{ia:staff}
In (\ref{eq:menu}), if every feasible action satisfies $d_{ia}=\varrho\,\ell_{ia}$ for a common $\varrho>0$, then $g_{ia}-\lambda\ell_{ia}-\mu d_{ia}=g_{ia}-(\lambda+\varrho\mu)\ell_{ia}$, and choices depend only on the combined multiplier. Action data alone cannot then separate funding scarcity from staffing scarcity. Separation requires actions whose cash and processing requirements vary independently, such as a buyout, which is cash-intensive but processing-light, and a documented modification, which is processing-intensive and requires relatively little cash, or a direct measure of one of the two constraints.

\subsection{Heterogeneous cash needs and the value of information}\label{ia:hetero}
Proposition~\ref{prop:info} fixes the cash need. With heterogeneous needs, the intermediary ranks by $(RQ-c)/\ell$, and the value of a signal depends on its joint distribution with cash needs. Two observations are immediate. If cash needs are independent of the signal, the intermediary still ranks by $Q$ within each cash-need class, so the funded set in each class is a top share of that class's posteriors. The signal changes the common shadow value of cash, however, and with it the funded share of each class, so the aggregate value of the signal need not be single-peaked in $K$; Proposition~\ref{prop:info} applies to each class holding its funded share fixed. If cash needs fall with the posterior, as in specification (\ref{eq:linear}) with $\omega>0$, the signal also identifies cheaper actions, and part of its value is a saving of cash. That component operates only through the budget, so it vanishes when the budget is slack.

\subsection{Displacement within the intermediary}\label{ia:displacement}
Under a binding budget, relief to one loan is funded by relief denied to another. With a constant cash need, the number of funded interventions is fixed locally, and better information about one group of borrowers changes who is helped without changing how many. With heterogeneous cash needs, the budget conserves dollars rather than counts:
\begin{equation}\label{eq:displacement}
\sum_{\text{groups } h} d\Big[\int_{i\in h}\ell_i x_i\,dF(i)\Big]=0.
\end{equation}
Treatment of one loan therefore affects others serviced by the same constrained intermediary. A loan-level estimate that ignores this interference can overstate the effect of an institution-wide change, such as the adoption of a new model or the arrival of new funding, and a credible evaluation must measure effects across the intermediary's whole book.

\subsection{Menus and the two shadow prices}\label{ia:menus}
Relief involves a menu. Let $a\in\mathcal A_i$ index continuation, repurchase, forbearance, modification, sale, and liquidation, as available for loan $i$. Each action has private value $g_{ia}$, cash need $\ell_{ia}$, and staff requirement $d_{ia}$. With aggregate cash $K$ and staff capacity $D$, the optimal action solves
\begin{equation}\label{eq:menu}
a_i^*\in\argmax_{a\in\mathcal A_i}\big\{g_{ia}-\lambda\ell_{ia}-\mu d_{ia}\big\},
\end{equation}
where $\lambda,\mu\ge 0$ are the two shadow prices. A decline in completed relief can reflect either one, and \citet{Agarwal2017} show that servicers' organizational capacity limited modifications under HAMP. The two constraints are observationally equivalent when staff and cash requirements are proportional across actions. If $d_{ia}=\varrho\,\ell_{ia}$ for all actions, choices depend only on $\lambda+\varrho\mu$. Separating them requires independent variation in cash and processing intensity or a direct measure of one constraint, which is why documentation and queue data are part of the mechanism rather than controls \citep{BandyopadhyayInside2026}; Section~\ref{ia:staff} states the separation condition and the actions that satisfy it.

\section{Dynamic Extension}\label{ia:dynamic}
Let $\mathbf z_t$ collect each loan's posterior, legal status, eligibility, and contractual cash-flow state. A finite-horizon intermediary chooses a vector of actions subject to current cash and staff constraints:
\begin{equation}\label{eq:bellman}
V_t(K_t,\mathbf z_t)=\max_{\mathbf a_t\in\mathcal F(K_t,\mathbf z_t)}\E_t\big[\Pi_t(\mathbf a_t,\mathbf z_t)+\delta\,V_{t+1}(K_{t+1},\mathbf z_{t+1})\big].
\end{equation}
Cash evolves through collected payments, advances, repurchases, operating costs, and financing. Information acquisition uses operating resources and shifts future posteriors. Transitions depend on the action and describe cure, renewed delinquency, completed modification, and liquidation. The static model is a one-period version that uses continuation-adjusted values.

Three features of the dynamic problem reinforce the static results. First, cash spent today reduces $K_{t+1}$ one-for-one while its return arrives with delay, so the shadow value of cash tends to be highest early in a delinquency wave, when advance obligations accumulate. Second, waiting has option value when information arrives over time, so a constrained intermediary can rationally postpone action on loans whose posteriors are near the threshold. Liquidity can therefore affect the timing of relief more than the relief eventually granted, which is why the empirical design measures outcomes at fixed horizons. Third, if the budget binds in each period and an action ties up its cash for $\tau$ periods before it is returned, the action uses the budget in each of those periods, so its cost in scarce cash is $\sum_{s<\tau}\delta^s\lambda_{t+s}\,\ell$, approximately $\tau\lambda\ell$ when multipliers are stable and discounting is small. Ranking by value per dollar then becomes ranking by value per dollar-period: a constrained intermediary passes over actions that tie up cash longest, even when they are worth more. \citet{BM2026} design dynamic, borrower-responsive household-finance policy by reinforcement learning. The extension here adds a budget shared across the portfolio, which creates interference across loans, so that the simulated value of a learned policy is not a causal estimate of its effect if a servicer adopted it institution-wide.

\section{Data Construction Protocol}\label{ia:data}
\subsection{Sources and the definition of action}
Ginnie Mae publishes monthly loan-level disclosure files for single-family pools. In these files, a loan removed from a pool under removal reason 2 (repurchase of a delinquent loan) identifies an exercised buyout. The files used in Section~\ref{sec:evidence} were assembled from the disclosures for \citet{BKS2026}: a vesting-date panel with twelve months of outcomes and the final status through December 2022, and the monthly histories of every loan that reached three missed payments. Tests of Propositions~\ref{prop:info}--\ref{prop:auc} require three additions: dated measures of issuer funding, a pre-decision ranking variable, and outcomes at fixed horizons after the decision.

\subsection{Canonical fields}
Table~\ref{tab:fields} lists the fields the protocol requires and what each must be verified against.
\begin{longtable}{p{.27\textwidth}p{.65\textwidth}}
\caption{Canonical Fields of the Data-Construction Protocol\label{tab:fields}}\\
\toprule Field & Definition and verification requirement\\\midrule\endfirsthead
\toprule Field & Definition and verification requirement\\\midrule\endhead
\texttt{loan\_id} & Stable identifier; document whether it survives repurchase, servicing transfer, and redelivery.\\
\texttt{report\_date} & Economic as-of date, distinct from the file's publication date.\\
\texttt{issuer\_id} & Issuer of record at the decision date, with a dated crosswalk for mergers and transfers.\\
\texttt{pool\_id} & Security at the date; a loan can pass through more than one pool.\\
\texttt{eligible\_date} & First date the action became available under the rule then in force, not the date the action was taken.\\
\texttt{buyout\_date} & Validated by removal reason; disappearance from a pool alone is not a buyout.\\
\texttt{upb}, \texttt{coupon} & Unpaid balance and note rate at the decision date.\\
\texttt{funding\_capacity} & Usable cash and undrawn committed facility capacity, measured before the decision; gross facility size is not availability.\\
\texttt{cash\_requirement} & Current cash required by the action net of immediately drawable financing; later proceeds are recorded separately.\\
\texttt{signal\_time} & Time stamp showing that the ranking information preceded the decision.\\
\texttt{action\_stage} & Contact, application, determination, offer, acceptance, execution, and completion as separately dated stages.\\
\texttt{outcome\_date} & Date of a verified outcome, distinguishing absence of evidence from evidence of no event.\\\bottomrule
\end{longtable}

\subsection{Population and attrition}
Report, in order, all source loans; loans in the date range; loans observed before eligibility; eligible loans; loans with a classified issuer; loans with funding data; loans with a usable pre-decision signal; and loans with observed outcomes at each horizon. Give both loan and issuer counts at each step, and analyze missing funding or signal data as selection into the sample. Revisions to the same loan-month are not independent observations, and a servicing transfer is a transition rather than a new delinquency spell. End-of-sample censoring and loss of coverage are different events, and a loan that leaves a Ginnie Mae pool can remain outstanding and delinquent outside it.

\subsection{Estimands and outcome hierarchy}
The primary institution-level estimand is the total effect of a funding change on delivered resolution within an $h$-month horizon for a cohort frozen before the change, allowing actions on other loans to respond. The allocation estimand is how the funding change shifts resolution across strata of pre-event information and cash requirement. The principal outcome is a completed, documented intervention within a common horizon; offers, reviews, processing time, repurchase, and redelivery are secondary and mechanism outcomes. Pending foreclosure, completed foreclosure, bankruptcy, and redelivery are distinct states and are not interchangeable welfare measures. Horizons are fixed before outcomes are examined, and confirmatory tests are distinguished from exploratory heterogeneity.

\section{Empirical Details}\label{ia:empirics}
\subsection{Sample, outcome, and issuer types}\label{ia:sample}
Each observation is a buyout option in the month the loan first reaches three missed payments, from January 2019 to September 2022, in single-family level-payment and reperforming pools of both Ginnie Mae programs. The construction drops loans outside the 50 states and the District of Columbia, loans first paying before 2003, loans with a single delinquent month surrounded by current months, loans with missing or implausible balance or loan-to-value ratio, and loans that cannot be matched to state house prices or Treasury rates. Exercise is removal under reason 2 before any cure, within three months of vesting; the file also records the month of removal within twelve months and the final removal reason through December 2022. Issuer charter types come from a classification of the issuers in Ginnie Mae's 2020 issuer file, extended by hand, from charter records, to the largest unclassified issuers of 2019--2022; unclassified issuers, housing finance agencies, and technology-based lenders are excluded from the main sample.

\subsection{Ranking index}
The index is estimated on the \eNTrain{} nonbank options that vested from 2016 to 2018 in the \eNTrainCells{} issuer-months in which some but not all options were exercised, by a logistic regression of exercise on standardized characteristics and a full set of issuer-month indicators (inverse regularization strength 10). The characteristics are the note rate minus the monthly mean of Freddie Mac's 30-year mortgage rate in the vesting month, the change in the ten-year Treasury rate since origination, log loan age, log balance, credit score and a missing indicator, current loan-to-value ratio, judicial-foreclosure state, VA and rural-housing indicators, pool issue-type indicators, first-time-buyer status, refinance and modification indicators, prior missed payments in the preceding year, and the annual insurance premium. Standardization and imputation use the training sample. The within-issuer-month AUC, averaged over cells with at least 50 options and both outcomes and weighted by cell size, is \eAUCMainNbNineteen, \eAUCMainNbTwenty, \eAUCMainNbTwentyOne, and \eAUCMainNbTwentyTwo{} for nonbank decisions in 2019, 2020, 2021, and 2022, and \eAUCMainDepNineteen, \eAUCMainDepTwenty, \eAUCMainDepTwentyOne, and \eAUCMainDepTwentyTwo{} for depository decisions. The mortgage spread alone has within-cell AUCs of \eAUCSPNbNineteen, \eAUCSPNbTwenty, \eAUCSPNbTwentyOne, and \eAUCSPNbTwentyTwo{} for nonbanks. Table~\ref{tab:robust} uses four alternatives: the same logit with the spread over the ten-year Treasury rate entered piecewise linearly with kinks at 2, 3, and 4 points, which ranks 2020 decisions less well because almost every 2020 option lies above the first kink; a gradient-boosted ranker (LightGBM, at most 15 leaves, at least 1,000 observations per leaf, learning rate 0.03, early stopping on held-out issuer-months) fit to the same decisions with each cell's log-odds of exercise as a fixed offset; the conditional logit fit to depository decisions; and a gradient-boosted classifier fit to pooled depository decisions without cell effects, the construction of an earlier version of this paper, which ranks nonbank decisions within cells poorly (within-cell AUC \eAUCPoolNbNineteen{} in 2019), because depositories exercised almost every option and their decisions reveal little about the order in which a constrained issuer funds options.

\subsection{Estimation}
The logit with cell fixed effects is estimated by block coordinate ascent: given the slope, each cell's intercept solves its score equation; given the intercepts, the slope takes a Newton step on the concentrated likelihood. Cells in which every option or no option was exercised carry no information about the slope and are dropped from that estimation. Standard errors use the cluster-robust sandwich on the concentrated score, clustered by issuer. For the fixed-ranking predictions, the intercept of each cell is chosen by bisection so that the mean predicted probability equals the observed funded share, holding the slope at its pre-shock estimate; cells with a funded share of zero or one have a predicted response of zero, as they do in the data. For the random-rationing benchmark, the fixed-ranking probabilities are computed at the issuer's own pre-period funded share and the cell's covariance is scaled by the ratio of the observed funded share to that share (or set equal to the fixed-ranking value when the observed share is higher). Level and relative responses aggregate within-cell covariances and variances across the cells of a period. Bootstrap standard errors resample issuers with replacement, re-estimate the pre-shock slope in each of 200 draws, and recompute all predictions. Cells are split by forbearance status because the index cannot rank options in and out of forbearance against each other: it is estimated before forbearance existed, and within issuer-months nonbanks exercised options on loans in forbearance \eNbFbCoefAbs{} percentage points less often than other options (Table~\ref{tab:forbearance}). Before May 2020 the flag does not exist and the cells are issuer-months. The bootstrap resamples the \eNbBootIssuers{} nonbank issuers and the \eDepBootIssuers{} depositories with replacement. In the files, a removal is recorded in the disclosure month in which the loan leaves the pool, and the vesting-month record is month zero, so exercise within three months covers months zero through three; exercise by June 30, 2020, is a removal recorded in the May or June 2020 file. Loans recorded as reperforming or still delinquent at the end of the twelve-month window (codes 8 and 9 in the construction) remain in the queue risk set.

\subsection{Robustness}
Table~\ref{tab:robust} repeats the freeze comparison for nonbanks with different ranking variables, decision cells, exercise windows, and samples. All indices are standardized over the depository and nonbank options of January 2019 to September 2022, as in Table~\ref{tab:empirics}. With issuer-month cells, the prediction treats options in and out of forbearance as ranked by the same index, although nonbanks funded almost none of the former. In every row that ranks on nonbank decisions, the freeze response lies at or above the random-rationing benchmark and at or below the fixed-ranking prediction, within a few hundredths, and no row overturns the conclusion that the freeze months alone do not separate them. The index estimated from depository decisions, which ranks nonbank decisions poorly, puts the response below both. The standard errors in this table come from 100 draws with a different seed from the 200 draws behind Table~\ref{tab:empirics}, which is why the baseline row's standard errors differ slightly.

\begin{table}[H]\centering
\caption{Robustness of the Freeze Test, Nonbank Issuers\label{tab:robust}}
\footnotesize
\setlength{\tabcolsep}{4pt}
\begin{tabular}{>{\raggedright\arraybackslash}p{.24\textwidth}ccccccc}\toprule
 & \multicolumn{2}{c}{Latent weight} & Funded & \multicolumn{2}{c}{Relative response} & \multicolumn{2}{c}{Freeze: actual minus}\\\cmidrule(lr){2-3}\cmidrule(lr){5-6}\cmidrule(lr){7-8}
Specification & Pre & Change, freeze & Freeze & Pre & Freeze & Fixed & Random\\\midrule
\RobustRows
\bottomrule
\end{tabular}
\begin{minipage}{0.97\textwidth}\footnotesize\setstretch{1.0}\smallskip
Pre is March 2019--February 2020 and freeze is March--June 2020; the funded share is for the freeze, in percent. The 12-month row uses the twelve-month removal window of the vesting file, which differs slightly from the calendar-month horizon of Figure~\ref{fig:queue} (\eNbFrzTwelve\% against \eNbFrzHtwelve\% for the freeze cohort). The last two columns report the freeze relative response minus the fixed-ranking prediction and minus the random-rationing benchmark, constructed as in Table~\ref{tab:empirics}, with standard errors from 100 bootstrap draws over issuers. Standard errors for the latent weight are clustered by issuer.
\end{minipage}
\end{table}

\subsection{Issuer-level results}
Table~\ref{tab:issuers} reports the freeze and recovery comparisons for the five nonbanks with the most options in the pre-shock period, among issuers that exercised at least 5\% of them, and for the remaining nonbanks as a group. Issuers are ranked ex ante, by pre-shock option counts; the 5\% floor excludes one large issuer that exercised almost none of its pre-shock options and so has no estimable pre-shock weight (it is in the last row). Each row re-estimates the pre-shock weight on that issuer's own decisions. The aggregate freeze shortfall from the fixed-ranking prediction in Table~\ref{tab:empirics} comes from issuer B; the other rows are at or above the prediction, except issuer A, which lies between the two benchmarks. Issuer-level responses carry no standard errors, and those of issuers whose funded share fell below 5\% rest on few exercises.

\begin{table}[H]\centering
\caption{The Largest Nonbanks in the Freeze and the Recovery\label{tab:issuers}}
\scriptsize
\setlength{\tabcolsep}{3pt}
\begin{tabular}{lccccccccccccc}\toprule
 & \multicolumn{4}{c}{Pre-shock, March 2019--February 2020} & \multicolumn{4}{c}{Freeze, March--June 2020} & \multicolumn{4}{c}{Recovery, January--June 2021}\\\cmidrule(lr){2-5}\cmidrule(lr){6-9}\cmidrule(lr){10-13}
Issuer & Options & Latent weight & Share & Actual & Share & Actual & Fixed & Random & Share & Actual & Fixed & Random\\\midrule
\IssuerRows\bottomrule
\end{tabular}
\begin{minipage}{0.97\textwidth}\footnotesize\setstretch{1.0}\smallskip
Nonbank issuers ranked by the number of options that vested in March 2019--February 2020, among issuers that exercised at least 5\% of them. Shares are funded shares in percent. Actual is the relative response; Fixed and Random are the fixed-ranking prediction and the random-rationing benchmark, constructed as in Table~\ref{tab:empirics} with the latent weight re-estimated on the issuer's own pre-shock decisions. The last row pools the other 78 nonbanks.
\end{minipage}
\end{table}

\subsection{Funding pressure within issuer types}\label{ia:funding}
Table~\ref{tab:funding} relates exercise to four measures of an issuer's own cash drain, each computed for the issuer's whole portfolio from the monthly disclosures: advances outstanding as a share of unpaid balance (in logs and in levels), advances outstanding in months of scheduled principal and interest, and the share of loans in forbearance. Advances are scheduled payments times months delinquent, summed over the issuer's loans. Each regression includes issuer and month fixed effects, weights issuer-months by the number of options or eligible loans, and clusters standard errors by issuer; the funded share and the relative response are computed from the decision cells of Section~\ref{sec:fixedrank}, and the monthly exercise rate from the eligible stock of Table~\ref{tab:stock}. In a cell-fixed-effects logit with index-by-month interactions, so that the advance term is identified from cross-issuer differences within a month, the latent weight on the index falls with a nonbank's own log advance burden (\eAdvNbBeta{} per log point, s.e.\ \eAdvNbBetaSE) and not significantly with a depository's (\eAdvDepBeta, \eAdvDepBetaSE). Adding log balance and its interaction with the burden to that logit gives index interactions of \eIdxBNbAdv{} (\eIdxBNbAdvSE) for nonbanks and \eIdxBDepAdv{} (\eIdxBDepAdvSE) for depositories, and the weight on balance does not vary with the burden at either type (\eBalNbAdv, s.e.\ \eBalNbAdvSE, and \eBalDepAdv, \eBalDepAdvSE). Two cross-sectional exposures to the 2020 shock, each issuer's May 2020 forbearance share and its 2019 state mix weighted by the February--April 2020 rise in state unemployment rates, have small dispersion across issuers (standard deviations of \eSSNbFbSD{} and \eSSNbLabSD{} percentage points among the \eSSNbIssuers{} nonbanks with at least 100 options in both periods) and are not significantly related to the change in an issuer's funded share from the pre-shock period to the freeze, although the estimates are imprecise: \eSSNbFb{} points per point of forbearance share (s.e.\ \eSSNbFbSE), or \eSSNbFbPerSD{} points per standard deviation of the exposure (\eSSNbFbPerSDSE), and \eSSNbLab{} per point of unemployment exposure (\eSSNbLabSE), or \eSSNbLabPerSD{} per standard deviation (\eSSNbLabPerSDSE), for nonbanks, against an aggregate fall of \eNbFrzDrop{} points; the depository coefficients are \eSSDepFb{} (\eSSDepFbSE) and \eSSDepLab{} (\eSSDepLabSE).

\paragraph{Distant exposure.} \citet{Aiello2022} identifies servicer constraints from a servicer's exposure to house-price returns in regions far from the focal loan. The analogue here is an issuer's exposure to the pandemic shock in states other than the loan's own: for a loan in state $s$, $E_{j,-s}=\sum_{s'\neq s} w_{js'}\Delta u_{s'}/\sum_{s'\neq s} w_{js'}$, where $w_{js'}$ is issuer $j$'s share of options vesting in state $s'$ from March to December 2019 and $\Delta u_{s'}$ the February--April 2020 rise in the state unemployment rate. Linear probability models of exercise within three months on $E_{j,-s}$ (in standard deviations of the full-portfolio exposure across issuers, \eDistNbEsd{} percentage points for nonbanks), the index, and log balance, with state-by-month-by-forbearance fixed effects and standard errors clustered by issuer, compare loans in the same state and month held by issuers with different distant exposure. For nonbanks the coefficient on exposure is \eDistNbFrzLevel{} percentage points (s.e.\ \eDistNbFrzLevelSE) in the freeze and \eDistNbRelLevel{} (\eDistNbRelLevelSE) in the release; for depositories, \eDistDepFrzLevel{} (\eDistDepFrzLevelSE) in the freeze. In the pre-shock placebo period, where the design predicts zero, the nonbank coefficient is \eDistNbPreLevel{} (\eDistNbPreLevelSE), nearly two standard errors from zero and larger in magnitude than in either 2020 period: issuers whose 2019 portfolios sat in the states the pandemic later hit hardest already exercised less in 2019, so the level coefficients cannot be read as effects of the 2020 shock, and relative to that baseline the freeze coefficient is higher, not lower. The informative result is the first stage. Used as an instrument for the issuer's log advance burden, exposure has none: the cluster-robust first-stage $F$ is \eDistNbFrzFsF{} for nonbanks in the freeze and \eDistNbRelFsF{} in the release. Nonbank issuers' portfolios are national, the dispersion of exposure across them is small, and the cash drain of 2020 did not scale with it. The design that identifies servicer constraints in the 2008 crisis therefore has no power here, which limits the design's ability to identify issuer-specific funding variation; it does not establish that the shock was identical across nonbanks.

\begin{table}[H]\centering
\caption{Exercise and the Issuer's Own Cash Drain\label{tab:funding}}
\scriptsize
\setlength{\tabcolsep}{3pt}
\begin{tabular}{>{\raggedright\arraybackslash}p{.30\textwidth}cccccc}\toprule
 & \multicolumn{3}{c}{Nonbanks} & \multicolumn{3}{c}{Depositories}\\\cmidrule(lr){2-4}\cmidrule(lr){5-7}
Measure & Funded share & Stock rate & Rel.\ resp. & Funded share & Stock rate & Rel.\ resp.\\\midrule
\FundingRows\bottomrule
\end{tabular}
\begin{minipage}{0.97\textwidth}\footnotesize\setstretch{1.0}\smallskip
Issuer-month regressions, January 2019--September 2022, with issuer and month fixed effects, on issuer-months with at least 20 vesting options (\eAdvNbN{} nonbank and \eAdvDepN{} depository issuer-months). Funded share is the share of options vesting in the month exercised within three months (in percentage points), weighted by options; stock rate is the monthly repurchase rate on the eligible stock (percentage points), on issuer-months with at least 50 eligible loans and weighted by eligible loans; rel.\ resp.\ is the relative response of exercise to the ranking index, defined only for issuer-months with some exercise. Coefficients are per unit of the measure in the row. Standard errors clustered by issuer in parentheses.
\end{minipage}
\end{table}

\paragraph{A second moment of the same shadow price.} Proposition~\ref{prop:allocation} has one multiplier doing two things: it sets how far down the ranking an issuer goes and it tilts the ranking against cash-intensive actions. If the 2020 collapse at a given nonbank reflected a rise in its own multiplier, the nonbanks whose funded share fell most should also have tilted most against balance. Adding to the cell-fixed-effects logit of Table~\ref{tab:balance} interactions of log balance and the index with each issuer-period's change in funded share (per ten percentage points, relative to the issuer's own pre-shock share), with period-specific weights absorbing the common shock, gives \eTwoMomNb{} (s.e.\ \eTwoMomNbSE) on the balance interaction for the \eTwoMomNbIssuers{} nonbanks in 2020--2021, and \eTwoMomDep{} (\eTwoMomDepSE) for depositories. The tilt did not scale with the size of an issuer's own collapse: like the advance burden and the exposure designs above, the cross-issuer comparison within nonbanks finds no issuer-specific component, and these estimates do not detect issuer-specific comovement between the two margins.

\paragraph{Origination intensity.} Issuers are matched to HMDA reporters by name, using the 2019 transmittal sheet, and by hand where the Ginnie Mae issuer name is a trade name. Where the Ginnie Mae issuer file and the transmittal sheet both give an address, the state agrees for 33 of 35 matched issuers, and the two exceptions are firms that moved their headquarters. Issuer numbers that share a legal entity identifier are combined, and the ratio is computed for the entity. Originations are loans with action taken equal to one in the public loan application register, summed in dollars over all loan types; purchases are action taken equal to six. Restricting the numerator to FHA, VA, and USDA loans gives a freeze coefficient of \fOrGov{} (s.e.\ \fOrGovSE; wild cluster bootstrap $p=\fOrGovP$). With the advance burden, log portfolio balance, and the pre-shock funded share each interacted with the freeze, the coefficient is \fOrAll{} (\fOrAllSE; $p=\fOrAllP$). The balance tilt of Section~\ref{sec:perdollar} does not vary with intensity across the 11 issuers with at least 20 freeze exercises (rank correlation 0.07), as it did not vary with the size of an issuer's own collapse. Depositories are not matched: their mortgage originations are a small part of a deposit-funded balance sheet, and the measure has no content for them.

\paragraph{Originations in 2020.} The constraint on buyouts did not carry over to new lending. Across the \fSpillN{} nonbank entities with HMDA originations in 2018 through 2020 and at least 50 options before the shock and in the freeze, the log change in originations from 2019 to 2020 has a coefficient of $-\fSpillB$ (s.e.\ \fSpillBSE) on the freeze fall in the issuer's exercise index, against $-\fSpillPl$ (\fSpillPlSE) for the change from 2018 to 2019. Originated loans are financed on warehouse lines and sold within weeks, so the two activities need not draw on identical funding facilities. The annual association does not establish that they never compete for internal liquidity.

\paragraph{Audited balances.} Table~\ref{tab:wedge} sets the balance of loans eligible for repurchase and not repurchased, as each firm reports it under ASC 860-50, against the unpaid balance of options that vested at the firm's Ginnie Mae issuer numbers in January through September of 2019 and of 2020 in this paper's data. Balances are read from the firms' annual reports and registration statements at the dates shown. A firm that kept exercising at its 2019 rate would show a balance that grows with the flow; the ratio of the two growth rates is above one at every firm in the table. At the one nonbank that reports the line quarterly it averaged \$\fPmPre{} billion before the shock, peaked at \$\fPmPeak{} billion in September 2020, and has averaged \$\fPmPost{} billion since mid-2022. One depository reports the same line. Its balance grew \fWdFlagBal{} times, and it is the depository whose exercise fell, from \fWdFlagPre\% to \fWdFlagPost\%. Across the five firms, the correlation between the log ratio of balance growth to flow growth and the change in the firm's exercise rate is $-\fWdCorr$.

\begin{table}[H]\centering
\caption{Unexercised Options on Audited Balance Sheets\label{tab:wedge}}
\setstretch{1.0}\small\setlength{\tabcolsep}{3.5pt}
\begin{tabular}{lllccccccc}\toprule
 & & & \multicolumn{2}{c}{Balance (\$bn)} & \multicolumn{2}{c}{Growth ($\times$)} & & \multicolumn{2}{c}{Exercised (\%)}\\
Firm & Type & As of & 2019 & 2020 & Balance & Flow & Ratio & 2019 & 2020\\\midrule
\WedgeRows\bottomrule
\end{tabular}
\begin{minipage}{0.97\textwidth}\footnotesize\setstretch{1.0}\smallskip
Balance is the reported balance of loans eligible for repurchase from Ginnie Mae pools and not repurchased, from Form 10-K (PennyMac Financial Services, Mr.\ Cooper Group, Flagstar Bancorp) and Form S-1/A (Caliber Home Loans, AmeriHome), at the date shown in 2019 and 2020. Flow is the unpaid balance of options vesting in January through September at the firm's issuer numbers. Exercised is the share of those options repurchased within three months of vesting.
\end{minipage}
\end{table}

\begin{table}[H]\centering\small
\caption{Nonbank Less Depository Exercise, Relative to the Pre-Shock Year\label{tab:typecontrast}}
\setstretch{1.0}
\begin{tabular}{lcccc}\toprule
 & (1) & (2) & (3) & (4)\\
 & Full sample & Within pool & From Jul 2019 & Placebo, Mar 2019\\\midrule
Nonbank $\times$ freeze & $-$\fDidFrz & $-$\fPool & $-$\fFlat & \\
 & (\fDidFrzSE) & (\fPoolSE) & (\fFlatSE) & \\
 & [\fDidP] & [\fPoolP] & [\fFlatP] & \\
Nonbank $\times$ release & $-$\fDidRel{} (\fDidRelSE) & & & \\
Nonbank $\times$ Jan--Jun 2021 & $-$\fDidRecOne{} (\fDidRecOneSE) & & & \\
Nonbank $\times$ Jul--Dec 2021 & $-$\fDidRecTwo{} (\fDidRecTwoSE) & & & \\
Nonbank $\times$ 2022 & $-$\fDidRate{} (\fDidRateSE) & & & \\
Nonbank $\times$ Mar--Jun 2019 & & & & \fPlac\\
 & & & & (\fPlacSE)\\
 & & & & [\fPlacP]\\\midrule
Fixed effects & Issuer, & Issuer, & Issuer, & Issuer,\\
 & state$\times$month & pool$\times$month & state$\times$month & state$\times$month\\
Window & 3/19--9/22 & 3/19--6/20 & 7/19--6/20 & 3/18--6/19\\
Options & \fDidN & \fPoolN & \fFlatN & \fPlacN\\
Issuers & \fDidG & \fPoolG & \fFlatG & \fPlacG\\\bottomrule
\end{tabular}
\begin{minipage}{0.97\textwidth}\footnotesize\setstretch{1.0}\smallskip
Linear probability models of exercise within three months of vesting, in percentage points, estimated on cell counts (weighted least squares on issuer-by-state-by-month or issuer-by-pool-by-month cells, weighted by options, which reproduces the loan-level estimates). Column 2 uses pool-months of multi-issuer pools that contain loans serviced by both types. Standard errors clustered by issuer in parentheses; wild cluster bootstrap $p$-values (Rademacher weights, null imposed, 999 draws; 499 in column 2) in brackets.
\end{minipage}
\end{table}

\subsection{Borrower characteristics from HMDA}\label{ia:hmda}
The Ginnie Mae disclosures do not report borrower income or race, and in the files for 2018 through 2022 the metropolitan-area field is blank, so a loan's location is known only by state. I attach borrower characteristics from the public Home Mortgage Disclosure Act loan files for 2018 through 2021, which report the note rate, term, and combined loan-to-value ratio. A Ginnie Mae loan originated in those years is paired with every originated first-lien FHA, VA, or RHS loan in HMDA with the same state, origination year, loan type, \$10,000 loan-amount interval, note rate, and term. A pair is kept if the combined loan-to-value ratios agree to 0.01, the loan purpose agrees, both records have one borrower or both have more than one, and the debt-to-income ratio, taken from Ginnie Mae's monthly new-issue loan-level file, falls in the interval or within one point of the integer that HMDA reports. The debt-to-income ratio is available for 81\% of the loans, and the condition is not applied when either source lacks it.

The public fields do not identify loans uniquely. Between 29\% and 40\% of loans, depending on the origination year, have at least one HMDA record that satisfies the conditions, the median number of such records is one, and 14\% to 15\% of loans are matched one to one. I therefore assign each loan the mean of each characteristic over its candidate records, and Table~\ref{tab:hmda} repeats the main estimate on the one-to-one subsample. The merge covers 11\% of nonbank options in the year before the shock and 15\% in the freeze (4\% and 6\% for depositories), because most delinquent loans in those months were originated before 2018, when HMDA did not report the fields used here. Among the loans covered, nonbank and depository borrowers are similar: income is 1.02 times the area median at both types, 41\% to 42\% have low or moderate income, 20\% and 18\% are Black, and 21\% and 17\% are Hispanic.

\begin{table}[H]\centering
\caption{Exercise in the Freeze by Borrower Characteristics\label{tab:hmda}}
\setstretch{1.0}\footnotesize\setlength{\tabcolsep}{3.5pt}
\begin{tabular}{lcccc}\toprule
\multicolumn{5}{l}{\emph{Panel A. Share of vested options exercised within three months (\%)}}\\
 & \multicolumn{2}{c}{Nonbank} & \multicolumn{2}{c}{Depository}\\
 & Before & Freeze & Before & Freeze\\\midrule
Income, lowest third & 42.6 & 5.4 & 64.8 & 75.3\\
Income, middle third & 41.6 & 5.0 & 61.6 & 72.3\\
Income, highest third & 37.2 & 4.6 & 57.8 & 69.1\\
Hispanic & 46.4 & 5.8 & 69.1 & 75.0\\
Not Hispanic & 38.6 & 4.5 & 60.1 & 71.0\\
Black & 39.5 & 4.9 & 61.2 & 69.9\\
Not Black & 40.2 & 4.8 & 61.7 & 72.1\\
Majority-minority tract & 42.8 & 5.1 & 66.3 & 72.7\\
Other tract & 38.6 & 4.6 & 59.6 & 71.0\\\midrule
\multicolumn{5}{l}{\emph{Panel B. Nonbank relative to depository (percentage points)}}\\
 & Before & Change in freeze & With balance & Options\\\midrule
Log income to area median (per s.d.) & 0.0 (0.3) & $-0.3$ (0.6) & $-0.4$ (1.0) & 77,666\\
Black & $-0.8$ (1.0) & $-0.4$ (1.2) & $-0.5$ (1.2) & 78,479\\
Hispanic & 12.6 (3.3) & $-10.6$ (3.3) & $-10.9$ (3.5) & 78,479\\
Hispanic, one-to-one matches & 15.1 (5.9) & $-13.7$ (6.1) & & 28,262\\
Majority-minority tract & 5.7 (2.3) & $-5.2$ (2.3) & $-5.5$ (2.5) & 82,681\\
Low- or moderate-income tract & 5.2 (1.6) & $-3.6$ (1.7) & $-3.7$ (1.4) & 82,596\\\bottomrule
\end{tabular}
\begin{minipage}{0.97\textwidth}\footnotesize\setstretch{1.0}\smallskip
The sample is vested options from January 2019 through June 2020 on loans originated in 2018 or later with at least one HMDA candidate record. Before is January 2019 through February 2020 and the freeze is March through June 2020. Income thirds are thirds of borrower income relative to the area median family income, set on the period before the shock. In Panel A, rates for race, ethnicity, and tract groups weight each option by the share of its candidate records in the group. Panel B reports, from a linear probability model with issuer-by-month effects, the coefficient on the characteristic interacted with a nonbank indicator (Before) and on the characteristic interacted with the nonbank indicator and the freeze (Change in freeze). With balance adds log unpaid balance with the same interactions; in the one-to-one row both columns include it. Standard errors, in parentheses, are clustered by issuer (113 issuers).
\end{minipage}
\end{table}

\subsection{A synthetic control for the type contrast}\label{ia:synth}
\begin{figure}[H]
\centering\includegraphics[width=0.72\textwidth]{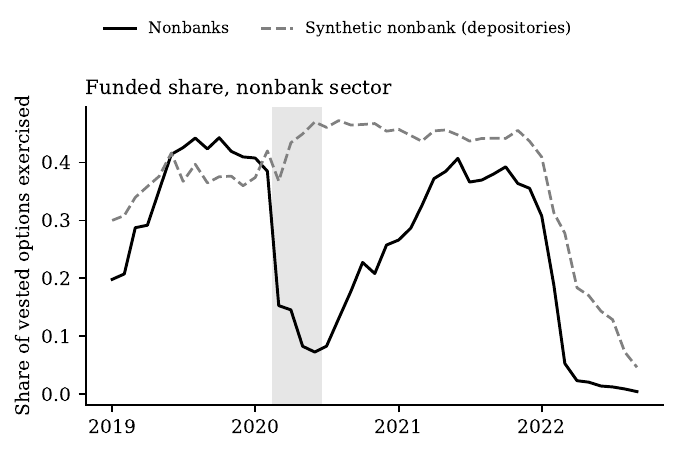}
\caption{Synthetic control for the nonbank sector's funded share.\label{fig:synth}}
\begin{minipage}{0.97\textwidth}\footnotesize\setstretch{1.0}
Monthly share of vested options exercised within three months by nonbank issuers (solid) and by a synthetic nonbank formed from depository issuers with nonnegative weights summing to one, chosen to match the nonbank series from January 2019 to February 2020 (dashed). Shading marks March--June 2020.
\end{minipage}
\end{figure}

The comparison between issuer types in Section~\ref{sec:fixedrank} can be made without assuming that the two types would have followed parallel paths. Following \citet{AbadieDiamondHainmueller2010}, Figure~\ref{fig:synth} compares the monthly funded share of the nonbank sector with a weighted combination of the \eSynDonorsM{} depositories with a consistently reported series, the weights chosen to match the nonbank series from January 2019 to February 2020. The synthetic series matches the pre-shock mean (\eSynMSecPreAct\% against \eSynMSecPreSyn\%, root mean squared error \eSynMSecRmspe{} points) and then diverges: the gap is \eSynMSecGapFrz{} points in the freeze, \eSynMSecGapRel{} in the release, \eSynMSecGapYone{} in 2021, and \eSynMSecGapYtwo{} in 2022. For the five largest nonbanks the freeze gaps are \eSynMAGapFrz, \eSynMBGapFrz, \eSynMCGapFrz, \eSynMDGapFrz, and \eSynMEGapFrz{} points. Inference is weak by construction. With \eSynMPlacebos{} donors, the smallest attainable placebo $p$-value is 0.05, and the one-sided placebo $p$-value for the sector's freeze gap is \eSynMSecP, because individual depositories' funded shares are volatile enough that several placebo gaps are as large; the value of the exercise is that the size and timing of the gap do not depend on a parallel-trends assumption. The gap in 2022 mixes the value shock with nonbanks' higher funding costs (Section~\ref{sec:unconstrained}).

\subsection{The monthly eligible stock}
The vesting-level test treats the decision at vesting. An issuer can exercise the option in any month in which the loan remains at least three payments delinquent and in its pool. Table~\ref{tab:stock} repeats the test on the \eStockN{} loan-months of that stock from January 2019 to September 2022, with the monthly exercise rate as the funded share and cells defined by issuer, month, and current forbearance status. Because nonbanks' monthly exercise rate is low in every period, the two benchmarks nearly coincide for them, and the table is informative about the ranking weight rather than about the rationing rule. For nonbanks the actual responses are close to the fixed-ranking computation except in the release, when the response and the latent weight both rise; for depositories the responses exceed the computation from 2021, and the weight rises in 2022. The index for a loan is its index at vesting, so for loans that vested years earlier the spread is measured against an older mortgage rate.

\begin{table}[H]\centering
\caption{The Fixed-Ranking Test on the Monthly Eligible Stock\label{tab:stock}}
\scriptsize
\setlength{\tabcolsep}{3.5pt}
\begin{tabular}{lcccccc}\toprule
 & Pre-shock & Freeze & Release & \multicolumn{2}{c}{Recovery} & Rate rise\\\cmidrule(lr){5-6}
 & 3/19--2/20 & 3/20--6/20 & 7/20--12/20 & 1/21--6/21 & 7/21--12/21 & 1/22--9/22\\\midrule
\StockRows\bottomrule
\end{tabular}
\begin{minipage}{0.97\textwidth}\footnotesize\setstretch{1.0}\smallskip
Every loan-month in which a loan is at least three payments delinquent and still in its pool, at depository and nonbank issuers. The ranking index is the loan's index at vesting. Cells are issuer by month by current forbearance status; the latent weight is estimated with cell fixed effects and period interactions, with standard errors clustered by issuer. Other definitions are as in Table~\ref{tab:empirics}.
\end{minipage}
\end{table}

\subsection{Limitations}
The index measures the ranking on hard information, and the fixed-ranking prediction assumes that the part of issuers' ranking it misses is logistic and stable. The index is estimated on 2016--2018 decisions and ranks later decisions less well as the loan population and the redelivery rules change. The freeze comparison is dominated by a few large issuers (Table~\ref{tab:issuers}), and a bootstrap over \eNbBootIssuers{} nonbank issuer clusters may understate uncertainty under such concentration; the freeze alone does not separate the fixed ranking from random rationing. All standard errors condition on the estimated index. The disclosures do not follow loans after they leave a pool, so outcomes after repurchase are not observed. The forbearance flag begins in May 2020. Issuer type classifies funding conditions but does not measure them. The comparisons across periods are not estimates of a causal effect of funding; they test whether the choices of constrained and unconstrained issuers are consistent with the model's rationing rule and its distinction between liquidity and value shocks, against the alternatives described in Section~\ref{sec:fixedrank}.

\section{Research Design for the Remaining Tests}\label{ia:design}
The tests that public disclosures cannot support require dated issuer funding variation, outcomes on both sides of the decision, and pre-decision borrower signals. This section sets out the design.

\paragraph{Risk set and outcomes.} The unit of analysis is a loan at the date an action becomes available, such as the vesting of the buyout option. For any funding event, the primary population is a cohort frozen on information dated before the event, so that the population is not conditioned on delinquency caused by the same shock. Outcomes are measured at fixed horizons: delivered modification, sustained performance, completed foreclosure, and redelivery. Loss of observation is recorded as such, and disappearance from a pool is not treated as default or cure. Internet Appendix~\ref{ia:data} gives the construction protocol.

\paragraph{Value and cash.} Empirical analogues of $g$ and $\ell$ use only information available at the decision: the difference in expected discounted cash flows between action and continuation, and current cash required net of immediately drawable financing. Expected recovery, holding period, funding rates, and the redelivery regime in force at the decision date enter separately. A risk score estimated on historical outcomes, $\hat q^{\text{obs}}_{it}=\widehat{\Pr}(Y_{i,t+h}=1\mid H_{it},S_{it})$, predicts outcomes under the historical policy. It is neither the success probability under intervention nor the treatment effect without further assumptions, and the tests below use it only as a ranking variable. Text or other soft signals must be time-stamped before the decision.

\paragraph{Funding variation and the information interaction.} Propositions~\ref{prop:info} and \ref{prop:reliance} imply that a funding shock changes the use of prior information with a sign that depends on the regime. With issuer funding variation $Z_{jt}$ and a prior signal $S_{i,t-1}$,
\begin{equation}\label{eq:allocation_emp}
\begin{split}
y_{ijt}={}&\alpha_j+\phi_{c(i),t}+\pi Z_{jt}+\gamma S_{i,t-1}+\theta\, Z_{jt}S_{i,t-1}+\theta_2\,Z_{jt}S_{i,t-1}\1\{m_{j0}>\hat m^*\}\\
&+f(H_{i,t-1},\hat g_{i,t-1},\hat\ell_{i,t-1})+\varepsilon_{ijt},
\end{split}
\end{equation}
where $y_{ijt}$ indicates that issuer $j$ acted on loan $i$ in month $t$, $\phi_{c(i),t}$ are risk-cell-by-month effects, $m_{j0}$ is the issuer's pre-event funded share, and $\hat m^*$ is the share of loans in the cell whose prior signal exceeds the cell mean, the switch point in Proposition~\ref{prop:reliance}(i). The model predicts $\theta>0$ and $\theta+\theta_2<0$: more funding raises reliance on information for issuers below the switch point and lowers it above. The prediction is a sign reversal at a location fixed in advance, which is harder to produce by specification search than a single interaction. Credible $Z_{jt}$ must shift usable funding without shifting loan values. Candidates include disruptions at an issuer's pre-existing facility lenders and mechanically determined facility limits, and each requires a verified first stage and an exclusion argument. The dated value shocks of Section~\ref{sec:unconstrained} supply the complementary comparison that does not require an excluded funding shock.

\paragraph{A within-nonbank funding shock: the FHFA advance cap.} The cross-issuer designs based on advances and pandemic exposure do not isolate a strong issuer-specific funding component (Section~\ref{ia:funding}). One dated shock varied across nonbanks for reasons unrelated to the value of repurchasing Ginnie Mae loans. On April 21, 2020, the FHFA limited servicers' obligation to advance principal and interest on GSE loans in forbearance to four months; Ginnie Mae imposed no such limit. A nonbank with a large GSE book therefore faced a smaller cash drain from mid-2020 onward, once its GSE forbearances passed four months, than a Ginnie-only nonbank, and the cap changed the value of no Ginnie Mae buyout option.

I measure exposure from the public disclosures. Fannie Mae's and Freddie Mac's monthly loan-level MBS disclosure files for December 2019 report the servicer of every loan in their pools; summing current balances by servicer and matching legal entities to the Ginnie Mae issuer file gives, for issuer $j$, $Z_j=G_j/(G_j+N_j)$, its December 2019 GSE balance relative to its GSE and Ginnie Mae balance. Where an issuer's GSE servicing sits with an affiliate under common ownership (PennyMac, New Residential), the affiliate's balance is included, since cash is fungible within the firm. Across the \eFhfaNbIssuers{} nonbanks, $Z_j$ averages \eFhfaZmean{} with a standard deviation of \eFhfaZsd, so the design has variation to work with.

It finds no funding effect. In issuer-month regressions with issuer and month fixed effects, the coefficient on $Z_j\times\1\{t\ge\text{May 2020}\}$ is \eFhfaNbB{} percentage points (s.e.\ \eFhfaNbSE; wild-cluster bootstrap $p=\eFhfaNbP$) for nonbanks through 2021 and \eFhfaDepB{} (\eFhfaDepSE) for depositories. By quarter, relative to the first quarter of 2020, the nonbank coefficients are \eFhfaEvQtwo{} (\eFhfaEvQtwoSE) in the second quarter of 2020, \eFhfaEvQthree{} (\eFhfaEvQthreeSE) in the third, when the cap began to release cash, and \eFhfaEvQfour{} (\eFhfaEvQfourSE) in the fourth, and between \eFhfaEvTwoOneLo{} and \eFhfaEvTwoOneHi{} in the quarters of 2021, against pre-shock coefficients close to zero. The sign after 2020 is the opposite of the one the cap predicts, and its timing, in 2021 when funding had recovered, points to business models rather than cash: the Ginnie Mae specialists, with low $Z_j$, are the servicers that expanded buyouts as redelivery became profitable in 2021. The issuer-level change in the weight on balance is unrelated to $Z_j$ (\eFhfaTiltB{}, permutation $p=\eFhfaTiltP$, \eFhfaTiltN{} issuers with enough decisions). The cap was small relative to the shock it addressed, it began to bind only in the second half of 2020, and exposure to it is entangled with specialization in Ginnie Mae servicing. The design that would separate them requires facility-level data on the terms of nonbanks' servicing-advance financing.

\paragraph{Total effects, not instrumented buyouts.} A funding shock changes several servicing margins at once, including buyout, modification, and liquidation timing, so it is not a valid instrument for any one of them. The primary borrower-level estimand is the total effect of the funding shock on the frozen cohort. Because of displacement within the intermediary, effects are reported across the whole book, weighted both by loan count and by balance. The disclosures follow a loan only while it is in a pool, and Ginnie Mae's terms of use permit analysis of its MBS, so the outcomes observed here are outcomes of loans in pools; outcomes after repurchase require servicer records.

\paragraph{Model evaluation.} Testing Proposition~\ref{prop:auc} requires outcomes for loans on both sides of the decision. The monthly files record whether loans left in pools later cure, prepay, or reach a claim. With outcomes for repurchased loans from servicer records, or bounds on them, scores trained on earlier cohorts can be evaluated at each issuer's realized funded share and their ranking compared with the AUC ranking. Where accuracy profiles cross, the model predicts that issuers with different funded shares should prefer different models.

\paragraph{Inference.} Standard errors should be clustered at the level at which funding varies. With few clusters, inference should use the wild cluster bootstrap \citep{CameronGelbachMiller2008}, and where parallel trends are invoked, sensitivity should follow \citet{RambachanRoth2023} using the full covariance matrix of the event-study coefficients.

\begin{table}[H]\centering
\caption{Further Implications Not Tested Here\label{tab:predictions_untested}}
\scriptsize\setstretch{1.0}
\begin{tabularx}{\textwidth}{>{\raggedright\arraybackslash}p{.23\textwidth}>{\raggedright\arraybackslash}p{.32\textwidth}Y}\toprule
Prediction & Observable implication & What a test requires\\\midrule
6. Model rankings depend on liquidity (Prop.~\ref{prop:auc}) & Out-of-sample value of scores at issuer-specific funded shares departs from the AUC ranking where accuracy profiles cross & Untested; requires outcomes on both sides of the decision\\
7. Recognized information expands funding (Prop.~\ref{prop:finance}) & Facility advance rates respond to verifiable model adoption or data sharing, not to internal adoption alone & Untested\\
8. Private selection can lower borrower surplus (Prop.~\ref{prop:welfare}) & Borrower gains from relief decline in the intermediary's predicted recovery & Self-cure evidence \citep{AdelinoGerardiWillen2013}; direct test requires borrower outcomes\\
9. Ex post compensation is inframarginal under binding funding (Prop.~\ref{prop:policy}) & Per-action incentives leave the cash committed to relief unchanged at constrained servicers, shift relief toward cheaper actions, and raise facility draws where funding is elastic & Staff capacity, the analogous constraint in Section~\ref{ia:menus}, limited HAMP modifications \citep{Agarwal2017}; untested\\\bottomrule
\end{tabularx}
\begin{minipage}{0.97\textwidth}\footnotesize\setstretch{1.0}\smallskip
The four implications of the model that the public disclosures cannot bear on; Table~\ref{tab:predictions} in the paper reports the five that they can. Proposition~\ref{prop:finance} is in Section~\ref{sec:information}; Propositions~\ref{prop:welfare} and \ref{prop:policy} are in Section~\ref{sec:welfare}.
\end{minipage}
\end{table}

\section{Numerical Verification}\label{ia:verify}
The script \texttt{code/model.py} generates Figures~\ref{fig:rationing}--\ref{fig:scores} and Tables~\ref{tab:uniform} and \ref{tab:scores} and writes a verification report, \texttt{model\_verification.json}, to the \texttt{data} folder. It asserts the following.
\begin{enumerate}
\item \emph{Propositions~\ref{prop:allocation} and \ref{prop:selection}.} Seventeen discretized linear programs with 4,000 posterior types reproduce the closed-form private value to within $1.3\times10^{-8}$. The funded share rises, and the success rate among funded opportunities and the shadow value of cash fall, monotonically in liquidity.
\item \emph{Proposition~\ref{prop:info}.} For posteriors distributed Beta(1.5,\,1) with $R=1$ and $c=0.35$, the decomposition into ranking and screening value holds to machine precision on a grid of 200,000 quantiles. $\Delta V$ is concave and peaks at $m=0.535=\Pr(Q>p)$ with value $0.1115=R\,\E\pos{Q-p}=\tfrac R2\E|Q-p|$, and $\Delta V(1)=\E\pos{c-RQ}=0.0290$.
\item \emph{Proposition~\ref{prop:reliance}.} With every action profitable, the slope of a linear probability regression of intervention on the posterior equals $\mathcal A(m)/(R\,\mathrm{Var}(Q))$ at three funded shares, the level slope peaks at $m^*$, and the slope relative to the base rate is nonincreasing in the funded share.
\item \emph{Corollary~\ref{cor:binary}.} The binary example peaks at 0.15 at $m=0.5$, vanishes at both endpoints, and exceeds an adoption cost of 0.06 exactly on $(0.2,0.8)$.
\item \emph{Proposition~\ref{prop:auc}.} The identity (\ref{eq:aucid}) holds for the two scores of Table~\ref{tab:scores} and for 200 randomly drawn discrete score distributions, with maximum absolute error $2.6\times10^{-10}$. A mean-preserving contraction of score B is dominated at every fundable share.
\item \emph{Proposition~\ref{prop:finance}.} With $k=1$, $L=0.6$, $K_0=0.2$, and score B, the fundable share is 0.286 under hard-information pricing and 0.385 under recognition; private value is 0.114 without the signal, 0.149 with it under hard-information pricing, and 0.269 under recognition.
\item \emph{Proposition~\ref{prop:welfare}.} In the example of Section~\ref{sec:welfare}, private value rises by 0.15, borrower surplus falls by 0.20, and total surplus falls by 0.05, while a planner using the same signal attains total surplus 0.50.
\item \emph{Proposition~\ref{prop:policy}.} In a linear program with 3,000 opportunities and heterogeneous cash needs, a flat ex post payment leaves cash deployed unchanged and raises the number of actions from 0.343 to 0.372; a payment proportional to cash need leaves the allocation unchanged; with a constant cash need, the payment leaves the number of actions unchanged. Under quadratic external-funding costs, a payment and an injection that produce the same funded share move external borrowing in opposite directions.
\end{enumerate}
These checks verify the stated numerical examples and their implementation; numerical checks are not substitutes for proofs or empirical replication. Separately, the scripts in \texttt{code/empirics} rebuild Figures~\ref{fig:tilt}, \ref{fig:empirics}, \ref{fig:queue}, and \ref{fig:shape}, Tables~\ref{tab:empirics}--\ref{tab:forbearance}, Tables~\ref{tab:robust}--\ref{tab:stock}, and every number in Sections~\ref{sec:data}--\ref{sec:forbearance}, including the index AUCs, from the source files.